\documentclass[reprint,amssymb, amsmath, aip,cha,singlecolumn]{revtex4-1}
\usepackage{graphicx}
\usepackage[caption = false]{subfig}
\usepackage{color}
\usepackage{epsfig}
\usepackage{ulem} 
\usepackage{ifpdf}
\usepackage{url}%
\usepackage{bm}
\usepackage{comment}
\usepackage[colorlinks=true,linkcolor=blue]{hyperref}%
\expandafter\ifx\csname package@font\endcsname\relax\else
 \expandafter\expandafter
 \expandafter\usepackage
 \expandafter\expandafter
 \expandafter{\csname package@font\endcsname}%
\fi

\begin{document}

\title{Review  On Laser Induced Breakdown spectroscopy: Methodology and Technical Developments}
\author{Jinto Thomas}
\email{jinto@ipr.res.in}
\affiliation{Institute for Plasma Research, Bhat, Gandhinagar,Gujarat, India, 382428}%
\affiliation{Homi Bhabha National Institute, Training School Complex, Anushaktinagar, Mumbai 400094, India}%
\author{Hem Chandra Joshi}
\email{hem_sup@yahoo.co.uk}
\affiliation{Institute for Plasma Research, Bhat, Gandhinagar,Gujarat, India, 382428}%
\affiliation{Homi Bhabha National Institute, Training School Complex, Anushaktinagar, Mumbai 400094, India}%

\date{\today}
\begin{abstract}
In this review we attempt to provide  a brief account of laser induced breakdown spectroscopy (LIBS) methodology and technological developments. We also summarise various methods adopted for exploiting LIBS. Besides, a brief overview of combination of LIBS in conjunction with other methods is also given.
\end{abstract}
\keywords{LIBS, Spectroscopy, Laser plasma}
\maketitle

\tableofcontents
\section{Introduction}\label{sec:intro}
In the recent years laser induced breakdown spectroscopy (LIBS) has found numerous applications encompassing various field\cite{Berlo2022}.  Simultaneously various LIBS techniques also emerged for better exploitation and interpretation of obtained data \cite{Fu_2020_Front,Daniel_2021_ASR}. New techniques like hand held and standoff LIBS have been developed and used\cite{SENESI2021106013}. Optimization of time window has been pointed out in ref. \cite{C8JA00415C} Exploiting delayed emission for LIBS indentation has also been suggested \cite{Garima_Jaas_2022}. \par 

From time to time review articles covering various aspects of LIBS have come up in the literature\cite{Hahn_2012_Appl_Spec, Hahn_2010_Appl_Spec, pasquini2007laser_JBC, singh2020laser, singh2007laser, miziolek2006laser,PabloAF,noll2012laser,ASR_Review_LIBS_1,ASR_LIBS_Application1, ASR_LIBS_Review_2,ASR_Tutorial_review_2020}.
 However they are focused on particular aspect or technique e.g. study of uranium containing compounds \cite{KAUTZ2021106283},  
  plasma facing components \cite{MAURYA2020152417,li2016review}, LIBS imaging \cite{JOLIVET201941}  element analysis of industrial materials\cite{app11199274} and  data analysis\cite{Dianxin_ApplSpecReview}, nano particles in LIBS \cite{D1JA00149C,DELLAGLIO2018105}, industrial applications \cite{C9AY02728A}, 
  underwater applications \cite{Ayumu_underwater}, food analysis\cite{molecules_libs_food} ,optical diagnostics and laser produced plasma\cite{Kautz:21,Hari_uranium}, cancer diagnostics and classification\cite{LIBS_Cancer}, 
  rapid CPVID detection \cite{Berlo2022} (Scientific reports 12, 2022, 1614) aerosol analysis\cite{Huview_ASR_2021} and geological samples \cite{Shujun_ASR_2015}. 
 Hand held and portable LIBS technique is reviewed in reference \cite{SENESI2021106013} . Technique used in LIBS quantification have also progressed \cite{FP_2022_Zhang} in the past. Ultra sensitive and multianalyte analysis of plasma plumes using laser induced fluorescence (LIF) has been reviewed in a recent article\cite{CHEUNG2022106473} .  Combination of other techniques e.g. 
 FTIR, Raman and Hyper spectral imaging (HI) have been elucidated  to gather detailed spatial information
 \cite{Ribeiro_2020_AO}
 \cite{HOLUB2022106487}
\cite{SUN2022106456} 
 \cite{D2JA00147K}
Hybrid LIBS-Raman-LIF been discussed in a recent review \cite{Dhanada_2021_ASR}.
\par
Despite these reports, efforts are continuously emerging to extend the LIBS detection range and its application in various platforms e.g. study of deposition on the tokamak first wall components\cite{LIBS_Nucl_Fusion2021} , analysis of hydrogen isotopes\cite{Kautz:21} hardness estimation etc. In this short review we attempt to briefly sketch salient features associated with LIBS which encompass phenomenological aspects to emerging applications and recent technical developments. Some new features e.g. colliding plasma, application of self-reversal in estimating isotopic abundance, filament induced LIBS and grating induced LIBS are also briefly covered. The overall goal of  this article to provide with first hand information to the LIBS community. We describe it in the following sections\par
\section{LIBS Methodology}\label{sec:Methodoloy}
\begin{figure}
\centering
\includegraphics[width=0.49\textwidth, trim=0cm 0cm 0cm 0cm, clip=true,angle=0]{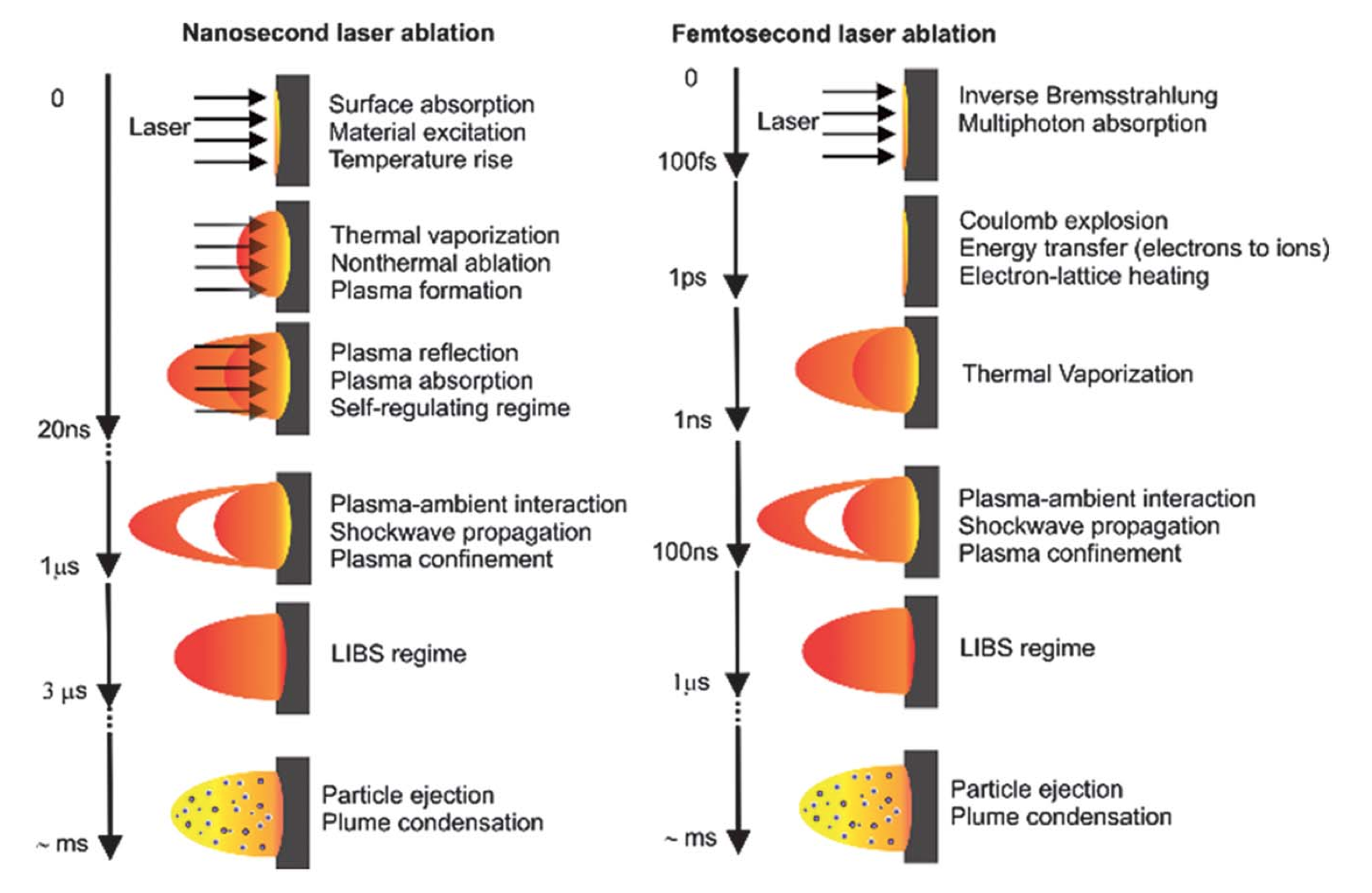}
\caption{\label{fig:Libs_Time_scale_ns_fs}  Approximate time scales of nanosecond and femtosecond energy absorption and laser ablation along with various processes happening during and after the laser pulse (adapted with permission from reference \cite{LIBS_Time_scale}).}
\end{figure}

When a substance is irradiated with a high power laser, the material is heated and results in melt which finally forms a plasma plume which consists of atom, ions and electrons. The mechanism and timescales of ablation process and subsequent plasma formation for short (fs) and long (ns) pulses are demonstrated in Fig~\ref{fig:Libs_Time_scale_ns_fs}. In ns-LIBS, at initial times in plasma formation the primary mechanisms are thermal vaporization and non-thermal evaporation whereas in case of fs LIBS, thermal evaporation occurs after Coulomb explosion, electron-ion energy transfer and heating of lattice by electrons which are followed by thermal vaporization. However, at later stages in case of ns LIBS, plasma reflection, absorption and reflection can occur whereas later stages for both ns and fs LIBS comprise of plasma-ambient interaction, shock wave propagation and confinement. Finally both are characterized by LIBS regime followed by plume condensation and particle ejection.
Plume hydrodynamics has been found to play an important role in molecular and nano-cluster formation. In the early stages, shock wave at the edge has been found to hinder molecular formation which takes place only after the shock wave collapse\cite{Hari_Anals_Chem_2016} . 
Regarding expansion behavior, a systematic study is carried out for fs, ps and ns ablations\cite{jinto_PhsD}. Differences in the propagation of the plume are observed for these three cases. For forward ablation of nickel thin film, in case of fs ablation, linear expansion is noticed for low background pressures which eventually shows shock wave like expansion at higher background pressures. For ps ablation, blast wave model describes the expansion at low pressures but a drag model appears appropriate for higher pressures. For fs ablation, effect of laser fluence on the emission characteristics in ultrafast laser produced copper was reported by Anoop et.al \cite{Anoop_JAP_2016}. At low to moderate fluences, neutral emission dominates but at higher fluences, ionic emission is predominant. Fast and slow components are also noticed in case of Zn (I)
 481 nm emission in ultrafast laser produced zinc plasma which are ascribed to neutral and recombination contributions to the emission \cite{Smijesh_JAP_2013} 
\subsection{Ablation types (Front ablation and Back ablation)}
\begin{figure}
\centering
\includegraphics[width=0.49\textwidth, trim=0cm 0cm 0cm 0cm, clip=true,angle=0]{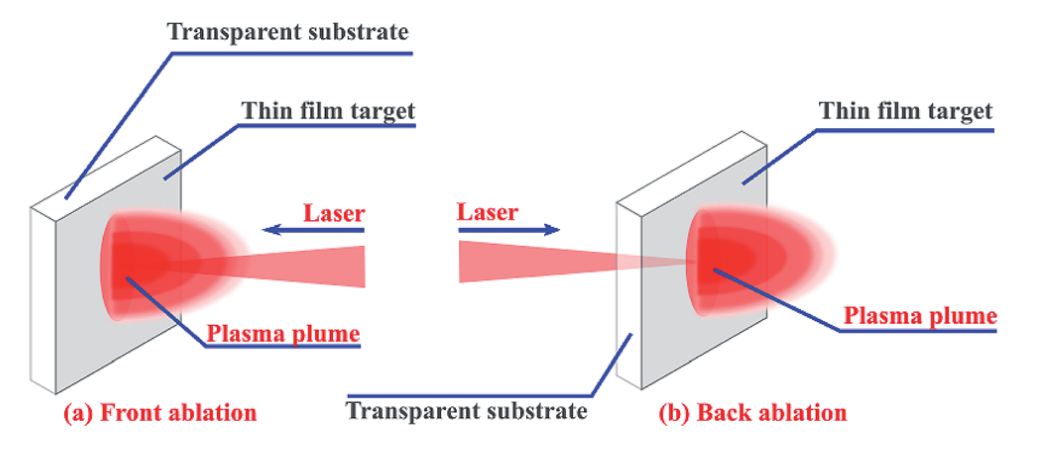}
\caption{\label{fig:Fron_rear}  Schematic diagram of laser produced plasma plume of thin film deposited on a transparent substrate in front ablation (FA) and back ablation (BA) geometries (adapted with permission from reference \cite{C9JA00158A}).
}
\end{figure}

\begin{figure}
\centering
\includegraphics[width=0.49\textwidth, trim=0cm 0cm 0cm 0cm, clip=true,angle=0]{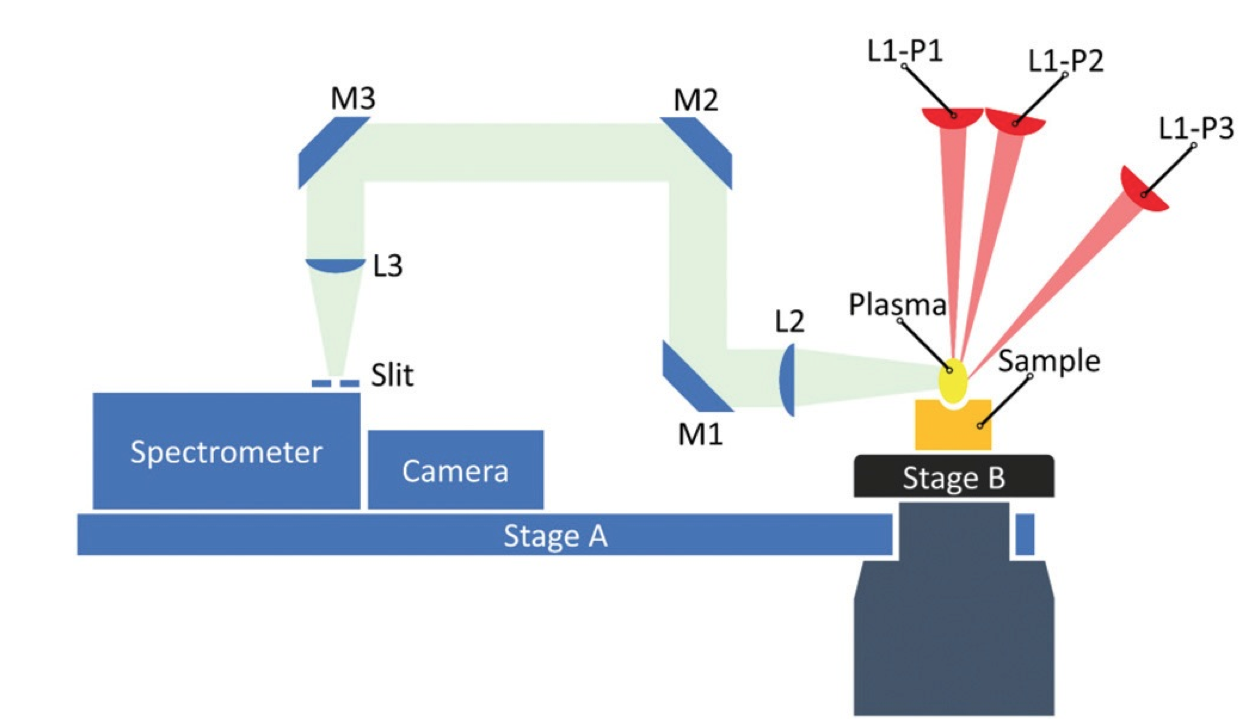}
\caption{\label{fig:Scheme_Tomo}  Schematic diagram of the tomographic system for demonstrating the effect of non-orthogonal ablation (adapted with permission from reference \cite{D1AN01292D}).
}
\end{figure}

In LIBS, basically the material can be ablated in three configurations viz. front, back and non-orthogonal ablation. Front and rear ablation geometries for thin film target are shown in Fig~\ref{fig:Fron_rear} and Fig~\ref{fig:Scheme_Tomo}. When laser is incident from the front side, it is termed as front ablation. On the other hand, when the laser is incident from backside, it is named as back or rear ablation. Notable differences in plume expansion geometry, composition of the plume and plume velocity are noticed. Higher velocity is obtained in case of front ablation as compared to the rear ablation\cite{Rear_Front_Alam}.   Moreover, spherical shock wave front is observed for both the cases, however, front side ablation has been found to have more excited state species as compared to rear ablation\cite{ESCOBAR2002} . Further, neutral species dominate in the rear 
ablation geometry\cite{C9JA00158A}. 

Besides these two geometries, laser ablation for different incident angles of laser beam has also been studied. Non orthogonal ablation has been found to increase inhomogeneity. It has been found to be composed of two parts; one following the ablation pulse and the other expanding along the sample normal. Moreover, the temporal evolution of the plasma, ionic and neutral emission and electron density and temperature have ben found to exhibit similar trends.
\subsection{Single pulse(SP)}
\begin{figure}
\centering
\includegraphics[width=0.49\textwidth, trim=0cm 0cm 0cm 0cm, clip=true,angle=0]{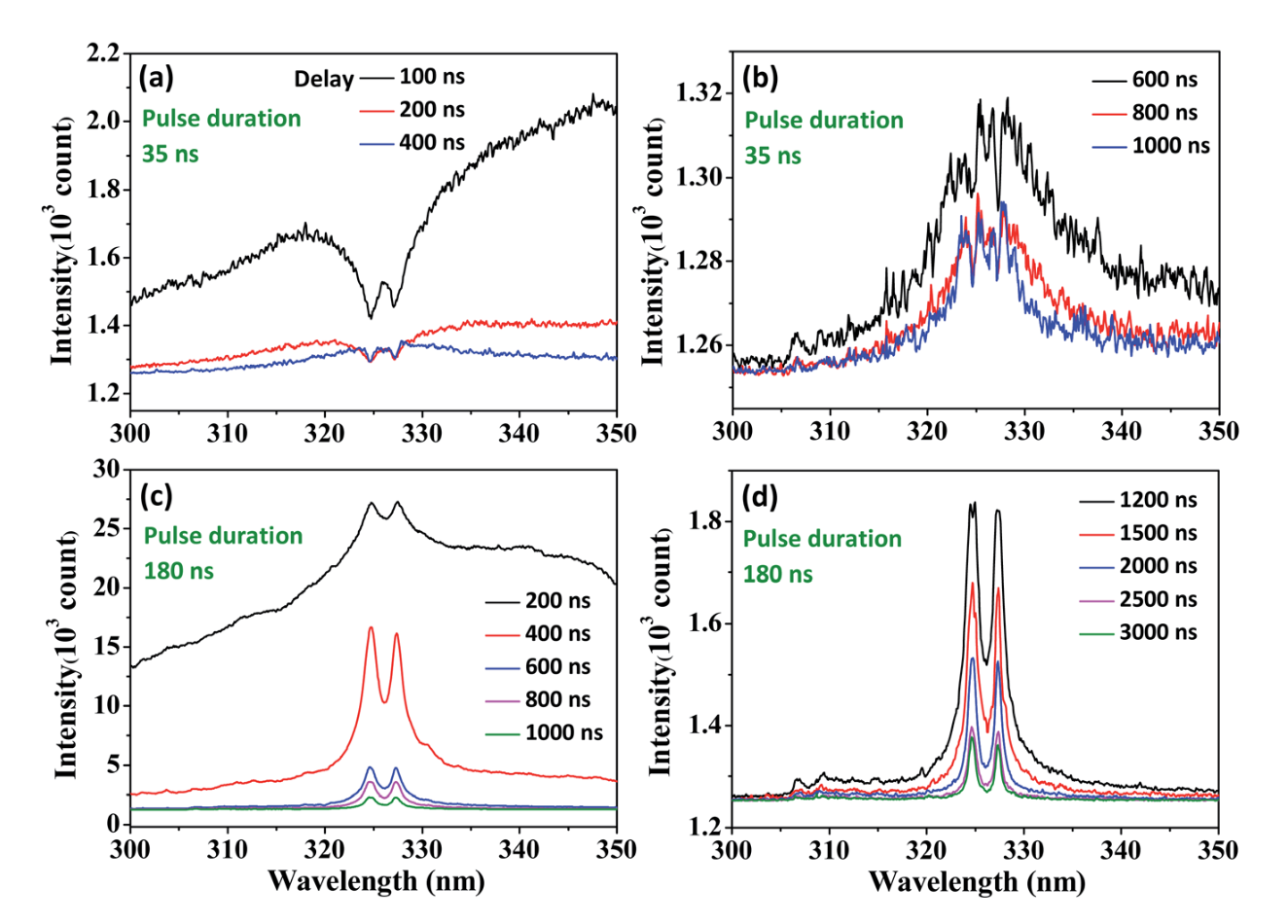}
\caption{\label{fig:LIBS_Cu}  Typical LIBS spectra of the atomic lines of Cu at varied delay times obtained at two pulse durations of 35ns  (a and b) and 180 ns (c and d) (adapted with permission from reference \cite{D1JA00151E} )   .
}
\end{figure}

In most of the LIBS experiments single pulsed laser is used for ablation. The pulse of the laser can vary from femtosecond to microsecond time\cite{D0JA00521E}. Enhancement in intensity with long ns pulses has been reported in submerged solids\cite{D0JA00521E} . Moreover, long ns pulses have been found to produce plasma with stronger emission and longer lifetime as compared to LIBS using short (35 ns) pulses. Figure~\ref{fig:LIBS_Cu} shows the LIBS studies using short (35 ns) and and long (180 ns) laser pulses for copper \cite{D1JA00151E}.

\subsection{Double pulse (DP)}

\begin{figure}
\centering
\includegraphics[width=0.49\textwidth, trim=0cm 0cm 0cm 0cm, clip=true,angle=0]{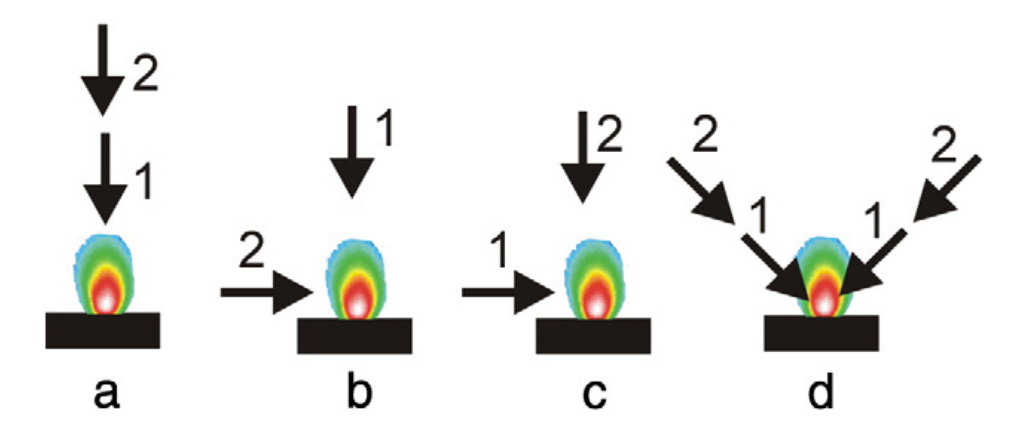}
\caption{\label{fig:DP_Geometries}  Various DP configurations (a) Collinear DP,   (b) Orthogonal reheating DP (c) orthogonal pre ablation DP  and (d)  dual pulsed cross beam (adapted with permission from reference \cite{DIWAKAR201365}).
}
\end{figure}
Double pulse (DP) LIBS has been found to enhance intensity of the atomic/ionic lines depending on inter pulse delay\cite{Wang_POP_2020} . A simple schematic of various configurations of DP LIBS is shown in Fig~\ref{fig:DP_Geometries}. The DP configuration can be collinear (a), orthogonal reheating (b) orthogonal pre ablation (c) or dual laser cross beam (d).
Further, DP LIBS can also have different configurations depending on lasers. It can have nano second + nanosecond (ns+ns), femtosecond+ femtosecond (fs+fs) femtosecond+ nanosecond (fs+ns) or nanosecond+ femtosecond (ns+fs) configurations. 
It has been found that the spectral intensity of copper plasma is higher in case of the configuration fs+ns \cite{Wang_POP_2020}.  The plasma temperature has been found to be lower whereas electron density is higher. It has been suggested that the second pulse re-excites the plasma resulting in enhanced spectral intensity.
Intensity enhancement has also been reported for orthogonal fs+fs DP LIBS. \cite{Nikolaos_Applied_Spectro} and is projected for LIBS imaging with better spatial resolution and spectro-chemical sensitivity. Effect of inverse Bremsstrahlung is reported for DP LIBS \cite{Siva_POP_2014}. In a recent work, effect of  DP ablation on the emission characteristics of plasma has been theoretically treated using hydrodynamic model \cite{D2JA00105E}. Enhancement in intensity is attributed to two mechanisms viz plasma-plasma coupling effect and pressure effect.
\par
\section{colliding Plasma}

\begin{figure}
\centering
\includegraphics[width=0.49\textwidth, trim=0cm 0cm 0cm 0cm, clip=true,angle=0]{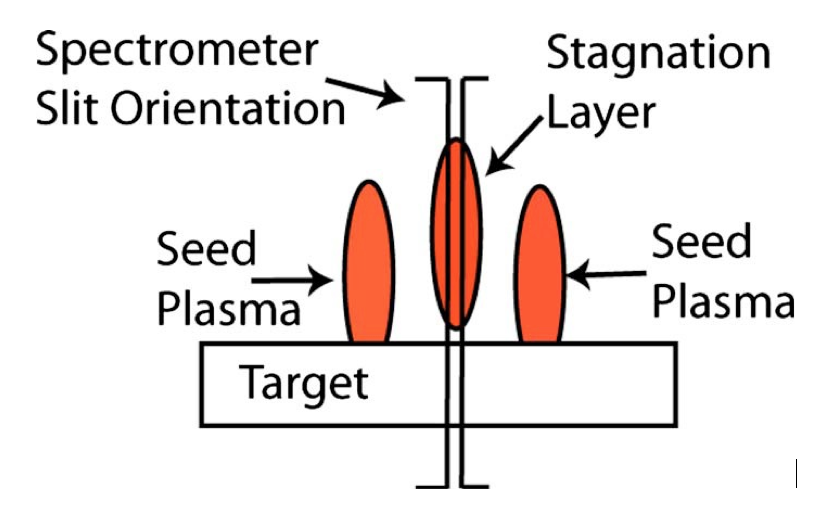}
\caption{\label{fig:Colliding_plasma_illustration}  Illustration of the orientation of the spectrometer slit with respect to the stagnation layer. This arrangement can provide one dimensional spatial resolution normal to the target along the stagnation layer (adapted with permission from reference \cite{Hough_JAP_2010}).
}
\end{figure}

When two laser produced plasmas (known as seed plasmas) are made to interact, an interaction zone is formed. This interaction zone is characterized by a stagnation layer. Colliding plasma schemes are useful in understanding plasma screening effects in fusion devices\cite{Shboul_POP_2014} . A simple setup for colliding plasma is shown in Fig~\ref{fig:Colliding_plasma_illustration}. The spectrometer slit is aligned along the propagation of stagnation layer.
However, the properties of the stagnation region depend on the relative orientation of the targets from which seed plasma are formed.  Two widely used target configurations- plane and wedge shaped for plasma collision studies. \par
Further, collisionality parameter
($\zeta =D/\lambda_{ii}$)  is defined to represent various scenarios where D is the separation between the two seed plasmas and $\lambda_{ii}$ is ion-ion mean free path defined by 

\begin{eqnarray}
\label{i-i_mean_freepath}
\lambda_{ii}(1\dashrightarrow2)=\frac{4\pi\epsilon_0^2m_i^2v_{12}^4}{e^4Z^4n_iln\Delta_{12}}
\end{eqnarray}

Collisionality parameter $\zeta >1$ indicates soft stagnation or interpenetration of the plumes species of seed plasma whereas $\zeta <1$indicates the condition of hard stagnation ( in this case collisions among the seed plasmas. will dominate and result in heated plasma.
As mean free path is highly dependent on $V_{12}$ and to a lesser extent on $m_i$, the collisionality condition can be engineered by the orientation of the seed plasmas with respect to each other and target material.
Further, the colliding plasma can be homogeneous or heterogeneous. If the seed plasmas are from the same element, it is called homogeneous whereas seed plasmas from different elements termed as heterogeneous
Signatures of enhanced neutral emission and molecular formation have been reported for the interaction zone \cite{Saxena_19}. Moreover properties of interaction zone are also modulated in the presence of magnetic field\cite{Alam_POP_2020}.  By observing the lines corresponding to trace elements, Tiwari et al \cite{TIWARI2022106411} showed that sensitivity can be optimized using colliding plasma in the presence of the magnetic field. In an other recent study Delaney et al \cite{DELANEY2022106430} studied the properties of stagnation layer formed in case of laterally colliding plasmas and annular plasmas and found that limit of detection (LOD) can be improved in case of colliding plasmas \cite{DELANEY2022106430}.
\section{Grating induced breakdown spectroscopy (GIBS)}
\begin{figure}
\centering
\includegraphics[width=0.49\textwidth, trim=0cm 0cm 0cm 0cm, clip=true,angle=0]{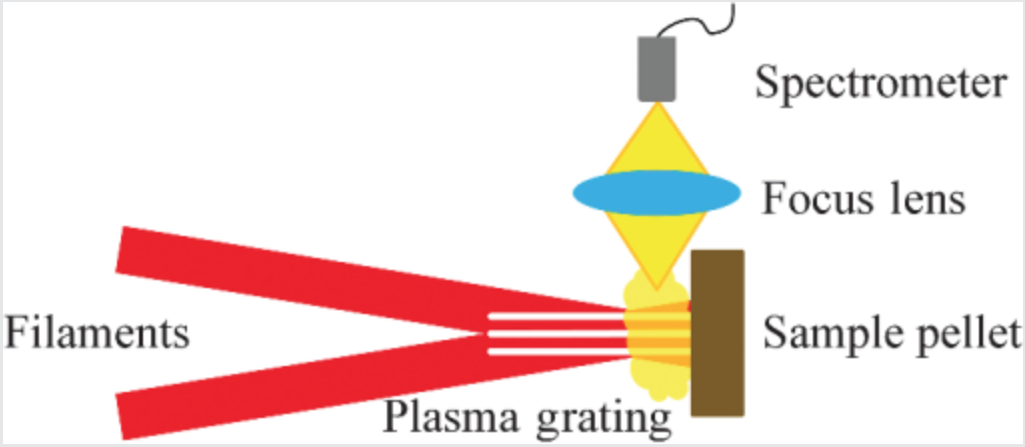}
\caption{\label{fig:GRAT_LIBS}  Experimental schematic diagram of GIBS (adapted with permission from reference \cite{GIBS_2020}).
}
\end{figure}

\begin{figure}
\centering
\includegraphics[width=0.49\textwidth, trim=0cm 0cm 0cm 0cm, clip=true,angle=0]{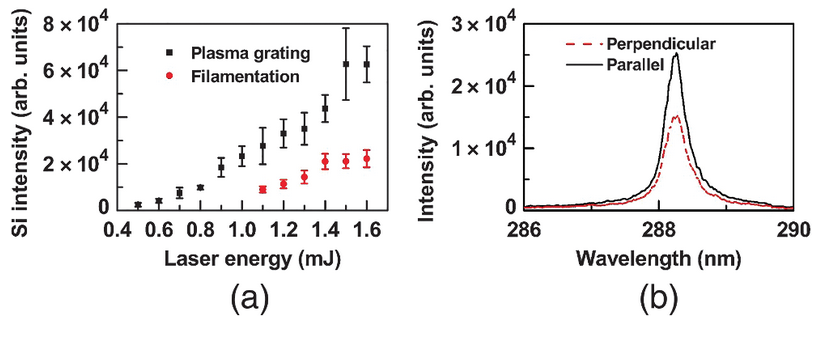}
\caption{\label{fig:FIBS_GIBS}  (a) Intensity of the Si 288.2 nm line as a function of the laser pulse energy detected with the FIBS and GIBS systems. (b) Intensity of the Si 288.2 nm line obtained by interaction of two beams with different polarizations. Intensity enhancement is evident in case of GIBS which also depends on polarization (adapted with permission from reference \cite{GIBS_2020} ).
}
\end{figure}

In conventional ns-LIBS technique,  plasma shielding affects its reproducibility, repeatability and signal to noise ratio. Interestingly in filament induced breakdown spectroscopy (FIBG) , remarkable property of the filaments to travel long distances independently of the diffraction limit makes it suitable for long range operation \cite{ROHWETTER20051025}. In case of FIBG, the problem of shielding is overcome but the power density profile is limited. These problems can be countered in plasma grating induced breakdown spectroscopy (GIBS)\cite{MOTTOROS2020329}. A simple illustration of FIBG is shown in Fig~\ref{fig:GRAT_LIBS}. Fig~\ref{fig:FIBS_GIBS} shows improvement intensity of Si 288.2 nm line with GIBS. .

\section{LIBS imaging/Confocal LIBS}
\begin{figure}
\centering
\includegraphics[width=0.49\textwidth, trim=0cm 0cm 0cm 0cm, clip=true,angle=0]{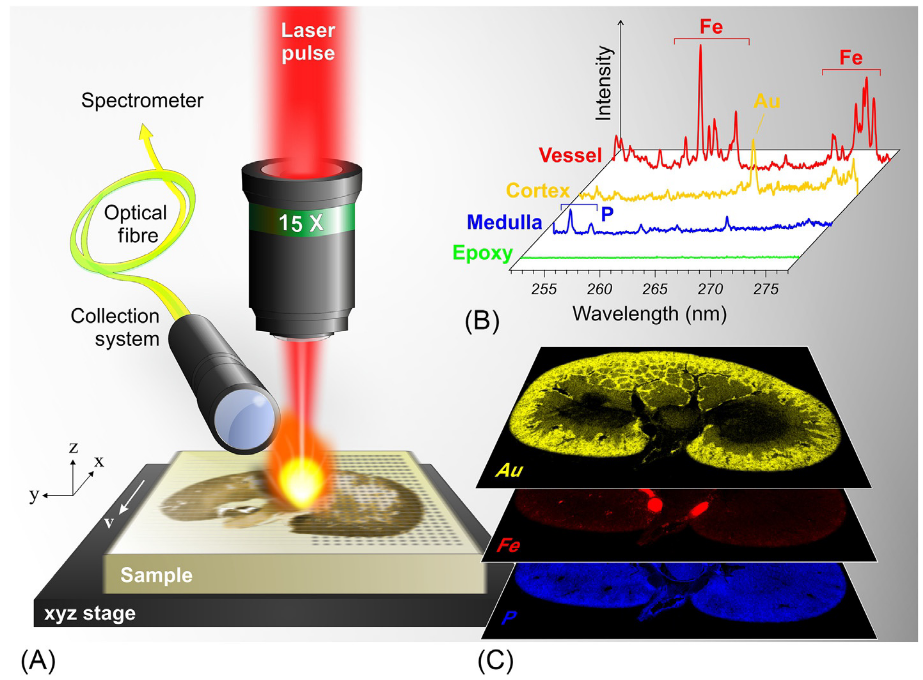}
\caption{\label{fig:Imaing_FIBS}  General protocol for LIBS imaging (A) Schematic view of the LIBS imaging setup with    
       main  components: the microscope objective is used to focus the laser pulse, the motorized 
       platform for moving the sample, and the detection system connected to a spectrometer via an 
      optical fiber. (B) Examples of single-shot emission spectra in the spectral range between 250 and 
       280 nm. (C) Sample relative-abundance images of Au (yellow), Fe (red), and P (blue) represented 
       using false color scales. (adapted with permission from reference \cite{MOTTOROS2020329})
}
\end{figure}

\begin{figure}
\centering
\includegraphics[width=0.49\textwidth, trim=0cm 0cm 0cm 0cm, clip=true,angle=0]{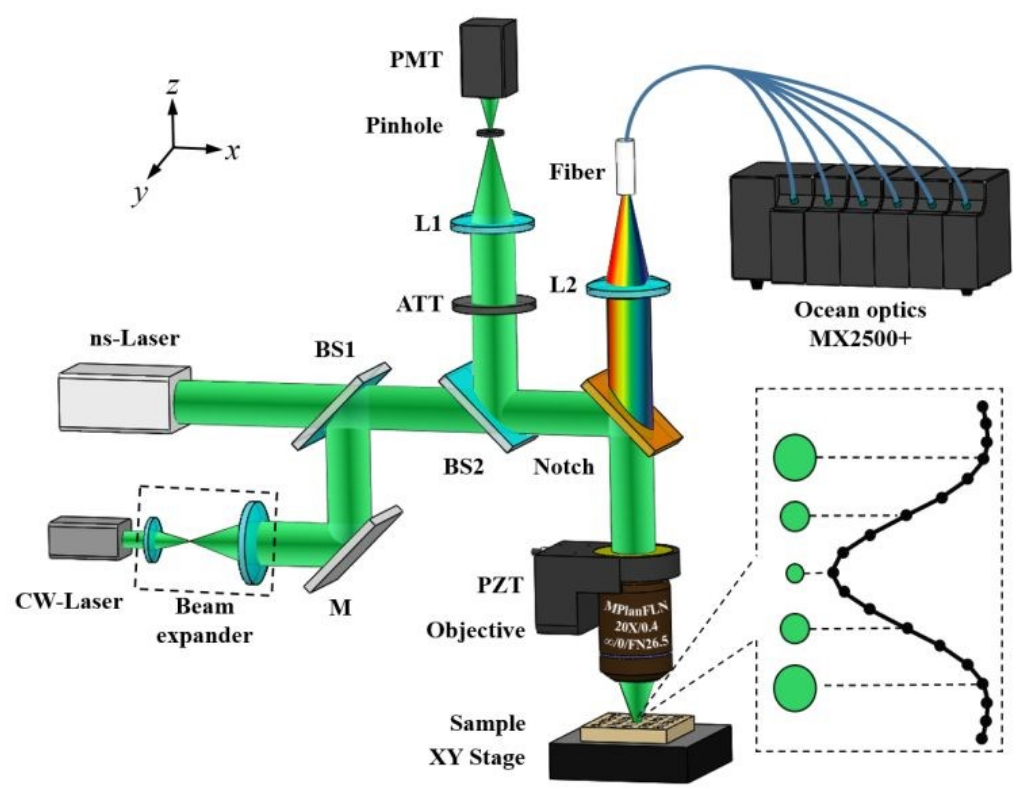}
\caption{\label{fig:Cofocal_LIBS}  Schematic of Confocal LIBS microscopy. The beam emitted from continuous-wave (CW) laser passes    through a beam expander and beam splitters BS1 and BS2, and then is focused onto the   sample by an objective. The reflected light transmits along the original light path and is reflected by the BS2, and then is finally detected by a photomultiplier tube (PMT) 
(adapted with permission from reference \cite{C9JA00387H}).
}
\end{figure}

Spatially resolved LIBS imaging has attracted considerable interest because of its importance in revealing elemental distribution in the sample. For improving lateral resolution in LIBS imaging techniques e.g. micro-LIBs, fs-LIBS and near field enhanced atomic emission spectroscopy have been proposed \cite{MOTTOROS2020329}. Figure~\ref{fig:Imaing_FIBS} demonstrates the general protocol for LIBS imaging whereas fig~\ref{fig:Cofocal_LIBS} shows confocal LIBS imaging setup.

\section{Self-Absorption, Optical Thickness}
LIBS based spectro-chemical analysis based on relation between the observed intensity of an emission line and analyte concentration and from analytical point of view a linear relationship is desired. However,
If optical thickness is large, 
part of the emitted radiation will be reabsorbed by the same species which is termed as self-absorption \cite{REZAEI2020105878}. Self-absorption coefficient (SA) is defined as the ratio of the actual intensity of an emission line to the theoretical intensity\cite{YANG2020163702} .Cases of self-absorption in case of homogeneous as well as inhomogeneous plasm are well described by Fatemeh Rezaei et al \cite{REZAEI2020105878} . A simplest check to ensure that
the plasma is thin is to check the  ratio of intensities of two lines which originate from same upper energy level or with upper energy levels with a small difference and if the following relation holds, it can be assumed to be optically thin.
\begin{eqnarray}
\label{Optical_thickness}
\frac{I_1}{I_2}=\frac{g_1A_1\lambda_2}{g_2A_2\lambda_1}
\end{eqnarray}
where  $I,g,A,\lambda$ with subscripts are the intensities, gaunt factor, transition probabilities and wavelengths respectively of the lines 1 and 2 under consideration.
Another way to is to estimate spectral absorption at the center of a line originating in levels i and j and can be given by 
\begin{eqnarray}
\label{Spectral_absorption}
k_{ji}(\lambda_0)=8.85\times10^{-13}f_{ji}\lambda_0^2n_iP_{ji}(\lambda_0)
\end{eqnarray}
where $k_{ji}$ ($cm^{-1}$) is absorption coefficient, $f_{ji}$ is the absorption oscillator strength, $\lambda_0$ is the wavelength and $P_{ji}$ ($\lambda_0$) is  normalized line profile at the center. For Lorentzian profile $P_{ji}$($\lambda_0$)= $1/(\pi\Delta\lambda_{1/2})$.

\section{Self-reversal of lines}
\begin{figure}
\centering
\includegraphics[width=0.49\textwidth, trim=0cm 0cm 0cm 0cm, clip=true,angle=0]{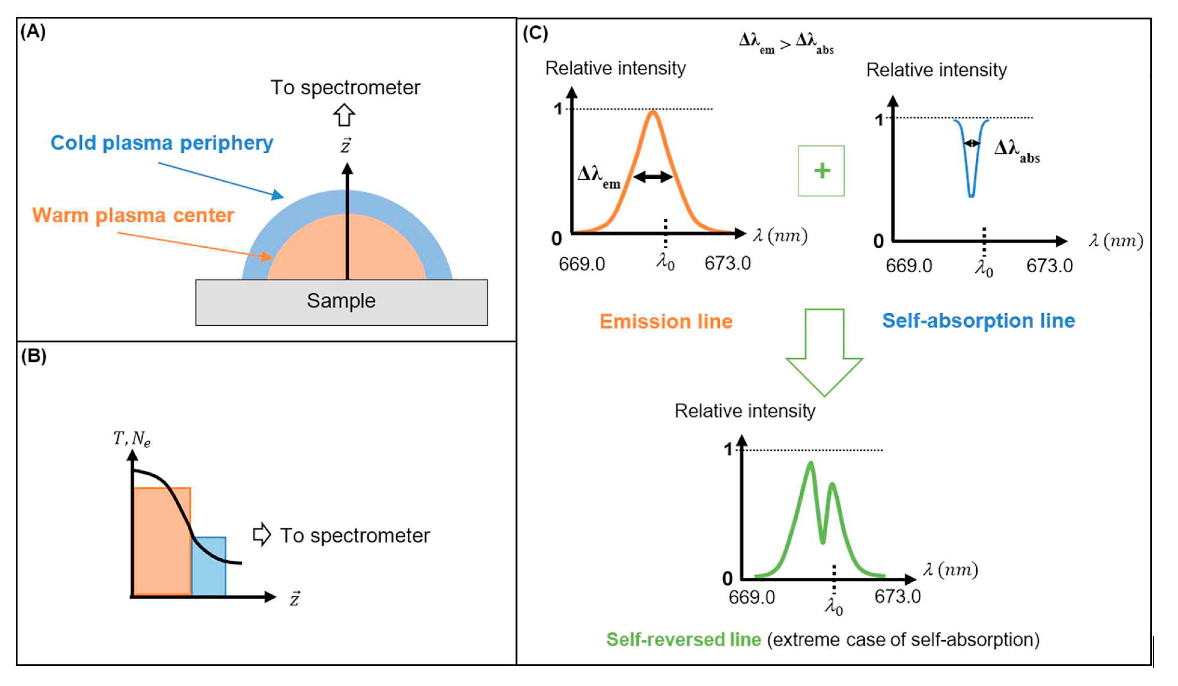}
\caption{\label{fig:Self_Reversal}  Figure (a, b) depicts how self-reversal can take place; (c,d) shows self-reversal in 670.8 nm resonance line of Lithium (Li I) (adapted with permission from reference 
\cite{TOUCHET2020105868}).
}
\end{figure}

\begin{figure}
\centering
\includegraphics[width=0.49\textwidth, trim=0cm 0cm 0cm 0cm, clip=true,angle=0]{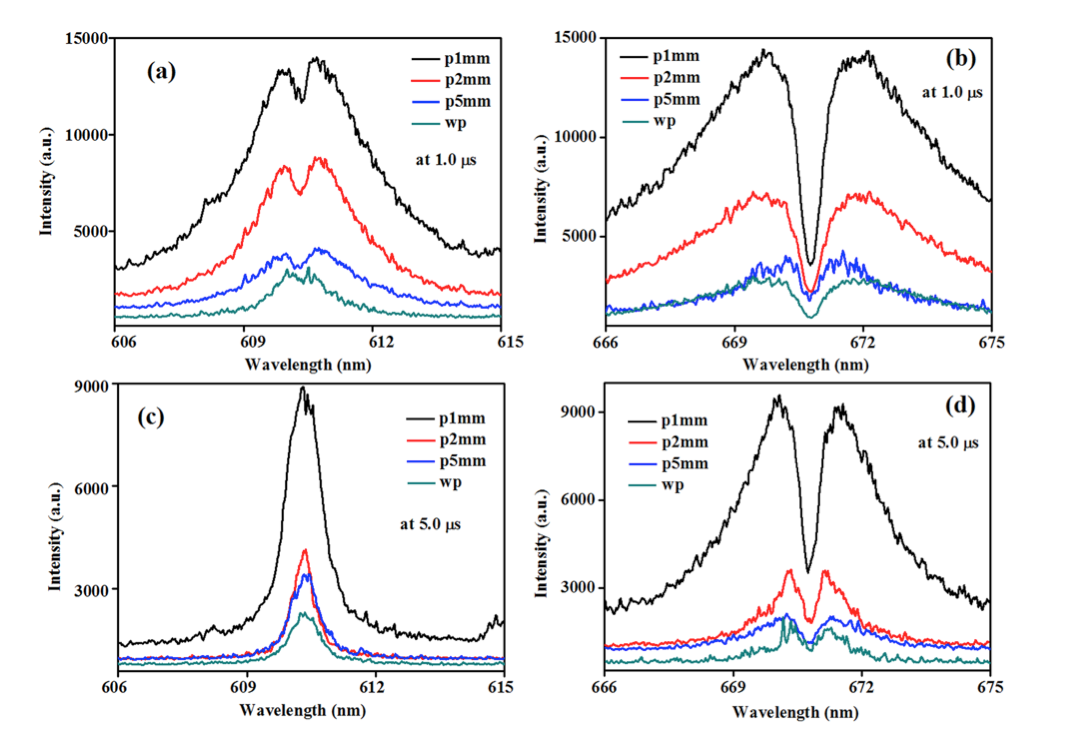}
\caption{\label{fig:Self_Reversal2}  Self-reversal in Lithium 670.8 nm and 610.3 nm lines in a confined geometry. More pronounced self-reversal is clearly evident for 670.8 nm resonance line (adapted with permission from reference
\cite{Bhupesh_POP_2013}).
}
\end{figure}

Self reversal results when the emission from the hot center is absorbed by the species present at cooler periphery or due to plasma inhomogeneity\cite{TOUCHET2020105868} as shown in figure~\ref {fig:Self_Reversal}. The emitted line shows dip at the peak due to absorption of the emitted line at rather hotter plasma center. A number of studies reported the presence of self-reversed lines under different plasma environments \cite{TOUCHET2020105868,Bhupesh_POP_2013,KUMAR2022127968,URBINA2022106489,Kautz_JAP_2021}. Strong self-reversal in Li 670.8 line is noticed when LIBS was performed in a confined geometry as shown in figure~\ref{fig:Self_Reversal2}. Moreover, it is dependent on the distance of the plate used to confine the plasma and time delay \cite{Bhupesh_POP_2013}. Strong self-reversal was also reported in laser produced plasma for Ba ionic line\cite{KUMAR2022127968}. Shock waves during breakdown had also been considered to play an important role in generating plasma inhomogeneities\cite{URBINA2022106489} . 
Laser induced fluorescence (LIF) of laser ablated filaments has been found to reduce the self-reversal features in the spectral profiles \cite{Kautz_JAP_2021}.

\begin{figure}
\centering
\includegraphics[width=0.49\textwidth, trim=0cm 0cm 0cm 0cm, clip=true,angle=0]{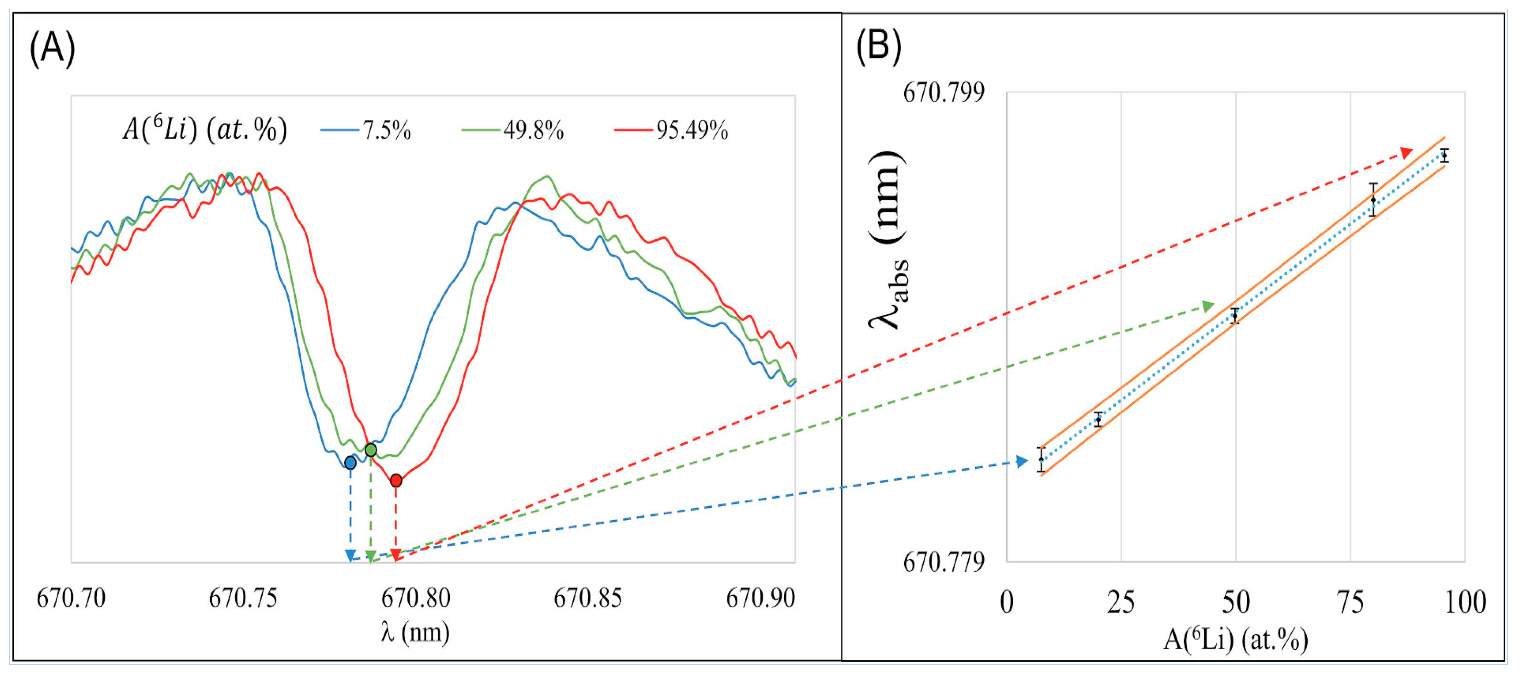}
\caption{\label{fig:Self_Reversal3}  Self reversal demonstrating Li isotopic dependence (adapted with permission from reference\cite{TOUCHET2020105868}).
}
\end{figure}
Self-reversal of lines in LIBS which was considered to be unwanted, appears to to be boon in disguise.   Estimation of isotopic abundance in case of lithium isotopes has been projected by exploiting the narrow (less effected by line broadening processes) self-reversed profile \cite{TOUCHET2020105868}. With further optimization and analysis of self-reversed lines, better quantitative estimate of isotopic abundance is expected. Figure~\ref{fig:Self_Reversal3} shows the isotopic dependence of self reversal profiles for lithium 670.8nm resonance line.

\section{Parameter estimation} The estimation of two important parameters viz electron density and    
electron temperature is important in understanding the plasma plume behavior as well as   
 estimation of elemental composition.
\subsection{Electron density} Electron density is mainly estimated from the Stark broadening of the lines and is well established in literature and will not be detailed in the present article. However, in some cases Stark parameters of certain lines are not available and hence need to be estimated. One of the adopted methods is to use an alloy with other metal with known Stark parameters and from the obtained density extract the unknown Stark parameters\cite{Mayo_2008}
  In another study Aragon et al. used fused glass samples to extract the Stark parameters for FeII and Ni II and Ti II with improved line to background ratios\cite{ARAGON201439,mnras_Manrique}.
   In some cases, these parameters are estimated from H$_\alpha$ line due to the presence of trace amount of water. 
   In case of nickel Stark parameters are estimated from the density obtained from  $H_\alpha$ line \cite{Jinto_POP,ElSherbini2006}. 
   Parameters for some lines of tungsten were estimated by using CII line (426.7 nm) in tungsten carbide plasma \cite{Nishijima_2015}.
   Cross calibration in a multi element plasma was also used to extract these parameters \cite{Liu2016}. Stark widths for UI and UII were estimated from the O I 799.19 nm line present as surface  impurity in a uranium metal target\cite{Hari_uranium}. Stark broadening coefficients for  Tantalum lines were extracted using extended C-sigma method \cite{POGGIALINI2020105829} from the information of Stark broadening coefficient for a known line.  However, these methods either may suffer from the interference from the presence of more elements on the plasma or the lines from the tracer impurities may be weak to provide sufficient intensity.
    Moreover presence of hydrogen, at low pressures presence of trace amount of water can give H$_\alpha$ line but may have low intensity. At the same time at higher air pressure (higher water content is expected), water is likely to contaminate the plasma as well as affect the plasma properties. From $H_\alpha$,  we can estimate Stark parameters of any element.\par
The spectral emission from LPP plasma are broadened due to various mechanisms, such as Stark, Doppler, van der Waals etc.\cite{Griem1974}.  The line shape of emission from LPP plasma varies depending on the broadening mechanisms. A Lorentzian profile is expected for the collisional broadening process such as Stark and van der Waal broadening. whereas Doppler broadening results in a Gaussian profile. Some times more than one mechanism results in the broadening and hence the lines shape will be a convolution of different profiles as discussed in details by Griem\cite{Griem1974}

In LPP, three primary mechanisms can contribute to the spectral line shape; Doppler, Stark and instrumental. FWHM due to Doppler broadening can be estimated by
\begin{eqnarray}
\label{dopler_broadening}
\Delta \lambda = 7.2\times10^{-7}\lambda_0\sqrt{\frac{T_e}{M}}
\end{eqnarray}

In case of LIBS, the main broadening mechanism is considered as Stark broadening as due to low temperature, contribution from Doppler is very small. 
 The Stark width is given by
\begin{eqnarray}
\label{stark_broadening}
\Delta \lambda_{1/2} = 2W(\frac{n_e}{10^{16}})A^0
\end{eqnarray}

Though the Stark broadening contribution is dominant in LPP plasma, the Doppler broadening may come significant at LPP emission of a plume expanding into a low background pressure and at a later time where the plasma density is not high.
Assuming instrumental width as Lorentzian, the actual Stark width can be deducted by subtracting the instrumental width from the fitted Lorentzian width.
Besides,  Stark broadening, Laser Thomson scattering and interferometry  can also be used to estimate electron density in laser produced plasmas \cite{Hari_RevModPhys.94.035002,Choudhary_interferometry}. It has been described in an earlier review article also\cite{Muraoka_2011}.

\subsection{Electron temperature} 
The estimation of plasma parameter using the OES is based on the Boltzmann and Saha equation. Boltzmann equation relates the ratio of population densities $N_j^z$ of excited energy levels and the number density $N^z$ with temperature T as
 \begin{eqnarray}
 \label{Bolt_eq}
 \frac{N_j^z}{N^z}=\frac{g_j^z \: exp(\frac{-E^z}{kT})}{U^z(T)}
 \end{eqnarray}
 where $z$ represents the ionization stages, $E_j^z$ and $g_j^z$ are the respective energy and degeneracy of the specified level, $U^z(T)$ is the partition function. 
 The Saha equation relates electron density $(N_{e})$ and temperature with, population density of different states $z$ and $z-1$ as follows
\begin{eqnarray}
\label{Saha_eq}
\frac{N_{e}N^z}{N^{z-1}}=\frac{2U^{z}(T)}{U^{z-1}(T)} \left(\frac{2\pi mkT}{h^2}\right)^{3/2} exp\left[\frac{-(E_\infty^{z-1}-\Delta E_\infty^{z-1})}{kT}\right]
\end{eqnarray}
where $E_\infty^{z-1} $ is the ionisation energy of species in charge state of $z-1$, $\Delta E_\infty^{z-1} $ is the correction in ionization energy due to the plasma interaction, $h$ is Planck?s constant $(6.626 \times 10^{-34}J.s)$, $m$ is the mass of the electron $(9.109\times10^{-31}Kg)$. All these parameters for most of the species are well known and available in data bases like NIST. The applicability of these equations largely depends on the validity of LTE conditions of plasma described in section~\ref{sec:LTE}.

Electron temperature can be estimated using the ratio of emission intensities of spectral lines of the same species and charge states using the Boltzman equation (Equation~\eqref{Bolt_eq}). Estimation of temperature is normally done
 either by taking the ratio of the emission intensity of two separate lines or using a Boltzmann plot method\cite{ARAGON2008893}. 
 In the case of  intensity ratios of line emissions, assuming local thermodynamic 
equilibrium (LTE) in the system, the temperature is estimated using the following equation which is derived from Boltzman equation,   
\begin{eqnarray}
\label{Temp_int_R}
\frac{I_1}{I_2}=\frac{g_1A_1\lambda_2}{g_2A_2\lambda_1}exp\left(\frac{-(E_1-E_2)}{k_bT_e}\right)
\end{eqnarray}
where$ \lambda_i, A_i, g_i, I_i  and E_i \: (i = 1,2)$ are the wavelength, transition probability, statistical 
weight, line intensity, and energy of the excited state respectively. 
 When only two lines are considered, the selection of these lines is very critical. To have a better accuracy, the upper state energy levels of the two lines under consideration have to be well separated. Moreover, the energy difference should be significantly larger than the plasma electron temperature to get an accurate estimation of electron temperature from line intensity ratio. Also, care has to be taken to make correction for the opacity of the plasma if it is present. A lack of consideration of these conditions can lead to wrong estimation of the temperature. Better estimate of electron temperature can be achieved by using spectral lines of two successive ionization states \cite{Harilal1998}.
Electron temperature can be estimated by eq.2 of this reference.

 A more general and accurate way for the estimation of temperature is possible by using the Boltzmann plot method. In this method, a number of lines can be used for the estimation of temperature. The equation for estimating temperature (equation~\eqref{Temp_int_R}) can be re written as
\begin{eqnarray}
 \ln{\bigg[\frac{I_{ij}\lambda_{ij}}{g_iA_{ij}}\bigg]}=\frac{-E_i}{KT_e} +C
  \label{BoltzmanPlot }
\end{eqnarray}                                                
where $I_{ij}$, $\lambda_{ij}$,$ A_{ij}$, $g_i $and $E_i$ are the spectral intensity, wavelength, transition probability and statistical weight of the upper state and upper state energy respectively.  
If a plot is made with $E_i$ on x axis and the left hand side(LHS) of equation \eqref{BoltzmanPlot } as the y axis, the slope of the graph will be equal to $\frac{ -1}{ KT_e}$ , from which one can easily estimate the plasma temperature. In this method, lines meeting the required conditions can be used, which improves the accuracy of temperature estimation.

Another method for the estimation of plasma temperature using spectroscopy is the line-to-continum ratio method \cite{MOON2012221}. The equation for this method is derived from Saha equation and from the expression for the integrated spectral emissivities of the respective lines it can be expresses as
\begin{eqnarray}
\label{Line_to_cont}
\frac{I_{1}}{\varepsilon_{c}}( \lambda)= \dfrac{2.0052 \times 10^{-5}A_{21} g_2\exp \left(\frac{E_{i}-\bigtriangleup E_{1}}{kT_{e}} \right) \exp \frac{-E_{2}}{kT_{exc}} }
{U_i \lambda_i T_e \left[\xi \left( 1 - \exp \left( \frac{-hc}{\lambda kT_{e}} \right)  \right)  +G\left( \exp \left( \frac{-hc}{\lambda kT_{e}} \right) \right) \right] }
\end{eqnarray}
where $I_1$ is the integrated intensity of emission line, $\varepsilon_{c}$  is the continuum emission coefficient, $A_{21}$ is the transition probability, $ g_2$ is the upper state statistical weight. $E_i$ is the ionization potential, $E_2$ is the upper state energy level, $\bigtriangleup E_{1}$ is the correction factor to inonization potential due to the plasma which can be neglected safely. $U_i$ is the partition function, $\xi$ is the free-bound continuum correction factor, G is the free free Gaunt factor.
Experimentally, $\varepsilon_{c}$ is measured closest to the line chosen. From the above equation we can calculate the electron temperature $T_{e}$.

As discussed in case of density, temperature can also be estimated using laser Thomson scattering \cite{Muraoka_2011}
\section{Local Thermodynamical Equilibrium}\label{sec:LTE}
For establishing that the levels are populated with Boltzmann distribution, the plasma is assumed is to be under local thermodynamic equilibrium (LTE). Macwhirter criterion defined below is taken as a necessary condition for the LTE condition
\begin{eqnarray}
\label{Macwhirter}
n_e\geq 1.6\times 10^{12}T_e^{0.5}\Delta E_{mn}^3
\end{eqnarray}
where $n_e$ is electron density in $cm^{-3}$, $T_e$ is electron temperature in Kelvin and $\Delta E_{mn}$ (eV) is the largest energy gap between the adjacent energy levels.
However for transient and inhomogeneous plasma. Christoferrati criteria\cite{CRISTOFORETTI} discussed below have to be checked.
In the case of transient plasma, like the case of a laser produced plasma, require to  be verified with the Christoferrati criteria in order to ensure the LTE situation. The Christoferrati criteria  state that the diffusion length $D_\lambda$ of atoms/ions, during a time period of the order of the relaxation time to the equilibrium, should be shorter than the variation length of temperature and electron number density in the plasma. The diffusion length is approximated as
\begin{eqnarray}
D_\lambda \approx 1.4\times 10^12 \times \left( \frac{(k_B T_e)^{3/4}}{N_e}\right) \times\left( \frac{\Delta E}{M_A f_{12}(g)} \right)^{1/2} 
\times e^{\Delta E/ 2k_B T_e}
\end{eqnarray}
where $k_B$ is the Boltzmann constant, $N_e$ is the electron number density, T is the plasma temperature, $M_A$ is the atomic mass of element,
$\Delta E$ is the energy difference in the upper and lower level, $f_{12}$ is the oscillator strength and g is the gaunt factor. 
$f_{12}$ is a dimensionless quantity of the probability of absorption or emission of electromagnetic radiation for a particular transition and $g$ is the correction factor to be used as an approximation to the classical calculation of emission results.
Similar to the variations in diffusion length, it is equally important that the  relaxation  time of  the  plasma  for establishing  the  thermodynamic  equilibrium  has to be  shorter  than  the  time  of  variation  of  plasma temperature  and  density\cite{CRISTOFORETTI}.
Typical laser produced plasma at its initial stage has density $\gg 10^{18} cm^{-3}$ and temperature of few eV, which meets the requirement of LTE condition. However, after a sizeable plasma expansion, the density falls rapidly which leads to non-LTE plasma conditions.

\section{Calibration free LIBS (CFLIBS)}
Convention LIBS analysis suffers from matrix effects and also requires reference sample for calibration curve. Hence calibration free LIBS (CF LIBS) approach was adopted. Details of this method are given in a review by Tongoni et al.  \cite{TOGNONI20101}  and also in recent reference of Zhang et al\cite{FP_2022_Zhang}.However, CFLIBS considers that (i) plasma plume represents the actual composition,  (ii) the plasma is in LTE condition in spatial as well as temporal window, (iii) plasma is considered to be homogeneous and (iv) the spectral lines under consideration are optically thin.  
\subsection{C Sigma graphs}
 Generalized curves of growth known as CSigma( $C \sigma$) graphs which include several lines of various elements of same ionic stage were suggested by Aragon et al. for LIBS\cite{ARAGON201490}. It is based on Saha, Boltzmann and radiative transfer equations under the assumption of local thermodynamic equilibrium (LTE). Further C-sigma graphs are based on the calculation of line cross section which allows estimation of self-absorption.
 
\subsection{ Internal reference for self-absorption correction (IRSAC)}
To overcome self-absorption in CF-LIBS, correction by internal reference has also been suggested \cite{Sun2009CorrectionOS} . Basically the line chosen as internal reference has lower energy level with higher energy or with low transition probability which is affected slightly by self –absorption. Based on this, the intensity of other lines can be corrected. Moreover, columnar density and standard reference line (CDSRL) method has been found to result in better accuracy than standard LIBS \cite{HU2021339008,Sun2009CorrectionOS}.

\section{Enhancement in signal/detection sensitivity}
Signal enhancement and subsequent increase in detection sensitivity has been a subject of wide investigations for better exploiting it in various applications. A recent review briefly describes some of the approaches attempted for it \cite{Fu_2020_Front}. The methods adopted of signal enhancement are DP LIBS, atmosphere control method, applying spatial constraint, application of magnetic and electric fields, microwave assisted LIBS (MALIBS), LIBS/ laser induced fluorescence combination (LIBS+LIF), nano particle enhanced LIBS( NELIBS).Sample temperature has also been found to enhance signal emission intensity\cite{C9JA00261H,C8JA00348C,Guo_AIP_Adva_2019, TAVASSOLI2009481}. Increase  
in signal emission is attributed to increased ablation rate.

\subsection{Effect of magnetic field} 
Introduction of magnetic field has been found to result in significant enhancement in the intensity of lines \cite{Rai_2003}. Various mechanisms e.g. confinement by the field, increase in electron impact excitation due to Joule heating and acting as a constraint been suggested for this enhancement\cite{JOSHI2010415,Fu_2020_Front}.

\subsection{Effect of electric field} 
Electric field assisted LIBS has been studied for the detection of chlorine and copper \cite{Ahmed2020}. In another work significant enhancement ($\approx8$ fold) is observed in case of copper lines \cite{AHMED_2021} . Fluctuations (contraction/expansion) in the laser produced plasma are suggested for observed intensity enhancement. Electric field effect on laser induced titanium  (Ti) plasma has also been investigated \cite{Asamoah2021}. Intensity enhancement in Ti lines has been found to depend on voltage biasing.

\subsection{Glow discharge LIBS (GDLIBS)} GDLIBS has been found to result in significant signal enhancement when compare with simple GD or LIBS\cite{Tereszchuk_2008} .GD-LIBS takes advantage of collisional excitation by exciting the material generated by LIBS \cite{TERESZCHUK2009378} .
Flame-enhanced LIBS (FELIBS): Enhancement in LIBS sensitivity was noticed  by producing laser plasma in the outer envelope of a neutral oxy-acetylene flame \cite{Liu_2014} . High temperature and low density plasma was observed before 4$\mu s$ which has been projected to be beneficial for enhancing LIBS sensitivity.
\subsection{Microwave Assisted LIBS (MALIBS)}
Interaction between microwave radiation and laser produced plasma has been studied in earlier works and significant increase in line intensity was noticed\cite{Liu_2010,Oba_2020}.Intensity enhancement and plasma sustainment in presence of microwave in air has been suggested to occur due to re-excitation and , of course, not due to absorption of the microwave\cite{Ikeda_22}.
\section{Surface Enhanced LIBS (SENLIBS)}
\begin{figure}
\centering
\includegraphics[width=0.49\textwidth, trim=0cm 0cm 0cm 0cm, clip=true,angle=0]{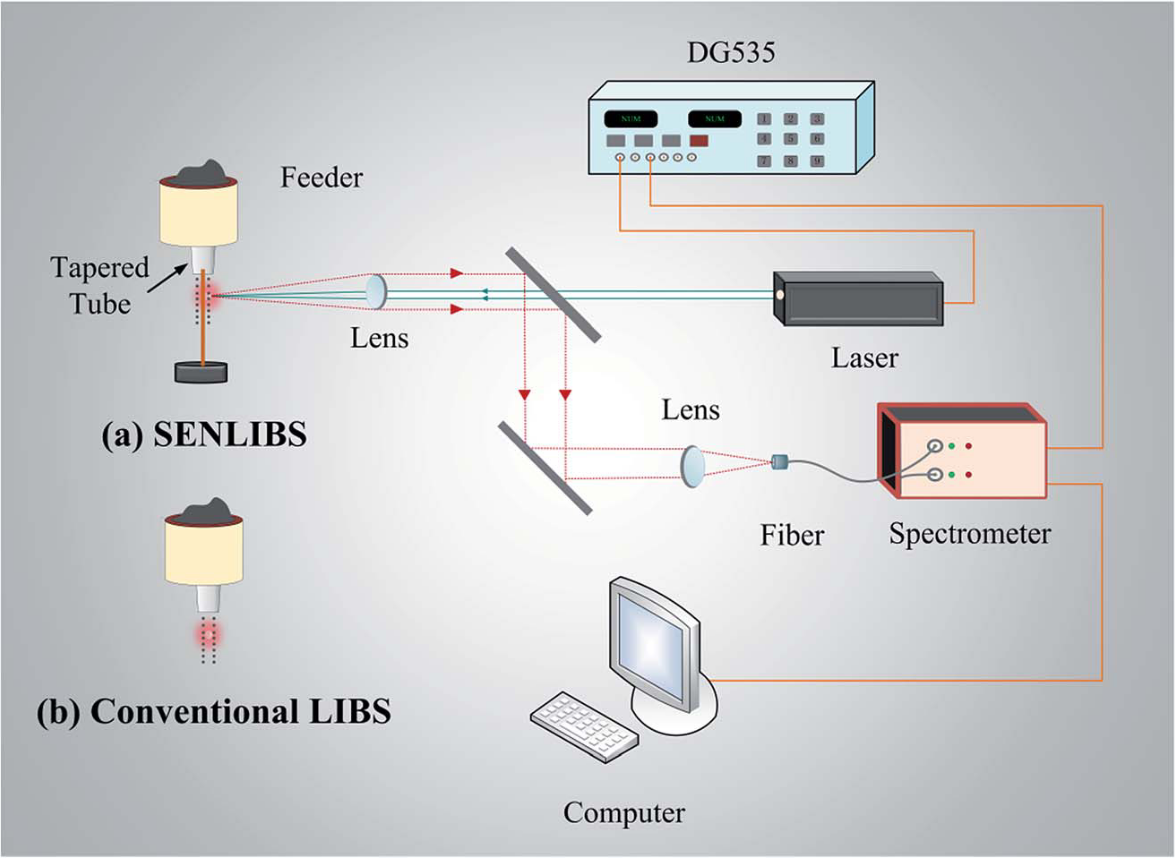}
\caption{\label{fig:SENLIBS}  Schematic diagram of (a) SENLIBS and (b) conventional LIBS (adapted with permission from reference\cite{C8JA00347E}).
}
\end{figure}

\begin{figure}
\centering
\includegraphics[width=0.49\textwidth, trim=0cm 0cm 0cm 0cm, clip=true,angle=0]{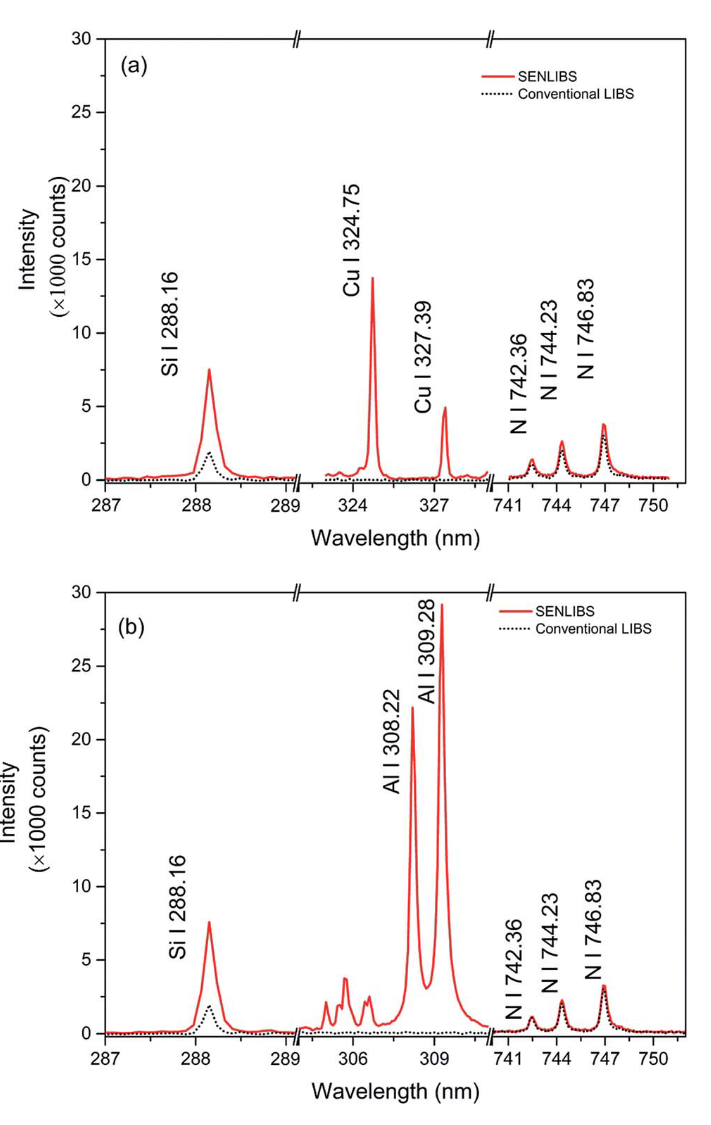}
\caption{\label{fig:Senlibs_Spectra}  Spectra showing Si I lines from silica in (a) copper rod and (b) in aluminium rod (adapted with permission from reference \cite{C8JA00347E}).
}
\end{figure}

In this  method metallic target is used to enhance and stabilise plasma for direct analysis of flow \cite{C8JA00347E,Yang2020}. It has been demonstrated that it has the potential of improving measurement sensitivity. Figure~\ref{fig:SENLIBS} shows the schematic diagram of SENLIBS and conventional LIBS.  Figure~\ref{fig:Senlibs_Spectra} shows substantial enhancement in the emission intensity for copper and aluminium lines. In both the experiments, it is worth mentioning that there is no significant variation in the background lines.

\section{LIF+ LIBS}
\begin{figure}
\centering
\includegraphics[width=0.49\textwidth, trim=0cm 0cm 0cm 0cm, clip=true,angle=0]{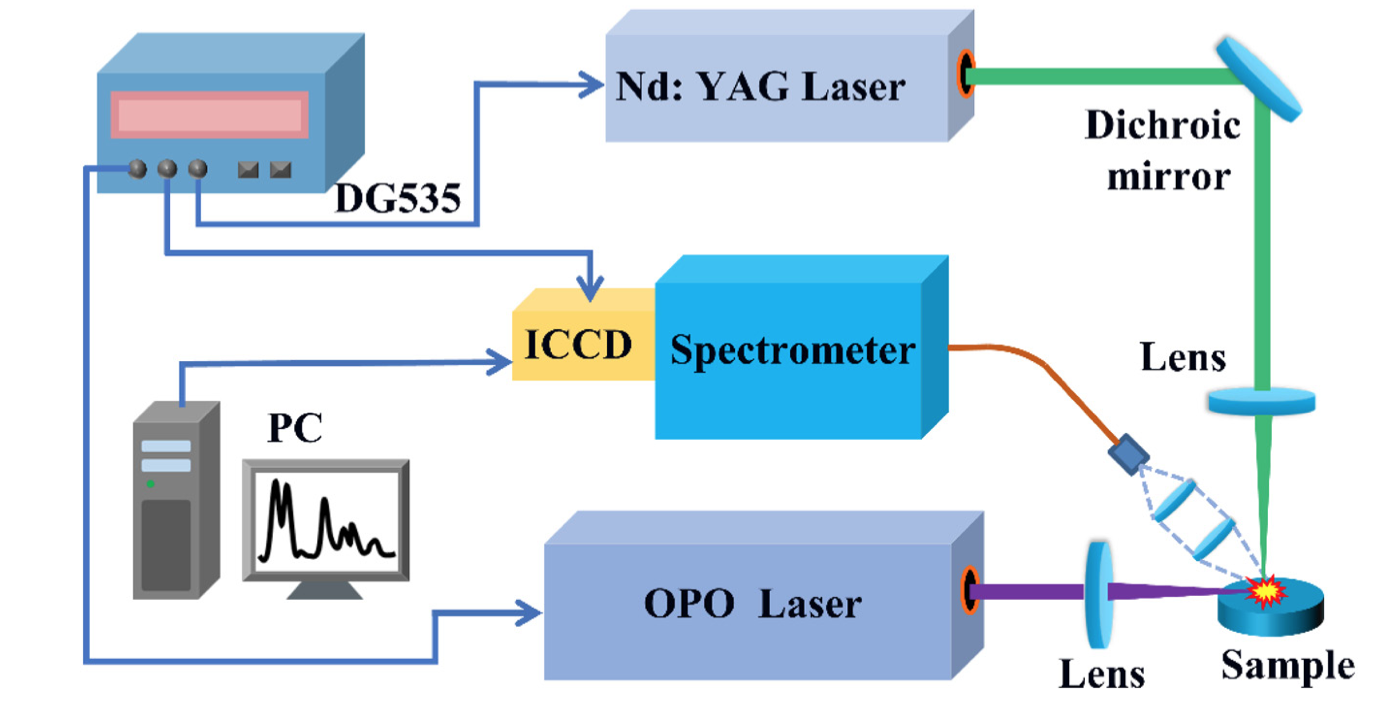}
\caption{\label{fig:LIBS_LIF}  Typical setup for LIBS+LIF combination (adapted with permission from reference 
\cite{Zhou21_LIBS_LIF}).
}
\end{figure}

\begin{figure}
\centering
\includegraphics[width=0.49\textwidth, trim=0cm 0cm 0cm 0cm, clip=true,angle=0]{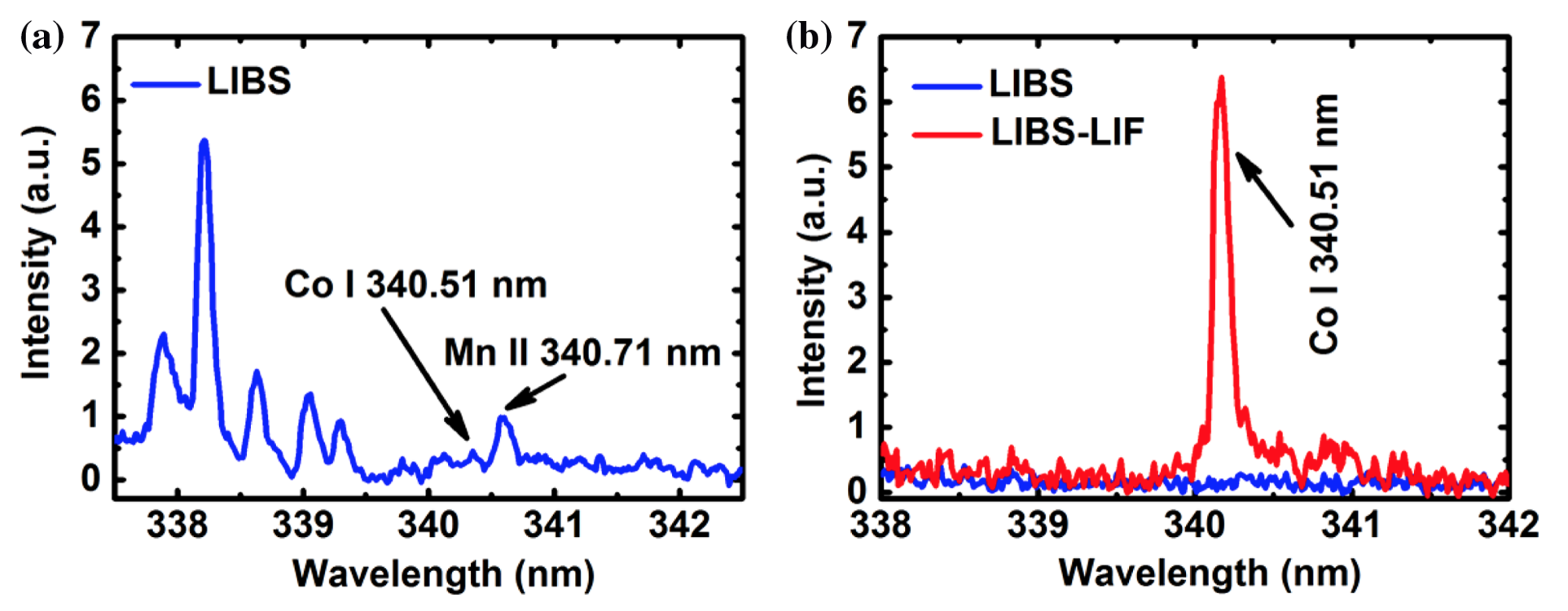}
\caption{\label{fig:Spec_LIF_LIBS}  Enhancement in Co (I) 340.51 nm line in LIBS+LIF (adapted with permission from reference
\cite{Zhou21_LIBS_LIF}).
}
\end{figure}

Combination of laser induced fluorescence with LIBS (LIF+LIBS) can enhance the intensity of lines and subsequently provide better detection sensitivity. A typical LIF+LIBS setup is shown in Fig~\ref{fig:LIBS_LIF}, and enhanced Co-I intensity is demonstrated in Fig~\ref{fig:Spec_LIF_LIBS} . Significant improvement in single shot Limit of detection is observed while combining LIBS and LIF . Moreover spectral  interference effect  which is problematic in conventional LIBS can be resolved with LIF +LIBS (adapted with permission from reference \cite{D2JA00120A}).

\section{Nanoparticle enhanced LIBS (NELIBS)} 
\begin{figure}
\centering
\includegraphics[width=0.49\textwidth, trim=0cm 0cm 0cm 0cm, clip=true,angle=0]{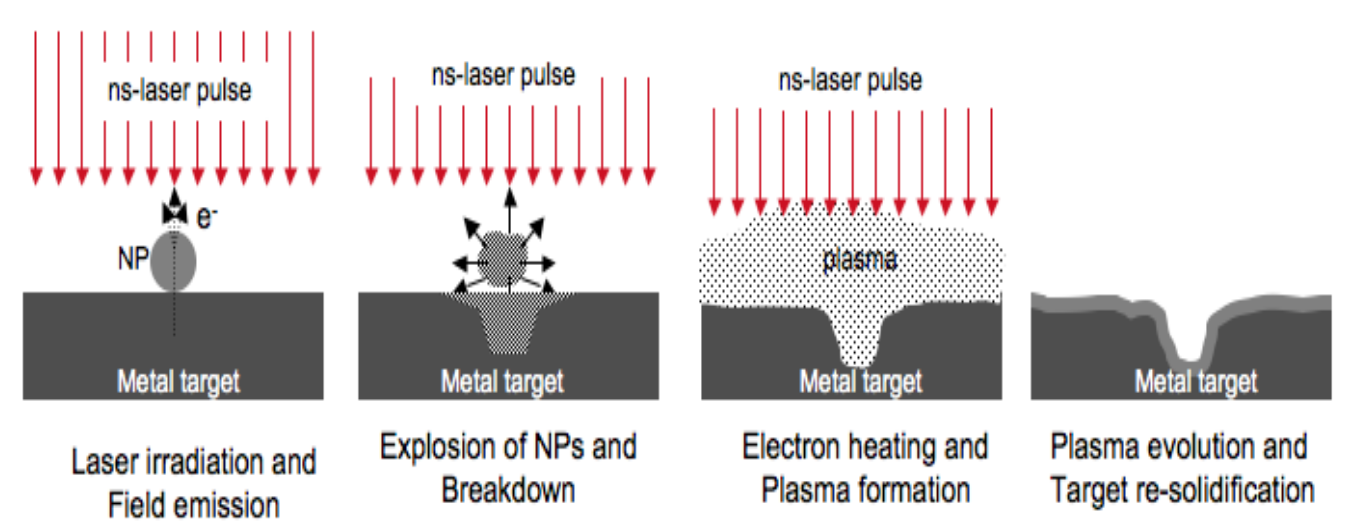}
\caption{\label{fig:NELIBS_schematic}  Schematic of the ablation process in NELIBS: (a) Laser irradiation and field emission,(b) explosion of NPs and breakdown, (c) electron heating and plasma formation, and (d) plasma evolution and target re-solidification (adapted with permission from reference
\cite{DEGIACOMO201419}).
}
\end{figure}

\begin{figure}
\centering
\includegraphics[width=0.49\textwidth, trim=0cm 0cm 0cm 0cm, clip=true,angle=0]{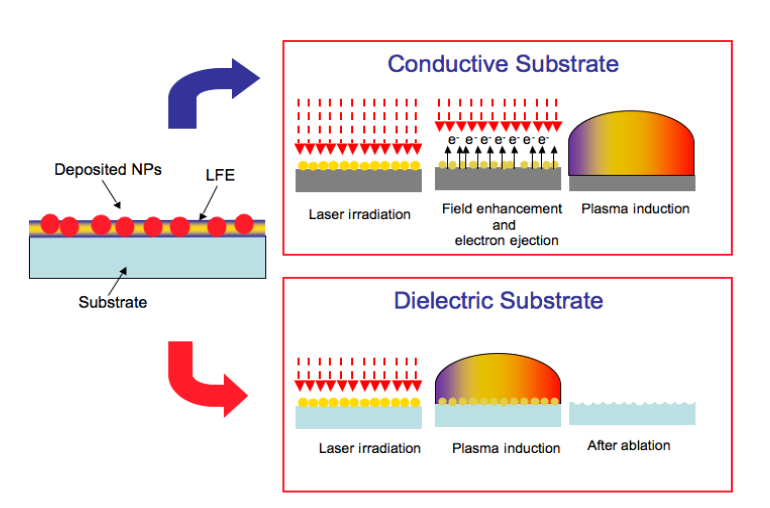}
\caption{\label{fig:NELIBS_1}  NP enhanced photo-ablation in metals and dielectrics (adapted with permission from reference
\cite{DELLAGLIO2018105}).
}
\end{figure}

Presence of nano-particles has been found to enhance LIBS intensity. In Fig~\ref{fig:NELIBS_schematic},\ref{fig:NELIBS_1}, processes associated in the presence of nano-particle in metallic target are shown. NELIBS has been found to have larger plasma volume and longer persistence in spite of similar plasma parameters. Production of more efficient seed electrons in comparison to conventional LIBS has been attributed to this enhancement.

\begin{figure}
\centering
\includegraphics[width=0.49\textwidth, trim=0cm 0cm 0cm 0cm, clip=true,angle=0]{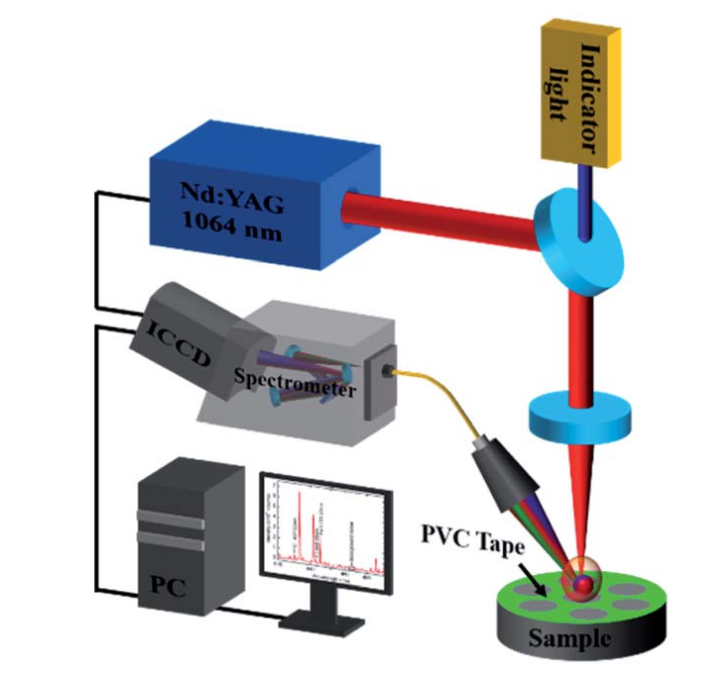}
\caption{\label{fig:NEMLIBS_scheme}  NEMLIBS set up (adapted with permission from reference
\cite{D0JA00528B}).
}
\end{figure}

\begin{figure}
\centering
\includegraphics[width=0.49\textwidth, trim=0cm 0cm 0cm 0cm, clip=true,angle=0]{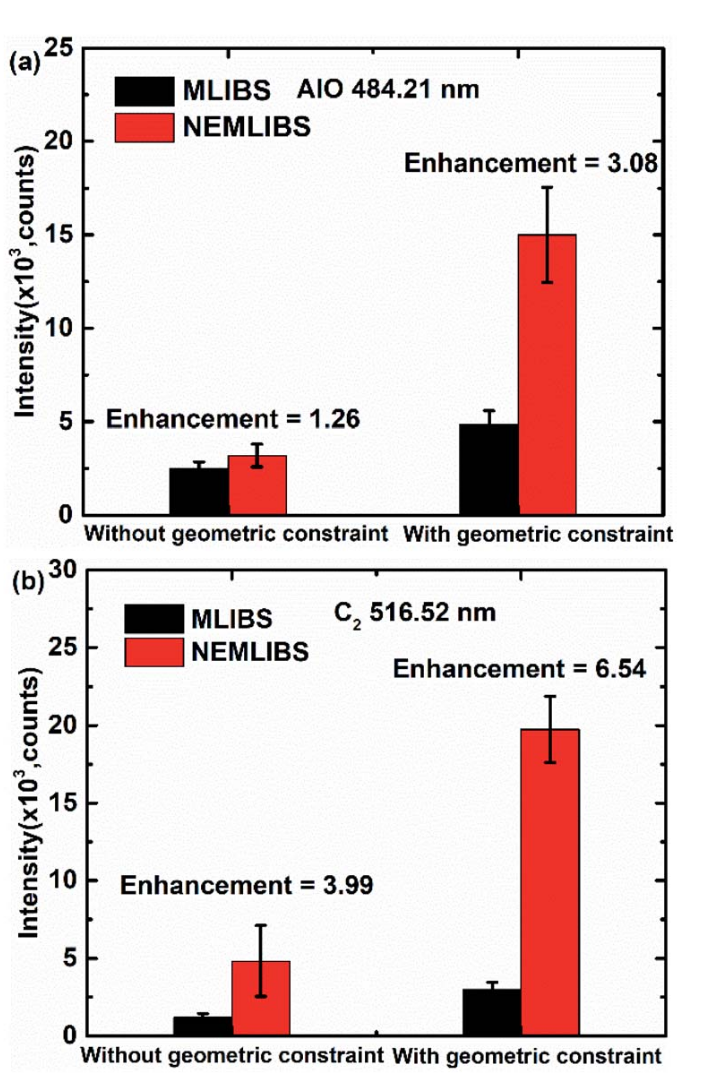}
\caption{\label{fig:MLIBS_NEMLIBS_Comparison}  Demonstration in enhancement  in NEMLIBS for AlO and C2
 with and without geometric constraint. (adapted with permission from reference \cite{D0JA00528B}).
}
\end{figure}

Nano-particle enhanced molecular LIBS (NEMLIBS) was recently reported by Tang et.al.\cite{D0JA00528B}. Geometric constraint was proposed to improve NEMLIBS. Moreover, larger spot size, higher laser energy and pre-ablation of sample are beneficial in NEMLIBS. A typical NEMLIBS set up is shown in figure~\ref{fig:NEMLIBS_scheme} and enhancement in molecular emission in the absence as well as in the presence of constraint is shown in Fig~\ref{fig:MLIBS_NEMLIBS_Comparison}.

\section{Spatial Constraint Method} 
\begin{figure}
\centering
\includegraphics[width=0.49\textwidth, trim=0cm 0cm 0cm 0cm, clip=true,angle=0]{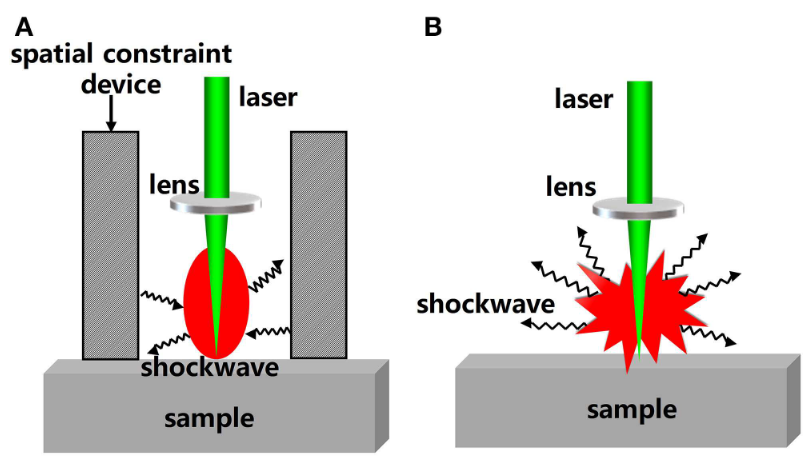}
\caption{\label{fig:Spatail_Constraint_Scheme}  Plasma plume evolution (A) in the presence of spatial constraint and (B) without spatial constraint  (adapted with permission from reference 
\cite{Fu_2020_Front}).
}
\end{figure}

Spatial constraint method: In this method a spatial constraint device is introduced at the periphery of the plasma\cite{Fu_2020_Front}. The generated shock wave will be reflected at the constraint, subsequently compressing the plasma. This will result in higher temperature and density which in turn will give enhanced signal. A schematic of spatial constraint method is depicted in figure~\ref{fig:Spatail_Constraint_Scheme}.

\section{Optically trapped LIBS (OTLIBS)} 
\begin{figure}
\centering
\includegraphics[width=0.49\textwidth, trim=0cm 0cm 0cm 0cm, clip=true,angle=0]{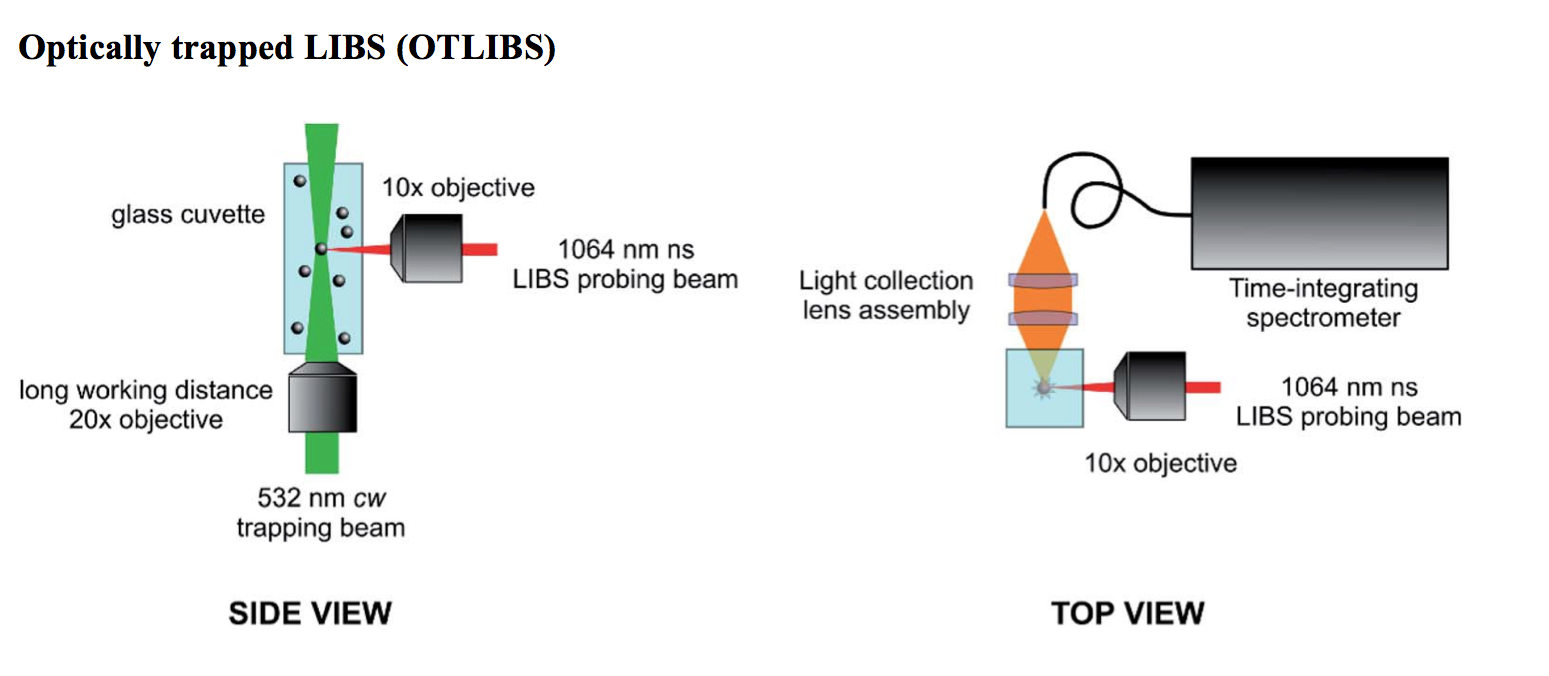}
\caption{\label{fig:Scheme_OTLIBS}  Scheme of an experimental arrangement for LIBS analysis of optical trapped single nano particles (adapted with permission from reference 
\cite{D1JA00149C} ).
}
\end{figure}

Optical catapulting-optical trapping LIBS:
In this technique spectral identification of  micro and nano sized particles is done by sequential optical catapulting, optical trapping and LIBS\cite{FORTES201478,acs.analchem.0c04827}. The details of optical trapping technique are reviewed in Galbács, et. al \cite{D1JA00149C}.
A typical arrangement for LIBS analysis of optically trapped single particle is illustrated in Fig ~\ref{fig:Scheme_OTLIBS}. The method has been demonstrated to have attogram-level detection sensitivity\cite{FORTES201478,PUROHIT201775}.

\section{Polarization Resolved LIBS (PRLIBS)}
Emission anisotropy in the expanding plasma plume has been studied because of its importance in deciphering electron distribution and self-generated electric and magnetic fields \cite{POP_anisotropy, JLA_Aghaba,Wubetu_2020,SHARMA20073113}  .
The degree of polarization at a particular wavelength is defined by 
\begin{eqnarray}
\label{polarization}
P_{\lambda}=\dfrac{I_H-I_V}{I_H+I_V}
\end{eqnarray}

\begin{figure}
\centering
\includegraphics[width=0.49\textwidth, trim=0cm 0cm 0cm 0cm, clip=true,angle=0]{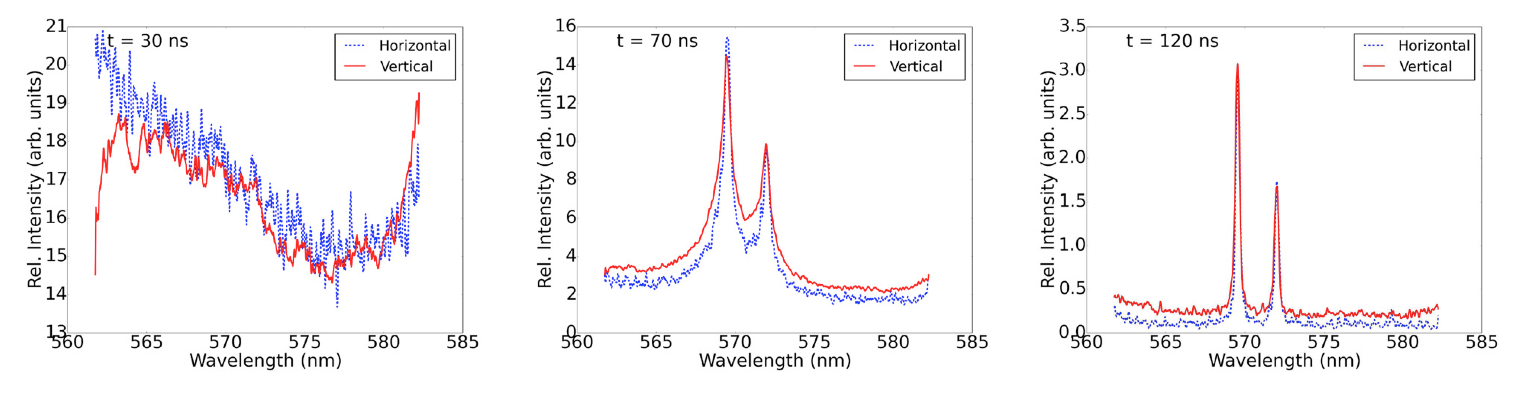}
\caption{\label{fig:Anisotropy_Al} Polarisation resolved spectra of $Al^{2+}$ for different stages of plasma evolution at a background pressure of $1\times10^{2}$mbar. The laser fluence was 550 J$/cm^2$ with an ICCD camera gate width of 10 ns (adapted with permission from reference 
\cite{Anisotropy_2017}
}
\end{figure}

Exploiting polarization resolved LIBS may be interesting in enhancing LIBS sensitivity \cite{Zhao_2014} as can be seen from figure~\ref{fig:Anisotropy_Al}

\section{RESONANT ENHANCED LIBS (RELIBS), RESONANT SURFACE ENHANCED LIBS (RSENILIBS)} 
\begin{figure}
\centering
\includegraphics[width=0.49\textwidth, trim=0cm 0cm 0cm 0cm, clip=true,angle=0]{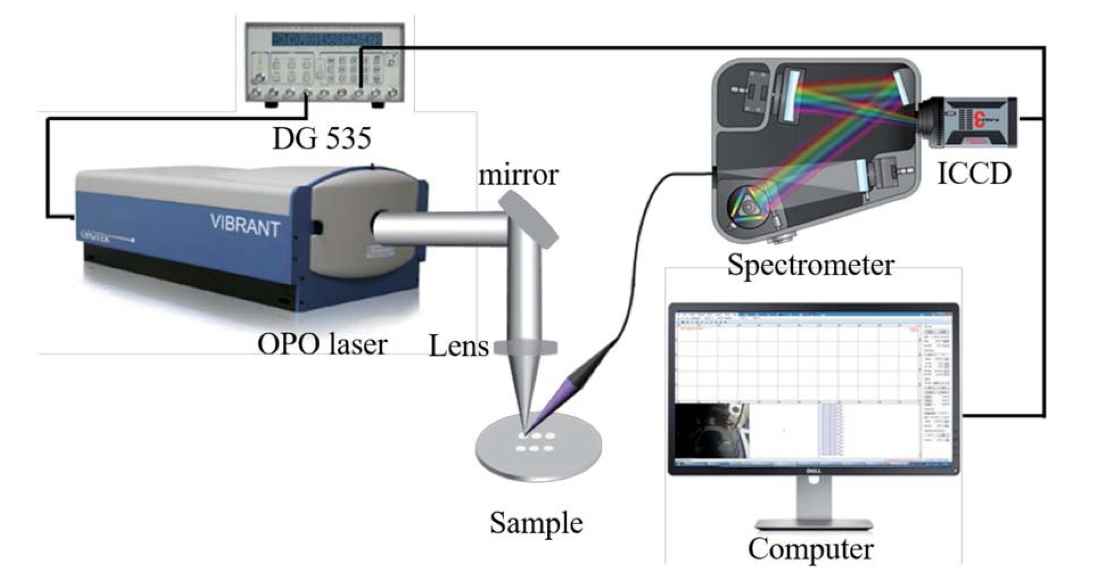}
\caption{\label{fig:R-SENLIBS_scheme} Schematic of R-SENILIBS set up  (adapted with permission from reference
\cite{D1JA00250C}).
}
\end{figure}

\begin{figure}
\centering
\includegraphics[width=0.49\textwidth, trim=0cm 0cm 0cm 0cm, clip=true,angle=0]{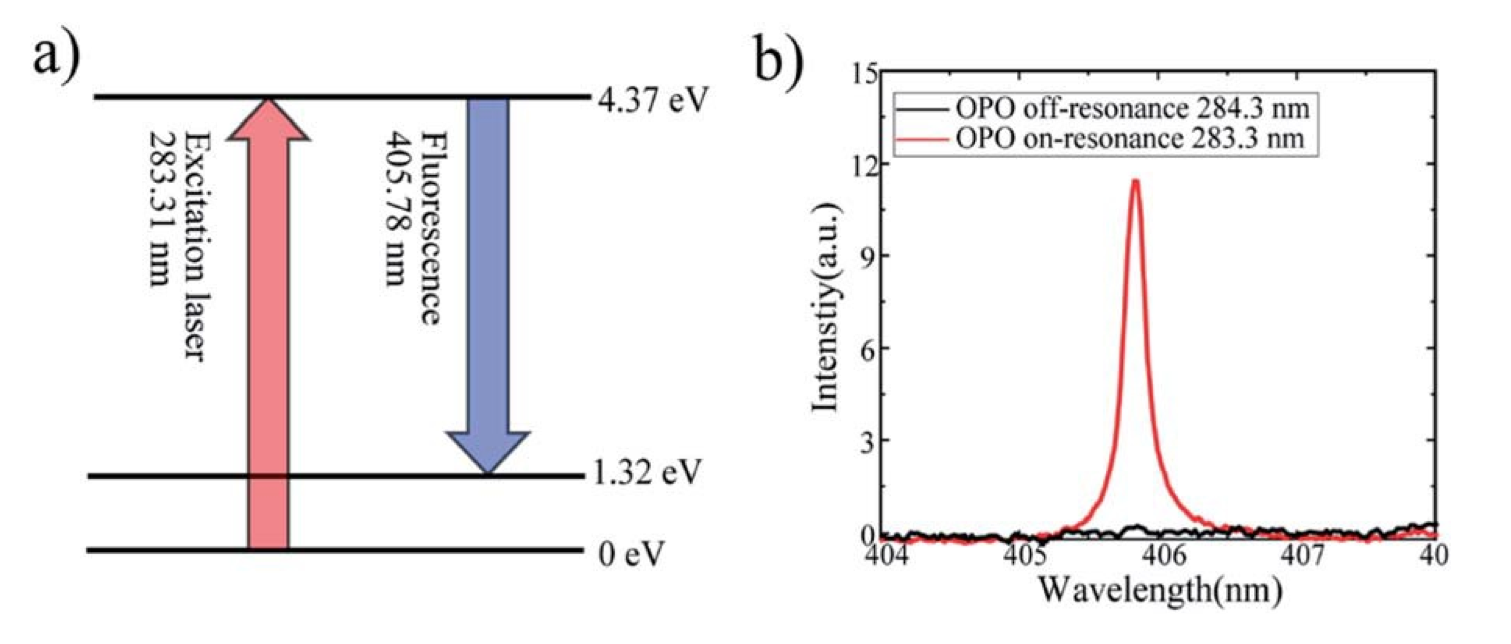}
\caption{\label{fig:Energy_level} (a)	Partial energy level diagram for resonant  excitation of Pb atoms and (b) off resonance and on resonance line spectra (adapted with permission from reference
\cite{D1JA00250C}).
}
\end{figure}

In RELIBS, the excitation laser is tuned to strong absorption line of one of the major species \cite{C3JA30308J}. The energy absorbed is  then distributed over all the elements in the plasma through collisions. A schematic of R-SENILIBS is given in figure~\ref{fig:R-SENLIBS_scheme}. Main advantage of RELIBS over LIBS-LIF is its ability to determine multiple species simultaneously. Concept of R-SENILIBS was reported to and was used to detect lead in water \cite{D1JA00250C}. In this method surface enhanced
LIBS is combined with resonance enhancement to improve the detection sensitivity. Figure~\ref{fig:Energy_level} shows the energy level diagram for resonant excitation in the case of Pb atoms. It can be seen that large enhancement in intensity is  observed in case of on resonance condition.

\section{Back Reflection Enhanced LIBS (BRELIBS)}
\begin{figure}
\centering
\includegraphics[width=0.49\textwidth, trim=0cm 0cm 0cm 0cm, clip=true,angle=0]{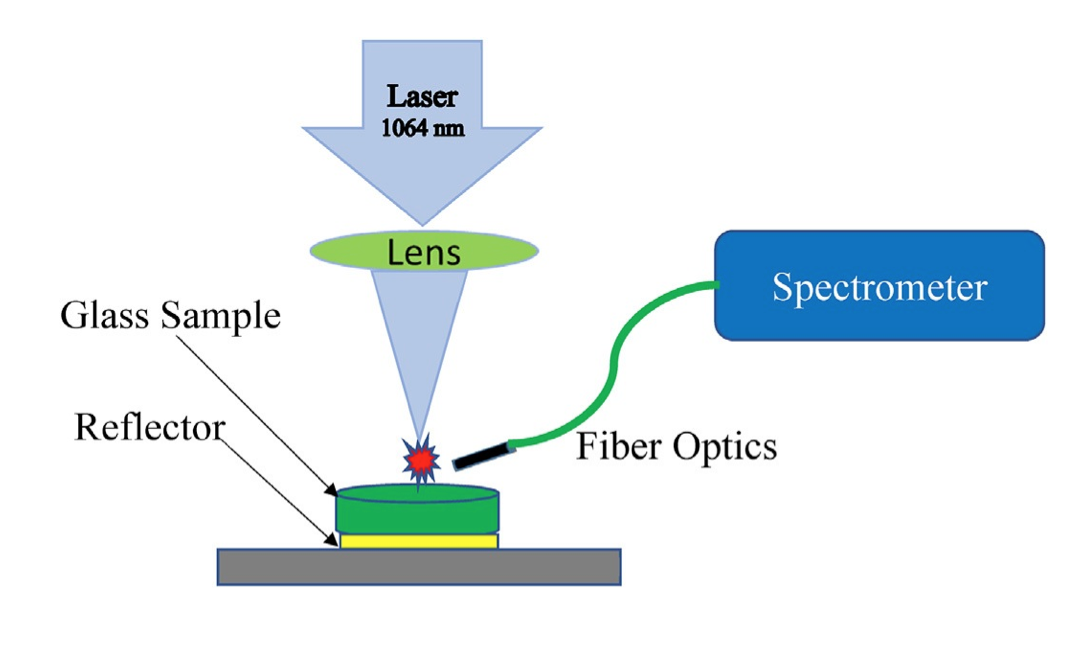}
\caption{\label{fig:BRELIBS} Schematic of back reflection enhanced LIBS (adapted with permission from reference
\cite{ABDELHARITH2021339024}).
}
\end{figure}

LIBS sensitivity is enhanced  using BRELIBS by using metallic reflectors behind transparent targets \cite{ABDELHARITH2021339024} as can be seen in figure~\ref{fig:BRELIBS}. In this method metallic reflectors behind target are used to enhance LIBS sensitivity. The reflected laser beam reheats the plasma resulting in enhanced
intensity. Further, the obtained LIBS spectrum shows pronounced increase in signal to noise ratio  (SNR). 

\section{Simultaneous LIBS Combination With Other Analytical Measurements}

In this section we briefly introduce other analytical methods which are simultaneously used with LIBS.

\subsection{LIBS-Raman}
Combination of Raman and LIBS enables one to study the  chemical composition in a broader context encompassing elemental
and morphological information \cite{HOLUB2022106487,SUN2022106456} . 
In earlier studies simultaneous Raman and LIBS measurements with a single ns laser with 1064 nm fundamental and 532 nm second harmonic were suggested by Sharma and coworkers\cite{SHARMA20071036,Samuel_appl_spec}. 
 However, combination of LIBS and Raman in a single platform has gained lot of attention\cite{Qingyu_ASR,ASR_hybrid_LIBS_RAMAN}.
  Spatial and temporal combining techniques have been suggested for this \cite{Lednev2018} .
   Potentially portable table top remote Raman-LIBS system has been reported recently\cite{MUHAMMEDSHAMEEM2022108264}. 

\subsection{LIBS-Laser Ablation- Inductively Coupled Plasma (LA-ICP)}
Meissner et.al.\cite{MEISSNER2004316} compared LIBS and Laser ablation-Inductively coupled mass spectroscopy (LA-ICP-MS) for the detection of trace elements in a solid matrix. LIBS is affected by matrix as well as self-absorption but does not depend on sample preparation. On the other hand, LA-ICP-MS detection limits are in principle lower  but the  preparation method strongly influences the measurements. 
Simultaneous LIBS and LA-ICP optical emission spectroscopy (LA-ICP-OES) have been reported for simultaneous analysis of the elements in asphaltenes\cite{Asphaltenes_LIBS} .
Elemental analysis was  done by LA-ICP-MS as  well. At the same time the aromatic/paraffinic nature was also determined by LIBS from H/C ratio.
Simultaneous LIBS and LA-ICP- mass spectroscopy (LA-ICP-MS) has been reported for spatially resolved mapping of major and trace elements Bastnasite rare earth ore \cite{LA-ICP-MS_JAAS}. 
The combination of two techniques 
provides complementary measurements that can be achieved with separate measurements due to low sensitivity or strong interferences.
Combined  LIBS/LA-ICP-MS has been reported for polymer alteration under corrosive conditions \cite{LA_ICP_MS_LIBS_SR}.
It has been demonstrated that LIBS/LA-ICP-MS is a powerful method for polymer characterization as well in the study of polymer degradation.

\subsection{ LIBS- X-Ray Fluorescence (XRF)}
Double pulse LIBS and micro-x-ray fluorescence (micro-XRF) were reported for characterizing materials \cite{Spie_LIBS_XRF}. They have found that LIBS is highly subjective to sample chemical and physical properties. Performance of portable LIBS and portable XRF device has been reported by  Rao et. al\cite{D1JA00404B}. While portable LIBS can give instantaneous measurement, its  accuracy is hampered by self-absorption. On the other hand, XRF measurements  have better limit of detection but the measurement consumes time.

\section{concluding remarks}

In this  brief review we sketch LIBS techniques from parameter estimation to the emerging  trends and also projected applications. In this article we have also introduced application aspects of some hitherto not exploited much phenomena e.g. colliding plasma,  self-reversal of lines and GIBS. Further, we believe that glimpses outlined in the present review  will provide a sound platform to the beginners also as it encompasses brief phenomenological aspects as well as recent developments in the field.

\section{References}

\begin{thebibliography}{140}%
\makeatletter
\providecommand \@ifxundefined [1]{%
 \@ifx{#1\undefined}
}%
\providecommand \@ifnum [1]{%
 \ifnum #1\expandafter \@firstoftwo
 \else \expandafter \@secondoftwo
 \fi
}%
\providecommand \@ifx [1]{%
 \ifx #1\expandafter \@firstoftwo
 \else \expandafter \@secondoftwo
 \fi
}%
\providecommand \natexlab [1]{#1}%
\providecommand \enquote  [1]{``#1''}%
\providecommand \bibnamefont  [1]{#1}%
\providecommand \bibfnamefont [1]{#1}%
\providecommand \citenamefont [1]{#1}%
\providecommand \href@noop [0]{\@secondoftwo}%
\providecommand \href [0]{\begingroup \@sanitize@url \@href}%
\providecommand \@href[1]{\@@startlink{#1}\@@href}%
\providecommand \@@href[1]{\endgroup#1\@@endlink}%
\providecommand \@sanitize@url [0]{\catcode `\\12\catcode `\$12\catcode
  `\&12\catcode `\#12\catcode `\^12\catcode `\_12\catcode `\%12\relax}%
\providecommand \@@startlink[1]{}%
\providecommand \@@endlink[0]{}%
\providecommand \url  [0]{\begingroup\@sanitize@url \@url }%
\providecommand \@url [1]{\endgroup\@href {#1}{\urlprefix }}%
\providecommand \urlprefix  [0]{URL }%
\providecommand \Eprint [0]{\href }%
\providecommand \doibase [0]{http://dx.doi.org/}%
\providecommand \selectlanguage [0]{\@gobble}%
\providecommand \bibinfo  [0]{\@secondoftwo}%
\providecommand \bibfield  [0]{\@secondoftwo}%
\providecommand \translation [1]{[#1]}%
\providecommand \BibitemOpen [0]{}%
\providecommand \bibitemStop [0]{}%
\providecommand \bibitemNoStop [0]{.\EOS\space}%
\providecommand \EOS [0]{\spacefactor3000\relax}%
\providecommand \BibitemShut  [1]{\csname bibitem#1\endcsname}%
\let\auto@bib@innerbib\@empty
\bibitem [{\citenamefont {Berlo}\ \emph {et~al.}(2022)\citenamefont {Berlo},
  \citenamefont {Xia}, \citenamefont {Zwillich}, \citenamefont {Gibbons},
  \citenamefont {Gaudiuso}, \citenamefont {Ewusi-Annan}, \citenamefont
  {Chiklis},\ and\ \citenamefont {Melikechi}}]{Berlo2022}%
  \BibitemOpen
  \bibfield  {author} {\bibinfo {author} {\bibfnamefont {K.}~\bibnamefont
  {Berlo}}, \bibinfo {author} {\bibfnamefont {W.}~\bibnamefont {Xia}}, \bibinfo
  {author} {\bibfnamefont {F.}~\bibnamefont {Zwillich}}, \bibinfo {author}
  {\bibfnamefont {E.}~\bibnamefont {Gibbons}}, \bibinfo {author} {\bibfnamefont
  {R.}~\bibnamefont {Gaudiuso}}, \bibinfo {author} {\bibfnamefont
  {E.}~\bibnamefont {Ewusi-Annan}}, \bibinfo {author} {\bibfnamefont {G.~R.}\
  \bibnamefont {Chiklis}}, \ and\ \bibinfo {author} {\bibfnamefont
  {N.}~\bibnamefont {Melikechi}},\ }\bibfield  {title} {\enquote {\bibinfo
  {title} {Laser induced breakdown spectroscopy for the rapid detection of
  sars-cov-2 immune response in plasma},}\ }\href {\doibase
  10.1038/s41598-022-05509-z} {\bibfield  {journal} {\bibinfo  {journal}
  {Scientific Reports}\ }\textbf {\bibinfo {volume} {12}},\ \bibinfo {pages}
  {1614} (\bibinfo {year} {2022})}\BibitemShut {NoStop}%
\bibitem [{\citenamefont {Fu}, \citenamefont {Li},\ and\ \citenamefont
  {Dong}(2020)}]{Fu_2020_Front}%
  \BibitemOpen
  \bibfield  {author} {\bibinfo {author} {\bibfnamefont {X.}~\bibnamefont
  {Fu}}, \bibinfo {author} {\bibfnamefont {G.}~\bibnamefont {Li}}, \ and\
  \bibinfo {author} {\bibfnamefont {D.}~\bibnamefont {Dong}},\ }\bibfield
  {title} {\enquote {\bibinfo {title} {Improving the detection sensitivity for
  laser-induced breakdown spectroscopy: A review},}\ }\href {\doibase
  10.3389/fphy.2020.00068} {\bibfield  {journal} {\bibinfo  {journal}
  {Frontiers in Physics}\ }\textbf {\bibinfo {volume} {8}},\ \bibinfo {pages}
  {68} (\bibinfo {year} {2020})}\BibitemShut {NoStop}%
\bibitem [{\citenamefont {Andrade}, \citenamefont {Pereira-Filho},\ and\
  \citenamefont {Amarasiriwardena}(2021{\natexlab{a}})}]{Daniel_2021_ASR}%
  \BibitemOpen
  \bibfield  {author} {\bibinfo {author} {\bibfnamefont {D.~F.}\ \bibnamefont
  {Andrade}}, \bibinfo {author} {\bibfnamefont {E.~R.}\ \bibnamefont
  {Pereira-Filho}}, \ and\ \bibinfo {author} {\bibfnamefont {D.}~\bibnamefont
  {Amarasiriwardena}},\ }\bibfield  {title} {\enquote {\bibinfo {title}
  {Current trends in laser-induced breakdown spectroscopy: a tutorial
  review},}\ }\href {\doibase 10.1080/05704928.2020.1739063} {\bibfield
  {journal} {\bibinfo  {journal} {Applied Spectroscopy Reviews}\ }\textbf
  {\bibinfo {volume} {56}},\ \bibinfo {pages} {98--114} (\bibinfo {year}
  {2021}{\natexlab{a}})},\ \Eprint
  {http://arxiv.org/abs/https://doi.org/10.1080/05704928.2020.1739063}
  {https://doi.org/10.1080/05704928.2020.1739063} \BibitemShut {NoStop}%
\bibitem [{\citenamefont {Senesi}, \citenamefont {Harmon},\ and\ \citenamefont
  {Hark}(2021)}]{SENESI2021106013}%
  \BibitemOpen
  \bibfield  {author} {\bibinfo {author} {\bibfnamefont {G.~S.}\ \bibnamefont
  {Senesi}}, \bibinfo {author} {\bibfnamefont {R.~S.}\ \bibnamefont {Harmon}},
  \ and\ \bibinfo {author} {\bibfnamefont {R.~R.}\ \bibnamefont {Hark}},\
  }\bibfield  {title} {\enquote {\bibinfo {title} {Field-portable and handheld
  laser-induced breakdown spectroscopy: Historical review, current status and
  future prospects},}\ }\href {\doibase
  https://doi.org/10.1016/j.sab.2020.106013} {\bibfield  {journal} {\bibinfo
  {journal} {Spectrochimica Acta Part B: Atomic Spectroscopy}\ }\textbf
  {\bibinfo {volume} {175}},\ \bibinfo {pages} {106013} (\bibinfo {year}
  {2021})}\BibitemShut {NoStop}%
\bibitem [{\citenamefont {Mal}\ \emph {et~al.}(2019)\citenamefont {Mal},
  \citenamefont {Junjuri}, \citenamefont {Gundawar},\ and\ \citenamefont
  {Khare}}]{C8JA00415C}%
  \BibitemOpen
  \bibfield  {author} {\bibinfo {author} {\bibfnamefont {E.}~\bibnamefont
  {Mal}}, \bibinfo {author} {\bibfnamefont {R.}~\bibnamefont {Junjuri}},
  \bibinfo {author} {\bibfnamefont {M.~K.}\ \bibnamefont {Gundawar}}, \ and\
  \bibinfo {author} {\bibfnamefont {A.}~\bibnamefont {Khare}},\ }\bibfield
  {title} {\enquote {\bibinfo {title} {Optimization of temporal window for
  application of calibration free-laser induced breakdown spectroscopy
  (cf-libs) on copper alloys in air employing a single line},}\ }\href
  {\doibase 10.1039/C8JA00415C} {\bibfield  {journal} {\bibinfo  {journal} {J.
  Anal. At. Spectrom.}\ }\textbf {\bibinfo {volume} {34}},\ \bibinfo {pages}
  {319--330} (\bibinfo {year} {2019})}\BibitemShut {NoStop}%
\bibitem [{\citenamefont {Arora}, \citenamefont {Thomas},\ and\ \citenamefont
  {Joshi}(2022)}]{Garima_Jaas_2022}%
  \BibitemOpen
  \bibfield  {author} {\bibinfo {author} {\bibfnamefont {G.}~\bibnamefont
  {Arora}}, \bibinfo {author} {\bibfnamefont {J.}~\bibnamefont {Thomas}}, \
  and\ \bibinfo {author} {\bibfnamefont {H.~C.}\ \bibnamefont {Joshi}},\
  }\bibfield  {title} {\enquote {\bibinfo {title} {On the delayed emission from
  a laser-produced aluminum plasma under an argon environment},}\ }\href
  {\doibase 10.1039/D2JA00065B} {\bibfield  {journal} {\bibinfo  {journal} {J.
  Anal. At. Spectrom.}\ }\textbf {\bibinfo {volume} {37}},\ \bibinfo {pages}
  {1119--1125} (\bibinfo {year} {2022})}\BibitemShut {NoStop}%
\bibitem [{\citenamefont {Hahn}\ and\ \citenamefont
  {Omenetto}(2012)}]{Hahn_2012_Appl_Spec}%
  \BibitemOpen
  \bibfield  {author} {\bibinfo {author} {\bibfnamefont {D.~W.}\ \bibnamefont
  {Hahn}}\ and\ \bibinfo {author} {\bibfnamefont {N.}~\bibnamefont
  {Omenetto}},\ }\bibfield  {title} {\enquote {\bibinfo {title} {Laser-induced
  breakdown spectroscopy (libs), part ii: Review of instrumental and
  methodological approaches to material analysis and applications to different
  fields},}\ }\href
  {http://www.osapublishing.org/as/abstract.cfm?URI=as-66-4-347} {\bibfield
  {journal} {\bibinfo  {journal} {Appl. Spectrosc.}\ }\textbf {\bibinfo
  {volume} {66}},\ \bibinfo {pages} {347--419} (\bibinfo {year}
  {2012})}\BibitemShut {NoStop}%
\bibitem [{\citenamefont {Hahn}\ and\ \citenamefont
  {Omenetto}(2010)}]{Hahn_2010_Appl_Spec}%
  \BibitemOpen
  \bibfield  {author} {\bibinfo {author} {\bibfnamefont {D.~W.}\ \bibnamefont
  {Hahn}}\ and\ \bibinfo {author} {\bibfnamefont {N.}~\bibnamefont
  {Omenetto}},\ }\bibfield  {title} {\enquote {\bibinfo {title} {Laser-induced
  breakdown spectroscopy (libs), part i: Review of basic diagnostics and
  plasma--particle interactions: Still-challenging issues within the analytical
  plasma community},}\ }\href
  {http://www.osapublishing.org/as/abstract.cfm?URI=as-64-12-335A} {\bibfield
  {journal} {\bibinfo  {journal} {Appl. Spectrosc.}\ }\textbf {\bibinfo
  {volume} {64}},\ \bibinfo {pages} {335A--366A} (\bibinfo {year}
  {2010})}\BibitemShut {NoStop}%
\bibitem [{\citenamefont {Pasquini}\ \emph {et~al.}(2007)\citenamefont
  {Pasquini}, \citenamefont {Cortez}, \citenamefont {Silva},\ and\
  \citenamefont {Gonzaga}}]{pasquini2007laser_JBC}%
  \BibitemOpen
  \bibfield  {author} {\bibinfo {author} {\bibfnamefont {C.}~\bibnamefont
  {Pasquini}}, \bibinfo {author} {\bibfnamefont {J.}~\bibnamefont {Cortez}},
  \bibinfo {author} {\bibfnamefont {L.}~\bibnamefont {Silva}}, \ and\ \bibinfo
  {author} {\bibfnamefont {F.~B.}\ \bibnamefont {Gonzaga}},\ }\bibfield
  {title} {\enquote {\bibinfo {title} {Laser induced breakdown spectroscopy},}\
  }\href@noop {} {\bibfield  {journal} {\bibinfo  {journal} {Journal of the
  Brazilian Chemical Society}\ }\textbf {\bibinfo {volume} {18}},\ \bibinfo
  {pages} {463--512} (\bibinfo {year} {2007})}\BibitemShut {NoStop}%
\bibitem [{\citenamefont {Singh}\ and\ \citenamefont
  {Thakur}(2020)}]{singh2020laser}%
  \BibitemOpen
  \bibfield  {author} {\bibinfo {author} {\bibfnamefont {J.~P.}\ \bibnamefont
  {Singh}}\ and\ \bibinfo {author} {\bibfnamefont {S.~N.}\ \bibnamefont
  {Thakur}},\ }\href@noop {} {\emph {\bibinfo {title} {Laser-induced breakdown
  spectroscopy}}}\ (\bibinfo  {publisher} {Elsevier},\ \bibinfo {year}
  {2020})\BibitemShut {NoStop}%
\bibitem [{\citenamefont {Singh}\ and\ \citenamefont
  {Thakur}(2007)}]{singh2007laser}%
  \BibitemOpen
  \bibfield  {author} {\bibinfo {author} {\bibfnamefont {J.~P.}\ \bibnamefont
  {Singh}}\ and\ \bibinfo {author} {\bibfnamefont {S.}~\bibnamefont {Thakur}},\
  }\href@noop {} {\emph {\bibinfo {title} {Laser-Induced Breakdown
  Spectroscopy}}}\ (\bibinfo  {publisher} {Elsevier},\ \bibinfo {year}
  {2007})\BibitemShut {NoStop}%
\bibitem [{\citenamefont {Miziolek}, \citenamefont {Palleschi},\ and\
  \citenamefont {Schechter}(2006)}]{miziolek2006laser}%
  \BibitemOpen
  \bibfield  {author} {\bibinfo {author} {\bibfnamefont {A.~W.}\ \bibnamefont
  {Miziolek}}, \bibinfo {author} {\bibfnamefont {V.}~\bibnamefont {Palleschi}},
  \ and\ \bibinfo {author} {\bibfnamefont {I.}~\bibnamefont {Schechter}},\
  }\href@noop {} {\emph {\bibinfo {title} {Laser induced breakdown
  spectroscopy}}}\ (\bibinfo  {publisher} {Cambridge university press},\
  \bibinfo {year} {2006})\BibitemShut {NoStop}%
\bibitem [{\citenamefont {Pablo}(2021)}]{PabloAF}%
  \BibitemOpen
  \bibfield  {author} {\bibinfo {author} {\bibfnamefont {A.~F.}\ \bibnamefont
  {Pablo}},\ }\href@noop {} {\emph {\bibinfo {title} {A Guide to Laser-induced
  Breakdown Spectroscopy}}}\ (\bibinfo  {publisher} {Nova Science Pub Inc},\
  \bibinfo {year} {2021})\BibitemShut {NoStop}%
\bibitem [{\citenamefont {Noll}(2012)}]{noll2012laser}%
  \BibitemOpen
  \bibfield  {author} {\bibinfo {author} {\bibfnamefont {R.}~\bibnamefont
  {Noll}},\ }\bibfield  {title} {\enquote {\bibinfo {title} {Laser-induced
  breakdown spectroscopy},}\ }in\ \href@noop {} {\emph {\bibinfo {booktitle}
  {Laser-Induced Breakdown Spectroscopy}}}\ (\bibinfo  {publisher} {Springer},\
  \bibinfo {year} {2012})\ pp.\ \bibinfo {pages} {7--15}\BibitemShut {NoStop}%
\bibitem [{\citenamefont {Zhang}\ \emph
  {et~al.}(2022{\natexlab{a}})\citenamefont {Zhang}, \citenamefont {Zhang},
  \citenamefont {Zhao}, \citenamefont {Chen}, \citenamefont {Ke}, \citenamefont
  {Xu},\ and\ \citenamefont {He}}]{ASR_Review_LIBS_1}%
  \BibitemOpen
  \bibfield  {author} {\bibinfo {author} {\bibfnamefont {D.}~\bibnamefont
  {Zhang}}, \bibinfo {author} {\bibfnamefont {H.}~\bibnamefont {Zhang}},
  \bibinfo {author} {\bibfnamefont {Y.}~\bibnamefont {Zhao}}, \bibinfo {author}
  {\bibfnamefont {Y.}~\bibnamefont {Chen}}, \bibinfo {author} {\bibfnamefont
  {C.}~\bibnamefont {Ke}}, \bibinfo {author} {\bibfnamefont {T.}~\bibnamefont
  {Xu}}, \ and\ \bibinfo {author} {\bibfnamefont {Y.}~\bibnamefont {He}},\
  }\bibfield  {title} {\enquote {\bibinfo {title} {A brief review of new data
  analysis methods of laser-induced breakdown spectroscopy: machine
  learning},}\ }\href {\doibase 10.1080/05704928.2020.1843175} {\bibfield
  {journal} {\bibinfo  {journal} {Applied Spectroscopy Reviews}\ }\textbf
  {\bibinfo {volume} {57}},\ \bibinfo {pages} {89--111} (\bibinfo {year}
  {2022}{\natexlab{a}})},\ \Eprint
  {http://arxiv.org/abs/https://doi.org/10.1080/05704928.2020.1843175}
  {https://doi.org/10.1080/05704928.2020.1843175} \BibitemShut {NoStop}%
\bibitem [{\citenamefont {Romppanen}\ \emph {et~al.}(2021)\citenamefont
  {Romppanen}, \citenamefont {Pölönen}, \citenamefont {Häkkänen},\ and\
  \citenamefont {Kaski}}]{ASR_LIBS_Application1}%
  \BibitemOpen
  \bibfield  {author} {\bibinfo {author} {\bibfnamefont {S.}~\bibnamefont
  {Romppanen}}, \bibinfo {author} {\bibfnamefont {I.}~\bibnamefont
  {Pölönen}}, \bibinfo {author} {\bibfnamefont {H.}~\bibnamefont
  {Häkkänen}}, \ and\ \bibinfo {author} {\bibfnamefont {S.}~\bibnamefont
  {Kaski}},\ }\bibfield  {title} {\enquote {\bibinfo {title} {Optimization of
  spodumene identification by statistical approach for laser-induced breakdown
  spectroscopy data of lithium pegmatite ores},}\ }\href {\doibase
  10.1080/05704928.2021.1963977} {\bibfield  {journal} {\bibinfo  {journal}
  {Applied Spectroscopy Reviews}\ }\textbf {\bibinfo {volume} {0}},\ \bibinfo
  {pages} {1--21} (\bibinfo {year} {2021})},\ \Eprint
  {http://arxiv.org/abs/https://doi.org/10.1080/05704928.2021.1963977}
  {https://doi.org/10.1080/05704928.2021.1963977} \BibitemShut {NoStop}%
\bibitem [{\citenamefont {Al-Najjar}\ \emph {et~al.}(2022)\citenamefont
  {Al-Najjar}, \citenamefont {Wudil}, \citenamefont {Ahmad}, \citenamefont
  {Al-Amoudi}, \citenamefont {Al-Osta},\ and\ \citenamefont
  {Gondal}}]{ASR_LIBS_Review_2}%
  \BibitemOpen
  \bibfield  {author} {\bibinfo {author} {\bibfnamefont {O.~A.}\ \bibnamefont
  {Al-Najjar}}, \bibinfo {author} {\bibfnamefont {Y.~S.}\ \bibnamefont
  {Wudil}}, \bibinfo {author} {\bibfnamefont {U.~F.}\ \bibnamefont {Ahmad}},
  \bibinfo {author} {\bibfnamefont {O.~S.~B.}\ \bibnamefont {Al-Amoudi}},
  \bibinfo {author} {\bibfnamefont {M.~A.}\ \bibnamefont {Al-Osta}}, \ and\
  \bibinfo {author} {\bibfnamefont {M.~A.}\ \bibnamefont {Gondal}},\ }\bibfield
   {title} {\enquote {\bibinfo {title} {Applications of laser induced breakdown
  spectroscopy in geotechnical engineering: a critical review of recent
  developments, perspectives and challenges},}\ }\href {\doibase
  10.1080/05704928.2022.2136192} {\bibfield  {journal} {\bibinfo  {journal}
  {Applied Spectroscopy Reviews}\ }\textbf {\bibinfo {volume} {0}},\ \bibinfo
  {pages} {1--37} (\bibinfo {year} {2022})},\ \Eprint
  {http://arxiv.org/abs/https://doi.org/10.1080/05704928.2022.2136192}
  {https://doi.org/10.1080/05704928.2022.2136192} \BibitemShut {NoStop}%
\bibitem [{\citenamefont {Andrade}, \citenamefont {Pereira-Filho},\ and\
  \citenamefont
  {Amarasiriwardena}(2021{\natexlab{b}})}]{ASR_Tutorial_review_2020}%
  \BibitemOpen
  \bibfield  {author} {\bibinfo {author} {\bibfnamefont {D.~F.}\ \bibnamefont
  {Andrade}}, \bibinfo {author} {\bibfnamefont {E.~R.}\ \bibnamefont
  {Pereira-Filho}}, \ and\ \bibinfo {author} {\bibfnamefont {D.}~\bibnamefont
  {Amarasiriwardena}},\ }\bibfield  {title} {\enquote {\bibinfo {title}
  {Current trends in laser-induced breakdown spectroscopy: a tutorial
  review},}\ }\href {\doibase 10.1080/05704928.2020.1739063} {\bibfield
  {journal} {\bibinfo  {journal} {Applied Spectroscopy Reviews}\ }\textbf
  {\bibinfo {volume} {56}},\ \bibinfo {pages} {98--114} (\bibinfo {year}
  {2021}{\natexlab{b}})},\ \Eprint
  {http://arxiv.org/abs/https://doi.org/10.1080/05704928.2020.1739063}
  {https://doi.org/10.1080/05704928.2020.1739063} \BibitemShut {NoStop}%
\bibitem [{\citenamefont {Kautz}\ \emph
  {et~al.}(2021{\natexlab{a}})\citenamefont {Kautz}, \citenamefont
  {Weerakkody}, \citenamefont {Finko}, \citenamefont {Curreli}, \citenamefont
  {Koroglu}, \citenamefont {Rose}, \citenamefont {Weisz}, \citenamefont
  {Crowhurst}, \citenamefont {Radousky}, \citenamefont {DeMagistris},
  \citenamefont {Sinha}, \citenamefont {Levin}, \citenamefont {Dreizin},
  \citenamefont {Phillips}, \citenamefont {Glumac},\ and\ \citenamefont
  {Harilal}}]{KAUTZ2021106283}%
  \BibitemOpen
  \bibfield  {author} {\bibinfo {author} {\bibfnamefont {E.~J.}\ \bibnamefont
  {Kautz}}, \bibinfo {author} {\bibfnamefont {E.~N.}\ \bibnamefont
  {Weerakkody}}, \bibinfo {author} {\bibfnamefont {M.~S.}\ \bibnamefont
  {Finko}}, \bibinfo {author} {\bibfnamefont {D.}~\bibnamefont {Curreli}},
  \bibinfo {author} {\bibfnamefont {B.}~\bibnamefont {Koroglu}}, \bibinfo
  {author} {\bibfnamefont {T.~P.}\ \bibnamefont {Rose}}, \bibinfo {author}
  {\bibfnamefont {D.~G.}\ \bibnamefont {Weisz}}, \bibinfo {author}
  {\bibfnamefont {J.~C.}\ \bibnamefont {Crowhurst}}, \bibinfo {author}
  {\bibfnamefont {H.~B.}\ \bibnamefont {Radousky}}, \bibinfo {author}
  {\bibfnamefont {M.}~\bibnamefont {DeMagistris}}, \bibinfo {author}
  {\bibfnamefont {N.}~\bibnamefont {Sinha}}, \bibinfo {author} {\bibfnamefont
  {D.~A.}\ \bibnamefont {Levin}}, \bibinfo {author} {\bibfnamefont {E.~L.}\
  \bibnamefont {Dreizin}}, \bibinfo {author} {\bibfnamefont {M.~C.}\
  \bibnamefont {Phillips}}, \bibinfo {author} {\bibfnamefont {N.~G.}\
  \bibnamefont {Glumac}}, \ and\ \bibinfo {author} {\bibfnamefont {S.~S.}\
  \bibnamefont {Harilal}},\ }\bibfield  {title} {\enquote {\bibinfo {title}
  {Optical spectroscopy and modeling of uranium gas-phase oxidation: Progress
  and perspectives},}\ }\href {\doibase
  https://doi.org/10.1016/j.sab.2021.106283} {\bibfield  {journal} {\bibinfo
  {journal} {Spectrochimica Acta Part B: Atomic Spectroscopy}\ }\textbf
  {\bibinfo {volume} {185}},\ \bibinfo {pages} {106283} (\bibinfo {year}
  {2021}{\natexlab{a}})}\BibitemShut {NoStop}%
\bibitem [{\citenamefont {Maurya}\ \emph {et~al.}(2020)\citenamefont {Maurya},
  \citenamefont {Marín-Roldán}, \citenamefont {Veis}, \citenamefont
  {Pathak},\ and\ \citenamefont {Sen}}]{MAURYA2020152417}%
  \BibitemOpen
  \bibfield  {author} {\bibinfo {author} {\bibfnamefont {G.~S.}\ \bibnamefont
  {Maurya}}, \bibinfo {author} {\bibfnamefont {A.}~\bibnamefont
  {Marín-Roldán}}, \bibinfo {author} {\bibfnamefont {P.}~\bibnamefont
  {Veis}}, \bibinfo {author} {\bibfnamefont {A.~K.}\ \bibnamefont {Pathak}}, \
  and\ \bibinfo {author} {\bibfnamefont {P.}~\bibnamefont {Sen}},\ }\bibfield
  {title} {\enquote {\bibinfo {title} {A review of the libs analysis for the
  plasma-facing components diagnostics},}\ }\href {\doibase
  https://doi.org/10.1016/j.jnucmat.2020.152417} {\bibfield  {journal}
  {\bibinfo  {journal} {Journal of Nuclear Materials}\ }\textbf {\bibinfo
  {volume} {541}},\ \bibinfo {pages} {152417} (\bibinfo {year}
  {2020})}\BibitemShut {NoStop}%
\bibitem [{\citenamefont {Li}\ \emph {et~al.}(2016)\citenamefont {Li},
  \citenamefont {Feng}, \citenamefont {Oderji}, \citenamefont {Luo},\ and\
  \citenamefont {Ding}}]{li2016review}%
  \BibitemOpen
  \bibfield  {author} {\bibinfo {author} {\bibfnamefont {C.}~\bibnamefont
  {Li}}, \bibinfo {author} {\bibfnamefont {C.-L.}\ \bibnamefont {Feng}},
  \bibinfo {author} {\bibfnamefont {H.~Y.}\ \bibnamefont {Oderji}}, \bibinfo
  {author} {\bibfnamefont {G.-N.}\ \bibnamefont {Luo}}, \ and\ \bibinfo
  {author} {\bibfnamefont {H.-B.}\ \bibnamefont {Ding}},\ }\bibfield  {title}
  {\enquote {\bibinfo {title} {Review of libs application in nuclear fusion
  technology},}\ }\href@noop {} {\bibfield  {journal} {\bibinfo  {journal}
  {Frontiers of Physics}\ }\textbf {\bibinfo {volume} {11}},\ \bibinfo {pages}
  {1--16} (\bibinfo {year} {2016})}\BibitemShut {NoStop}%
\bibitem [{\citenamefont {Jolivet}\ \emph {et~al.}(2019)\citenamefont
  {Jolivet}, \citenamefont {Leprince}, \citenamefont {Moncayo}, \citenamefont
  {Sorbier}, \citenamefont {Lienemann},\ and\ \citenamefont
  {Motto-Ros}}]{JOLIVET201941}%
  \BibitemOpen
  \bibfield  {author} {\bibinfo {author} {\bibfnamefont {L.}~\bibnamefont
  {Jolivet}}, \bibinfo {author} {\bibfnamefont {M.}~\bibnamefont {Leprince}},
  \bibinfo {author} {\bibfnamefont {S.}~\bibnamefont {Moncayo}}, \bibinfo
  {author} {\bibfnamefont {L.}~\bibnamefont {Sorbier}}, \bibinfo {author}
  {\bibfnamefont {C.-P.}\ \bibnamefont {Lienemann}}, \ and\ \bibinfo {author}
  {\bibfnamefont {V.}~\bibnamefont {Motto-Ros}},\ }\bibfield  {title} {\enquote
  {\bibinfo {title} {Review of the recent advances and applications of
  libs-based imaging},}\ }\href {\doibase
  https://doi.org/10.1016/j.sab.2018.11.008} {\bibfield  {journal} {\bibinfo
  {journal} {Spectrochimica Acta Part B: Atomic Spectroscopy}\ }\textbf
  {\bibinfo {volume} {151}},\ \bibinfo {pages} {41--53} (\bibinfo {year}
  {2019})}\BibitemShut {NoStop}%
\bibitem [{\citenamefont {Pedarnig}\ \emph {et~al.}(2021)\citenamefont
  {Pedarnig}, \citenamefont {Trautner}, \citenamefont {Gr{\"u}nberger},
  \citenamefont {Giannakaris}, \citenamefont {Eschlb{\"o}ck-Fuchs},\ and\
  \citenamefont {Hofstadler}}]{app11199274}%
  \BibitemOpen
  \bibfield  {author} {\bibinfo {author} {\bibfnamefont {J.~D.}\ \bibnamefont
  {Pedarnig}}, \bibinfo {author} {\bibfnamefont {S.}~\bibnamefont {Trautner}},
  \bibinfo {author} {\bibfnamefont {S.}~\bibnamefont {Gr{\"u}nberger}},
  \bibinfo {author} {\bibfnamefont {N.}~\bibnamefont {Giannakaris}}, \bibinfo
  {author} {\bibfnamefont {S.}~\bibnamefont {Eschlb{\"o}ck-Fuchs}}, \ and\
  \bibinfo {author} {\bibfnamefont {J.}~\bibnamefont {Hofstadler}},\ }\href
  {\doibase 10.3390/app11199274} {\enquote {\bibinfo {title} {Review of element
  analysis of industrial materials by in-line laser---induced breakdown
  spectroscopy (libs)},}\ } (\bibinfo {year} {2021})\BibitemShut {NoStop}%
\bibitem [{\citenamefont {Zhang}\ \emph {et~al.}(2020)\citenamefont {Zhang},
  \citenamefont {Zhang}, \citenamefont {Zhao}, \citenamefont {Chen},
  \citenamefont {Ke}, \citenamefont {Xu},\ and\ \citenamefont
  {He}}]{Dianxin_ApplSpecReview}%
  \BibitemOpen
  \bibfield  {author} {\bibinfo {author} {\bibfnamefont {D.}~\bibnamefont
  {Zhang}}, \bibinfo {author} {\bibfnamefont {H.}~\bibnamefont {Zhang}},
  \bibinfo {author} {\bibfnamefont {Y.}~\bibnamefont {Zhao}}, \bibinfo {author}
  {\bibfnamefont {Y.}~\bibnamefont {Chen}}, \bibinfo {author} {\bibfnamefont
  {C.}~\bibnamefont {Ke}}, \bibinfo {author} {\bibfnamefont {T.}~\bibnamefont
  {Xu}}, \ and\ \bibinfo {author} {\bibfnamefont {Y.}~\bibnamefont {He}},\
  }\bibfield  {title} {\enquote {\bibinfo {title} {A brief review of new data
  analysis methods of laser-induced breakdown spectroscopy: machine
  learning},}\ }\href {\doibase 10.1080/05704928.2020.1843175} {\bibfield
  {journal} {\bibinfo  {journal} {Applied Spectroscopy Reviews}\ }\textbf
  {\bibinfo {volume} {0}},\ \bibinfo {pages} {1--23} (\bibinfo {year}
  {2020})},\ \Eprint
  {http://arxiv.org/abs/https://doi.org/10.1080/05704928.2020.1843175}
  {https://doi.org/10.1080/05704928.2020.1843175} \BibitemShut {NoStop}%
\bibitem [{\citenamefont {Galbács}\ \emph {et~al.}(2021)\citenamefont
  {Galbács}, \citenamefont {Kéri}, \citenamefont {Kohut}, \citenamefont
  {Veres},\ and\ \citenamefont {Geretovszky}}]{D1JA00149C}%
  \BibitemOpen
  \bibfield  {author} {\bibinfo {author} {\bibfnamefont {G.}~\bibnamefont
  {Galbács}}, \bibinfo {author} {\bibfnamefont {A.}~\bibnamefont {Kéri}},
  \bibinfo {author} {\bibfnamefont {A.}~\bibnamefont {Kohut}}, \bibinfo
  {author} {\bibfnamefont {M.}~\bibnamefont {Veres}}, \ and\ \bibinfo {author}
  {\bibfnamefont {Z.}~\bibnamefont {Geretovszky}},\ }\bibfield  {title}
  {\enquote {\bibinfo {title} {Nanoparticles in analytical laser and plasma
  spectroscopy – a review of recent developments in methodology and
  applications},}\ }\href {\doibase 10.1039/D1JA00149C} {\bibfield  {journal}
  {\bibinfo  {journal} {J. Anal. At. Spectrom.}\ }\textbf {\bibinfo {volume}
  {36}},\ \bibinfo {pages} {1826--1872} (\bibinfo {year} {2021})}\BibitemShut
  {NoStop}%
\bibitem [{\citenamefont {Dell'Aglio}, \citenamefont {Alrifai},\ and\
  \citenamefont {{De Giacomo}}(2018)}]{DELLAGLIO2018105}%
  \BibitemOpen
  \bibfield  {author} {\bibinfo {author} {\bibfnamefont {M.}~\bibnamefont
  {Dell'Aglio}}, \bibinfo {author} {\bibfnamefont {R.}~\bibnamefont {Alrifai}},
  \ and\ \bibinfo {author} {\bibfnamefont {A.}~\bibnamefont {{De Giacomo}}},\
  }\bibfield  {title} {\enquote {\bibinfo {title} {Nanoparticle enhanced laser
  induced breakdown spectroscopy (nelibs), a first review},}\ }\href {\doibase
  https://doi.org/10.1016/j.sab.2018.06.008} {\bibfield  {journal} {\bibinfo
  {journal} {Spectrochimica Acta Part B: Atomic Spectroscopy}\ }\textbf
  {\bibinfo {volume} {148}},\ \bibinfo {pages} {105--112} (\bibinfo {year}
  {2018})}\BibitemShut {NoStop}%
\bibitem [{\citenamefont {Legnaioli}\ \emph {et~al.}(2020)\citenamefont
  {Legnaioli}, \citenamefont {Campanella}, \citenamefont {Poggialini},
  \citenamefont {Pagnotta}, \citenamefont {Harith}, \citenamefont
  {Abdel-Salam},\ and\ \citenamefont {Palleschi}}]{C9AY02728A}%
  \BibitemOpen
  \bibfield  {author} {\bibinfo {author} {\bibfnamefont {S.}~\bibnamefont
  {Legnaioli}}, \bibinfo {author} {\bibfnamefont {B.}~\bibnamefont
  {Campanella}}, \bibinfo {author} {\bibfnamefont {F.}~\bibnamefont
  {Poggialini}}, \bibinfo {author} {\bibfnamefont {S.}~\bibnamefont
  {Pagnotta}}, \bibinfo {author} {\bibfnamefont {M.~A.}\ \bibnamefont
  {Harith}}, \bibinfo {author} {\bibfnamefont {Z.~A.}\ \bibnamefont
  {Abdel-Salam}}, \ and\ \bibinfo {author} {\bibfnamefont {V.}~\bibnamefont
  {Palleschi}},\ }\bibfield  {title} {\enquote {\bibinfo {title} {Industrial
  applications of laser-induced breakdown spectroscopy: a review},}\ }\href
  {\doibase 10.1039/C9AY02728A} {\bibfield  {journal} {\bibinfo  {journal}
  {Anal. Methods}\ }\textbf {\bibinfo {volume} {12}},\ \bibinfo {pages}
  {1014--1029} (\bibinfo {year} {2020})}\BibitemShut {NoStop}%
\bibitem [{\citenamefont {MATSUMOTO}\ and\ \citenamefont
  {SAKKA}(2021)}]{Ayumu_underwater}%
  \BibitemOpen
  \bibfield  {author} {\bibinfo {author} {\bibfnamefont {A.}~\bibnamefont
  {MATSUMOTO}}\ and\ \bibinfo {author} {\bibfnamefont {T.}~\bibnamefont
  {SAKKA}},\ }\bibfield  {title} {\enquote {\bibinfo {title} {A review of
  underwater laser-induced breakdown spectroscopy of submerged solids},}\
  }\href {\doibase 10.2116/analsci.20R007} {\bibfield  {journal} {\bibinfo
  {journal} {Analytical Sciences}\ }\textbf {\bibinfo {volume} {37}},\ \bibinfo
  {pages} {1061--1072} (\bibinfo {year} {2021})}\BibitemShut {NoStop}%
\bibitem [{\citenamefont {Stefas}\ \emph {et~al.}(2021)\citenamefont {Stefas},
  \citenamefont {Gyftokostas}, \citenamefont {Nanou}, \citenamefont
  {Kourelias},\ and\ \citenamefont {Couris}}]{molecules_libs_food}%
  \BibitemOpen
  \bibfield  {author} {\bibinfo {author} {\bibfnamefont {D.}~\bibnamefont
  {Stefas}}, \bibinfo {author} {\bibfnamefont {N.}~\bibnamefont {Gyftokostas}},
  \bibinfo {author} {\bibfnamefont {E.}~\bibnamefont {Nanou}}, \bibinfo
  {author} {\bibfnamefont {P.}~\bibnamefont {Kourelias}}, \ and\ \bibinfo
  {author} {\bibfnamefont {S.}~\bibnamefont {Couris}},\ }\bibfield  {title}
  {\enquote {\bibinfo {title} {Laser-induced breakdown spectroscopy: An
  efficient tool for food science and technology (from the analysis of martian
  rocks to the analysis of olive oil, honey, milk, and other natural earth
  products)},}\ }\href {\doibase 10.3390/molecules26164981} {\bibfield
  {journal} {\bibinfo  {journal} {Molecules}\ }\textbf {\bibinfo {volume} {26}}
  (\bibinfo {year} {2021}),\ 10.3390/molecules26164981}\BibitemShut {NoStop}%
\bibitem [{\citenamefont {Kautz}\ \emph
  {et~al.}(2021{\natexlab{b}})\citenamefont {Kautz}, \citenamefont {Devaraj},
  \citenamefont {Senor},\ and\ \citenamefont {Harilal}}]{Kautz:21}%
  \BibitemOpen
  \bibfield  {author} {\bibinfo {author} {\bibfnamefont {E.~J.}\ \bibnamefont
  {Kautz}}, \bibinfo {author} {\bibfnamefont {A.}~\bibnamefont {Devaraj}},
  \bibinfo {author} {\bibfnamefont {D.~J.}\ \bibnamefont {Senor}}, \ and\
  \bibinfo {author} {\bibfnamefont {S.~S.}\ \bibnamefont {Harilal}},\
  }\bibfield  {title} {\enquote {\bibinfo {title} {Hydrogen isotopic analysis
  of nuclear reactor materials using ultrafast laser-induced breakdown
  spectroscopy},}\ }\href {\doibase 10.1364/OE.412351} {\bibfield  {journal}
  {\bibinfo  {journal} {Opt. Express}\ }\textbf {\bibinfo {volume} {29}},\
  \bibinfo {pages} {4936--4946} (\bibinfo {year}
  {2021}{\natexlab{b}})}\BibitemShut {NoStop}%
\bibitem [{\citenamefont {Burger}\ \emph {et~al.}(2019)\citenamefont {Burger},
  \citenamefont {Skrodzki}, \citenamefont {Jovanovic}, \citenamefont
  {Phillips},\ and\ \citenamefont {Harilal}}]{Hari_uranium}%
  \BibitemOpen
  \bibfield  {author} {\bibinfo {author} {\bibfnamefont {M.}~\bibnamefont
  {Burger}}, \bibinfo {author} {\bibfnamefont {P.~J.}\ \bibnamefont
  {Skrodzki}}, \bibinfo {author} {\bibfnamefont {I.}~\bibnamefont {Jovanovic}},
  \bibinfo {author} {\bibfnamefont {M.~C.}\ \bibnamefont {Phillips}}, \ and\
  \bibinfo {author} {\bibfnamefont {S.~S.}\ \bibnamefont {Harilal}},\
  }\bibfield  {title} {\enquote {\bibinfo {title} {Laser-produced uranium
  plasma characterization and stark broadening measurements},}\ }\href
  {\doibase 10.1063/1.5099643} {\bibfield  {journal} {\bibinfo  {journal}
  {Physics of Plasmas}\ }\textbf {\bibinfo {volume} {26}},\ \bibinfo {pages}
  {093103} (\bibinfo {year} {2019})},\ \Eprint
  {http://arxiv.org/abs/https://doi.org/10.1063/1.5099643}
  {https://doi.org/10.1063/1.5099643} \BibitemShut {NoStop}%
\bibitem [{\citenamefont {Khan}\ \emph {et~al.}(2022)\citenamefont {Khan},
  \citenamefont {Wang}, \citenamefont {Idrees}, \citenamefont {Xiangli},
  \citenamefont {Teng}, \citenamefont {Cui}, \citenamefont {Zhao},
  \citenamefont {Wei},\ and\ \citenamefont {Abrar}}]{LIBS_Cancer}%
  \BibitemOpen
  \bibfield  {author} {\bibinfo {author} {\bibfnamefont {M.~N.}\ \bibnamefont
  {Khan}}, \bibinfo {author} {\bibfnamefont {Q.}~\bibnamefont {Wang}}, \bibinfo
  {author} {\bibfnamefont {B.~S.}\ \bibnamefont {Idrees}}, \bibinfo {author}
  {\bibfnamefont {W.}~\bibnamefont {Xiangli}}, \bibinfo {author} {\bibfnamefont
  {G.}~\bibnamefont {Teng}}, \bibinfo {author} {\bibfnamefont {X.}~\bibnamefont
  {Cui}}, \bibinfo {author} {\bibfnamefont {Z.}~\bibnamefont {Zhao}}, \bibinfo
  {author} {\bibfnamefont {K.}~\bibnamefont {Wei}}, \ and\ \bibinfo {author}
  {\bibfnamefont {M.}~\bibnamefont {Abrar}},\ }\bibfield  {title} {\enquote
  {\bibinfo {title} {A review on laser-induced breakdown spectroscopy in
  different cancers diagnosis and classification},}\ }\href {\doibase
  10.3389/fphy.2022.821057} {\bibfield  {journal} {\bibinfo  {journal}
  {Frontiers in Physics}\ }\textbf {\bibinfo {volume} {10}} (\bibinfo {year}
  {2022}),\ 10.3389/fphy.2022.821057}\BibitemShut {NoStop}%
\bibitem [{\citenamefont {Ji}\ \emph {et~al.}(2021)\citenamefont {Ji},
  \citenamefont {Ding}, \citenamefont {Zhang}, \citenamefont {Hu},\ and\
  \citenamefont {Zhong}}]{Huview_ASR_2021}%
  \BibitemOpen
  \bibfield  {author} {\bibinfo {author} {\bibfnamefont {H.}~\bibnamefont
  {Ji}}, \bibinfo {author} {\bibfnamefont {Y.}~\bibnamefont {Ding}}, \bibinfo
  {author} {\bibfnamefont {L.}~\bibnamefont {Zhang}}, \bibinfo {author}
  {\bibfnamefont {Y.}~\bibnamefont {Hu}}, \ and\ \bibinfo {author}
  {\bibfnamefont {X.}~\bibnamefont {Zhong}},\ }\bibfield  {title} {\enquote
  {\bibinfo {title} {Review of aerosol analysis by laser-induced breakdown
  spectroscopy},}\ }\href {\doibase 10.1080/05704928.2020.1780604} {\bibfield
  {journal} {\bibinfo  {journal} {Applied Spectroscopy Reviews}\ }\textbf
  {\bibinfo {volume} {56}},\ \bibinfo {pages} {193--220} (\bibinfo {year}
  {2021})},\ \Eprint
  {http://arxiv.org/abs/https://doi.org/10.1080/05704928.2020.1780604}
  {https://doi.org/10.1080/05704928.2020.1780604} \BibitemShut {NoStop}%
\bibitem [{\citenamefont {Qiao}\ \emph {et~al.}(2015)\citenamefont {Qiao},
  \citenamefont {Ding}, \citenamefont {Tian}, \citenamefont {Yao},\ and\
  \citenamefont {Yang}}]{Shujun_ASR_2015}%
  \BibitemOpen
  \bibfield  {author} {\bibinfo {author} {\bibfnamefont {S.}~\bibnamefont
  {Qiao}}, \bibinfo {author} {\bibfnamefont {Y.}~\bibnamefont {Ding}}, \bibinfo
  {author} {\bibfnamefont {D.}~\bibnamefont {Tian}}, \bibinfo {author}
  {\bibfnamefont {L.}~\bibnamefont {Yao}}, \ and\ \bibinfo {author}
  {\bibfnamefont {G.}~\bibnamefont {Yang}},\ }\bibfield  {title} {\enquote
  {\bibinfo {title} {A review of laser-induced breakdown spectroscopy for
  analysis of geological materials},}\ }\href {\doibase
  10.1080/05704928.2014.911746} {\bibfield  {journal} {\bibinfo  {journal}
  {Applied Spectroscopy Reviews}\ }\textbf {\bibinfo {volume} {50}},\ \bibinfo
  {pages} {1--26} (\bibinfo {year} {2015})},\ \Eprint
  {http://arxiv.org/abs/https://doi.org/10.1080/05704928.2014.911746}
  {https://doi.org/10.1080/05704928.2014.911746} \BibitemShut {NoStop}%
\bibitem [{\citenamefont {Zhang}\ \emph
  {et~al.}(2022{\natexlab{b}})\citenamefont {Zhang}, \citenamefont {Ou},
  \citenamefont {Wang}, \citenamefont {Lin}, \citenamefont {Lv}, \citenamefont
  {Qin}, \citenamefont {Li}, \citenamefont {Yang}, \citenamefont {Zhao},\ and\
  \citenamefont {Zhang}}]{FP_2022_Zhang}%
  \BibitemOpen
  \bibfield  {author} {\bibinfo {author} {\bibfnamefont {N.}~\bibnamefont
  {Zhang}}, \bibinfo {author} {\bibfnamefont {T.}~\bibnamefont {Ou}}, \bibinfo
  {author} {\bibfnamefont {M.}~\bibnamefont {Wang}}, \bibinfo {author}
  {\bibfnamefont {Z.}~\bibnamefont {Lin}}, \bibinfo {author} {\bibfnamefont
  {C.}~\bibnamefont {Lv}}, \bibinfo {author} {\bibfnamefont {Y.}~\bibnamefont
  {Qin}}, \bibinfo {author} {\bibfnamefont {J.}~\bibnamefont {Li}}, \bibinfo
  {author} {\bibfnamefont {H.}~\bibnamefont {Yang}}, \bibinfo {author}
  {\bibfnamefont {N.}~\bibnamefont {Zhao}}, \ and\ \bibinfo {author}
  {\bibfnamefont {Q.}~\bibnamefont {Zhang}},\ }\bibfield  {title} {\enquote
  {\bibinfo {title} {A brief review of calibration-free laser-induced breakdown
  spectroscopy},}\ }\href {\doibase 10.3389/fphy.2022.887171} {\bibfield
  {journal} {\bibinfo  {journal} {Frontiers in Physics}\ }\textbf {\bibinfo
  {volume} {10}} (\bibinfo {year} {2022}{\natexlab{b}}),\
  10.3389/fphy.2022.887171}\BibitemShut {NoStop}%
\bibitem [{\citenamefont {Cheung}(2022)}]{CHEUNG2022106473}%
  \BibitemOpen
  \bibfield  {author} {\bibinfo {author} {\bibfnamefont {N.-H.}\ \bibnamefont
  {Cheung}},\ }\bibfield  {title} {\enquote {\bibinfo {title} {Laser-induced
  plume fluorescence for ultrasensitive and simultaneous multianalyte analysis
  – a review},}\ }\href {\doibase https://doi.org/10.1016/j.sab.2022.106473}
  {\bibfield  {journal} {\bibinfo  {journal} {Spectrochimica Acta Part B:
  Atomic Spectroscopy}\ }\textbf {\bibinfo {volume} {194}},\ \bibinfo {pages}
  {106473} (\bibinfo {year} {2022})}\BibitemShut {NoStop}%
\bibitem [{\citenamefont {Ribeiro}\ \emph {et~al.}(2020)\citenamefont
  {Ribeiro}, \citenamefont {Senesi}, \citenamefont {Cabral}, \citenamefont
  {Cena}, \citenamefont {Marangoni}, \citenamefont {Kiefer},\ and\
  \citenamefont {Nicolodelli}}]{Ribeiro_2020_AO}%
  \BibitemOpen
  \bibfield  {author} {\bibinfo {author} {\bibfnamefont {M.~C.~S.}\
  \bibnamefont {Ribeiro}}, \bibinfo {author} {\bibfnamefont {G.~S.}\
  \bibnamefont {Senesi}}, \bibinfo {author} {\bibfnamefont {J.~S.}\
  \bibnamefont {Cabral}}, \bibinfo {author} {\bibfnamefont {C.}~\bibnamefont
  {Cena}}, \bibinfo {author} {\bibfnamefont {B.~S.}\ \bibnamefont {Marangoni}},
  \bibinfo {author} {\bibfnamefont {C.}~\bibnamefont {Kiefer}}, \ and\ \bibinfo
  {author} {\bibfnamefont {G.}~\bibnamefont {Nicolodelli}},\ }\bibfield
  {title} {\enquote {\bibinfo {title} {Evaluation of rice varieties using libs
  and ftir techniques associated with pca and machine learning algorithms},}\
  }\href {\doibase 10.1364/AO.409029} {\bibfield  {journal} {\bibinfo
  {journal} {Appl. Opt.}\ }\textbf {\bibinfo {volume} {59}},\ \bibinfo {pages}
  {10043--10048} (\bibinfo {year} {2020})}\BibitemShut {NoStop}%
\bibitem [{\citenamefont {Holub}\ \emph {et~al.}(2022)\citenamefont {Holub},
  \citenamefont {Porizka}, \citenamefont {Kizovsky}, \citenamefont {Prochazka},
  \citenamefont {Samek},\ and\ \citenamefont {Kaiser}}]{HOLUB2022106487}%
  \BibitemOpen
  \bibfield  {author} {\bibinfo {author} {\bibfnamefont {D.}~\bibnamefont
  {Holub}}, \bibinfo {author} {\bibfnamefont {P.}~\bibnamefont {Porizka}},
  \bibinfo {author} {\bibfnamefont {M.}~\bibnamefont {Kizovsky}}, \bibinfo
  {author} {\bibfnamefont {D.}~\bibnamefont {Prochazka}}, \bibinfo {author}
  {\bibfnamefont {O.}~\bibnamefont {Samek}}, \ and\ \bibinfo {author}
  {\bibfnamefont {J.}~\bibnamefont {Kaiser}},\ }\bibfield  {title} {\enquote
  {\bibinfo {title} {The potential of combining laser-induced breakdown
  spectroscopy and raman spectroscopy data for the analysis of wood samples},}\
  }\href {\doibase https://doi.org/10.1016/j.sab.2022.106487} {\bibfield
  {journal} {\bibinfo  {journal} {Spectrochimica Acta Part B: Atomic
  Spectroscopy}\ }\textbf {\bibinfo {volume} {195}},\ \bibinfo {pages} {106487}
  (\bibinfo {year} {2022})}\BibitemShut {NoStop}%
\bibitem [{\citenamefont {Sun}\ \emph {et~al.}(2022)\citenamefont {Sun},
  \citenamefont {Song}, \citenamefont {Lin},\ and\ \citenamefont
  {Gao}}]{SUN2022106456}%
  \BibitemOpen
  \bibfield  {author} {\bibinfo {author} {\bibfnamefont {H.}~\bibnamefont
  {Sun}}, \bibinfo {author} {\bibfnamefont {C.}~\bibnamefont {Song}}, \bibinfo
  {author} {\bibfnamefont {X.}~\bibnamefont {Lin}}, \ and\ \bibinfo {author}
  {\bibfnamefont {X.}~\bibnamefont {Gao}},\ }\bibfield  {title} {\enquote
  {\bibinfo {title} {Identification of meat species by combined laser-induced
  breakdown and raman spectroscopies},}\ }\href {\doibase
  https://doi.org/10.1016/j.sab.2022.106456} {\bibfield  {journal} {\bibinfo
  {journal} {Spectrochimica Acta Part B: Atomic Spectroscopy}\ }\textbf
  {\bibinfo {volume} {194}},\ \bibinfo {pages} {106456} (\bibinfo {year}
  {2022})}\BibitemShut {NoStop}%
\bibitem [{\citenamefont {Sandoval-Munoz}\ \emph {et~al.}(2022)\citenamefont
  {Sandoval-Munoz}, \citenamefont {Velasquez}, \citenamefont {Alvarez},
  \citenamefont {Perez}, \citenamefont {Velasquez}, \citenamefont {Torres},
  \citenamefont {Sbarbaro-Hofer}, \citenamefont {Motto-Ros},\ and\
  \citenamefont {Yanez}}]{D2JA00147K}%
  \BibitemOpen
  \bibfield  {author} {\bibinfo {author} {\bibfnamefont {C.}~\bibnamefont
  {Sandoval-Munoz}}, \bibinfo {author} {\bibfnamefont {G.}~\bibnamefont
  {Velasquez}}, \bibinfo {author} {\bibfnamefont {J.}~\bibnamefont {Alvarez}},
  \bibinfo {author} {\bibfnamefont {F.}~\bibnamefont {Perez}}, \bibinfo
  {author} {\bibfnamefont {M.}~\bibnamefont {Velasquez}}, \bibinfo {author}
  {\bibfnamefont {S.}~\bibnamefont {Torres}}, \bibinfo {author} {\bibfnamefont
  {D.}~\bibnamefont {Sbarbaro-Hofer}}, \bibinfo {author} {\bibfnamefont
  {V.}~\bibnamefont {Motto-Ros}}, \ and\ \bibinfo {author} {\bibfnamefont
  {J.}~\bibnamefont {Yanez}},\ }\bibfield  {title} {\enquote {\bibinfo {title}
  {Enhanced elemental and mineralogical imaging of cu-mineralized rocks by
  coupling ?-libs and hsi},}\ }\href {\doibase 10.1039/D2JA00147K} {\bibfield
  {journal} {\bibinfo  {journal} {J. Anal. At. Spectrom.}\ }\textbf {\bibinfo
  {volume} {37}},\ \bibinfo {pages} {1981--1993} (\bibinfo {year}
  {2022})}\BibitemShut {NoStop}%
\bibitem [{\citenamefont {S}\ \emph {et~al.}(2021{\natexlab{a}})\citenamefont
  {S}, \citenamefont {George}, \citenamefont {Kartha}, \citenamefont
  {Chidangil},\ and\ \citenamefont {K}}]{Dhanada_2021_ASR}%
  \BibitemOpen
  \bibfield  {author} {\bibinfo {author} {\bibfnamefont {D.~V.}\ \bibnamefont
  {S}}, \bibinfo {author} {\bibfnamefont {S.~D.}\ \bibnamefont {George}},
  \bibinfo {author} {\bibfnamefont {V.~B.}\ \bibnamefont {Kartha}}, \bibinfo
  {author} {\bibfnamefont {S.}~\bibnamefont {Chidangil}}, \ and\ \bibinfo
  {author} {\bibfnamefont {U.~V.}\ \bibnamefont {K}},\ }\bibfield  {title}
  {\enquote {\bibinfo {title} {Hybrid libs-raman-lif systems for multi-modal
  spectroscopic applications: a topical review},}\ }\href {\doibase
  10.1080/05704928.2020.1800486} {\bibfield  {journal} {\bibinfo  {journal}
  {Applied Spectroscopy Reviews}\ }\textbf {\bibinfo {volume} {56}},\ \bibinfo
  {pages} {463--491} (\bibinfo {year} {2021}{\natexlab{a}})},\ \Eprint
  {http://arxiv.org/abs/https://doi.org/10.1080/05704928.2020.1800486}
  {https://doi.org/10.1080/05704928.2020.1800486} \BibitemShut {NoStop}%
\bibitem [{\citenamefont {van~der Meiden}\ \emph {et~al.}(2021)\citenamefont
  {van~der Meiden}, \citenamefont {Almaviva}, \citenamefont {Butikova},
  \citenamefont {Dwivedi}, \citenamefont {Gasior}, \citenamefont {Gromelski},
  \citenamefont {Hakola}, \citenamefont {Jiang}, \citenamefont {J{\~{o}}gi},
  \citenamefont {Karhunen}, \citenamefont {Kubkowska}, \citenamefont {Laan},
  \citenamefont {Maddaluno}, \citenamefont {Mar{\'{\i}}n-Rold{\'{a}}n},
  \citenamefont {Paris}, \citenamefont {Piip}, \citenamefont
  {Pisar{\v{c}}{\'{\i}}k}, \citenamefont {Sergienko}, \citenamefont {Veis},
  \citenamefont {Veis}, \citenamefont {Brezinsek},\ and\ \citenamefont {the
  EUROfusion WP PFC~Team}}]{LIBS_Nucl_Fusion2021}%
  \BibitemOpen
  \bibfield  {author} {\bibinfo {author} {\bibfnamefont {H.}~\bibnamefont
  {van~der Meiden}}, \bibinfo {author} {\bibfnamefont {S.}~\bibnamefont
  {Almaviva}}, \bibinfo {author} {\bibfnamefont {J.}~\bibnamefont {Butikova}},
  \bibinfo {author} {\bibfnamefont {V.}~\bibnamefont {Dwivedi}}, \bibinfo
  {author} {\bibfnamefont {P.}~\bibnamefont {Gasior}}, \bibinfo {author}
  {\bibfnamefont {W.}~\bibnamefont {Gromelski}}, \bibinfo {author}
  {\bibfnamefont {A.}~\bibnamefont {Hakola}}, \bibinfo {author} {\bibfnamefont
  {X.}~\bibnamefont {Jiang}}, \bibinfo {author} {\bibfnamefont
  {I.}~\bibnamefont {J{\~{o}}gi}}, \bibinfo {author} {\bibfnamefont
  {J.}~\bibnamefont {Karhunen}}, \bibinfo {author} {\bibfnamefont
  {M.}~\bibnamefont {Kubkowska}}, \bibinfo {author} {\bibfnamefont
  {M.}~\bibnamefont {Laan}}, \bibinfo {author} {\bibfnamefont {G.}~\bibnamefont
  {Maddaluno}}, \bibinfo {author} {\bibfnamefont {A.}~\bibnamefont
  {Mar{\'{\i}}n-Rold{\'{a}}n}}, \bibinfo {author} {\bibfnamefont
  {P.}~\bibnamefont {Paris}}, \bibinfo {author} {\bibfnamefont
  {K.}~\bibnamefont {Piip}}, \bibinfo {author} {\bibfnamefont {M.}~\bibnamefont
  {Pisar{\v{c}}{\'{\i}}k}}, \bibinfo {author} {\bibfnamefont {G.}~\bibnamefont
  {Sergienko}}, \bibinfo {author} {\bibfnamefont {M.}~\bibnamefont {Veis}},
  \bibinfo {author} {\bibfnamefont {P.}~\bibnamefont {Veis}}, \bibinfo {author}
  {\bibfnamefont {S.}~\bibnamefont {Brezinsek}}, \ and\ \bibinfo {author}
  {\bibnamefont {the EUROfusion WP PFC~Team}},\ }\bibfield  {title} {\enquote
  {\bibinfo {title} {Monitoring of tritium and impurities in the first wall of
  fusion devices using a {LIBS} based diagnostic},}\ }\href {\doibase
  10.1088/1741-4326/ac31d6} {\bibfield  {journal} {\bibinfo  {journal} {Nuclear
  Fusion}\ }\textbf {\bibinfo {volume} {61}},\ \bibinfo {pages} {125001}
  (\bibinfo {year} {2021})}\BibitemShut {NoStop}%
\bibitem [{\citenamefont {LaHaye}\ \emph {et~al.}(2013)\citenamefont {LaHaye},
  \citenamefont {Harilal}, \citenamefont {Diwakar},\ and\ \citenamefont
  {Hassanein}}]{LIBS_Time_scale}%
  \BibitemOpen
  \bibfield  {author} {\bibinfo {author} {\bibfnamefont {N.~L.}\ \bibnamefont
  {LaHaye}}, \bibinfo {author} {\bibfnamefont {S.~S.}\ \bibnamefont {Harilal}},
  \bibinfo {author} {\bibfnamefont {P.~K.}\ \bibnamefont {Diwakar}}, \ and\
  \bibinfo {author} {\bibfnamefont {A.}~\bibnamefont {Hassanein}},\ }\bibfield
  {title} {\enquote {\bibinfo {title} {The effect of laser pulse duration on
  icp-ms signal intensity{,} elemental fractionation{,} and detection limits in
  fs-la-icp-ms},}\ }\href {\doibase 10.1039/C3JA50200G} {\bibfield  {journal}
  {\bibinfo  {journal} {J. Anal. At. Spectrom.}\ }\textbf {\bibinfo {volume}
  {28}},\ \bibinfo {pages} {1781--1787} (\bibinfo {year} {2013})}\BibitemShut
  {NoStop}%
\bibitem [{\citenamefont {Harilal}\ \emph {et~al.}(2016)\citenamefont
  {Harilal}, \citenamefont {Brumfield}, \citenamefont {Cannon},\ and\
  \citenamefont {Phillips}}]{Hari_Anals_Chem_2016}%
  \BibitemOpen
  \bibfield  {author} {\bibinfo {author} {\bibfnamefont {S.~S.}\ \bibnamefont
  {Harilal}}, \bibinfo {author} {\bibfnamefont {B.~E.}\ \bibnamefont
  {Brumfield}}, \bibinfo {author} {\bibfnamefont {B.~D.}\ \bibnamefont
  {Cannon}}, \ and\ \bibinfo {author} {\bibfnamefont {M.~C.}\ \bibnamefont
  {Phillips}},\ }\bibfield  {title} {\enquote {\bibinfo {title} {Shock wave
  mediated plume chemistry for molecular formation in laser ablation
  plasmas},}\ }\href {\doibase 10.1021/acs.analchem.5b04136} {\bibfield
  {journal} {\bibinfo  {journal} {Analytical Chemistry}\ }\textbf {\bibinfo
  {volume} {88}},\ \bibinfo {pages} {2296--2302} (\bibinfo {year} {2016})},\
  \bibinfo {note} {pMID: 26732866},\ \Eprint
  {http://arxiv.org/abs/https://doi.org/10.1021/acs.analchem.5b04136}
  {https://doi.org/10.1021/acs.analchem.5b04136} \BibitemShut {NoStop}%
\bibitem [{\citenamefont {Thomas}\ \emph {et~al.}(2019)\citenamefont {Thomas},
  \citenamefont {Joshi}, \citenamefont {Kumar},\ and\ \citenamefont
  {Philip}}]{jinto_PhsD}%
  \BibitemOpen
  \bibfield  {author} {\bibinfo {author} {\bibfnamefont {J.}~\bibnamefont
  {Thomas}}, \bibinfo {author} {\bibfnamefont {H.~C.}\ \bibnamefont {Joshi}},
  \bibinfo {author} {\bibfnamefont {A.}~\bibnamefont {Kumar}}, \ and\ \bibinfo
  {author} {\bibfnamefont {R.}~\bibnamefont {Philip}},\ }\bibfield  {title}
  {\enquote {\bibinfo {title} {Pulse width dependent dynamics of laser-induced
  plasma from a ni thin film},}\ }\href {\doibase 10.1088/1361-6463/aaff43}
  {\bibfield  {journal} {\bibinfo  {journal} {Journal of Physics D: Applied
  Physics}\ }\textbf {\bibinfo {volume} {52}},\ \bibinfo {pages} {135201}
  (\bibinfo {year} {2019})}\BibitemShut {NoStop}%
\bibitem [{\citenamefont {Anoop}\ \emph {et~al.}(2016)\citenamefont {Anoop},
  \citenamefont {Harilal}, \citenamefont {Philip}, \citenamefont {Bruzzese},\
  and\ \citenamefont {Amoruso}}]{Anoop_JAP_2016}%
  \BibitemOpen
  \bibfield  {author} {\bibinfo {author} {\bibfnamefont {K.~K.}\ \bibnamefont
  {Anoop}}, \bibinfo {author} {\bibfnamefont {S.~S.}\ \bibnamefont {Harilal}},
  \bibinfo {author} {\bibfnamefont {R.}~\bibnamefont {Philip}}, \bibinfo
  {author} {\bibfnamefont {R.}~\bibnamefont {Bruzzese}}, \ and\ \bibinfo
  {author} {\bibfnamefont {S.}~\bibnamefont {Amoruso}},\ }\bibfield  {title}
  {\enquote {\bibinfo {title} {Laser fluence dependence on emission dynamics of
  ultrafast laser induced copper plasma},}\ }\href {\doibase 10.1063/1.4967313}
  {\bibfield  {journal} {\bibinfo  {journal} {Journal of Applied Physics}\
  }\textbf {\bibinfo {volume} {120}},\ \bibinfo {pages} {185901} (\bibinfo
  {year} {2016})},\ \Eprint
  {http://arxiv.org/abs/https://doi.org/10.1063/1.4967313}
  {https://doi.org/10.1063/1.4967313} \BibitemShut {NoStop}%
\bibitem [{\citenamefont {Smijesh}\ and\ \citenamefont
  {Philip}(2013)}]{Smijesh_JAP_2013}%
  \BibitemOpen
  \bibfield  {author} {\bibinfo {author} {\bibfnamefont {N.}~\bibnamefont
  {Smijesh}}\ and\ \bibinfo {author} {\bibfnamefont {R.}~\bibnamefont
  {Philip}},\ }\bibfield  {title} {\enquote {\bibinfo {title} {Emission
  dynamics of an expanding ultrafast-laser produced zn plasma under different
  ambient pressures},}\ }\href {\doibase 10.1063/1.4820575} {\bibfield
  {journal} {\bibinfo  {journal} {Journal of Applied Physics}\ }\textbf
  {\bibinfo {volume} {114}},\ \bibinfo {pages} {093301} (\bibinfo {year}
  {2013})},\ \Eprint {http://arxiv.org/abs/https://doi.org/10.1063/1.4820575}
  {https://doi.org/10.1063/1.4820575} \BibitemShut {NoStop}%
\bibitem [{\citenamefont {Mondal}, \citenamefont {Singh},\ and\ \citenamefont
  {Kumar}(2018)}]{Rear_Front_Alam}%
  \BibitemOpen
  \bibfield  {author} {\bibinfo {author} {\bibfnamefont {A.}~\bibnamefont
  {Mondal}}, \bibinfo {author} {\bibfnamefont {R.~K.}\ \bibnamefont {Singh}}, \
  and\ \bibinfo {author} {\bibfnamefont {A.}~\bibnamefont {Kumar}},\ }\bibfield
   {title} {\enquote {\bibinfo {title} {Effect of ablation geometry on the
  dynamics, composition, and geometrical shape of thin film plasma},}\ }\href
  {\doibase 10.1063/1.4991469} {\bibfield  {journal} {\bibinfo  {journal}
  {Physics of Plasmas}\ }\textbf {\bibinfo {volume} {25}},\ \bibinfo {pages}
  {013517} (\bibinfo {year} {2018})},\ \Eprint
  {http://arxiv.org/abs/https://doi.org/10.1063/1.4991469}
  {https://doi.org/10.1063/1.4991469} \BibitemShut {NoStop}%
\bibitem [{\citenamefont {Escobar-Alarcón}\ \emph {et~al.}(2002)\citenamefont
  {Escobar-Alarcón}, \citenamefont {Camps}, \citenamefont {Haro-Poniatowski},
  \citenamefont {Villagran},\ and\ \citenamefont {Sanchez}}]{ESCOBAR2002}%
  \BibitemOpen
  \bibfield  {author} {\bibinfo {author} {\bibfnamefont {L.}~\bibnamefont
  {Escobar-Alarcón}}, \bibinfo {author} {\bibfnamefont {E.}~\bibnamefont
  {Camps}}, \bibinfo {author} {\bibfnamefont {E.}~\bibnamefont
  {Haro-Poniatowski}}, \bibinfo {author} {\bibfnamefont {M.}~\bibnamefont
  {Villagran}}, \ and\ \bibinfo {author} {\bibfnamefont {C.}~\bibnamefont
  {Sanchez}},\ }\bibfield  {title} {\enquote {\bibinfo {title}
  {Characterization of rear- and front-side laser ablation plasmas for
  thin-film deposition},}\ }\href {\doibase
  https://doi.org/10.1016/S0169-4332(02)00356-2} {\bibfield  {journal}
  {\bibinfo  {journal} {Applied Surface Science}\ }\textbf {\bibinfo {volume}
  {197-198}},\ \bibinfo {pages} {192--196} (\bibinfo {year}
  {2002})}\BibitemShut {NoStop}%
\bibitem [{\citenamefont {Mondal}, \citenamefont {Singh},\ and\ \citenamefont
  {Joshi}(2019)}]{C9JA00158A}%
  \BibitemOpen
  \bibfield  {author} {\bibinfo {author} {\bibfnamefont {A.}~\bibnamefont
  {Mondal}}, \bibinfo {author} {\bibfnamefont {R.~K.}\ \bibnamefont {Singh}}, \
  and\ \bibinfo {author} {\bibfnamefont {H.~C.}\ \bibnamefont {Joshi}},\
  }\bibfield  {title} {\enquote {\bibinfo {title} {Neutral and ion composition
  of laser produced lithium plasma plume in front and back ablation of thin
  film},}\ }\href {\doibase 10.1039/C9JA00158A} {\bibfield  {journal} {\bibinfo
   {journal} {J. Anal. At. Spectrom.}\ }\textbf {\bibinfo {volume} {34}},\
  \bibinfo {pages} {1822--1828} (\bibinfo {year} {2019})}\BibitemShut {NoStop}%
\bibitem [{\citenamefont {Képeš}\ \emph {et~al.}(2021)\citenamefont
  {Képeš}, \citenamefont {Gornushkin}, \citenamefont {Pořízka},\ and\
  \citenamefont {Kaiser}}]{D1AN01292D}%
  \BibitemOpen
  \bibfield  {author} {\bibinfo {author} {\bibfnamefont {E.}~\bibnamefont
  {Képeš}}, \bibinfo {author} {\bibfnamefont {I.}~\bibnamefont {Gornushkin}},
  \bibinfo {author} {\bibfnamefont {P.}~\bibnamefont {Pořízka}}, \ and\
  \bibinfo {author} {\bibfnamefont {J.}~\bibnamefont {Kaiser}},\ }\bibfield
  {title} {\enquote {\bibinfo {title} {Spatiotemporal spectroscopic
  characterization of plasmas induced by non-orthogonal laser ablation},}\
  }\href {\doibase 10.1039/D0AN01996H} {\bibfield  {journal} {\bibinfo
  {journal} {Analyst}\ }\textbf {\bibinfo {volume} {146}},\ \bibinfo {pages}
  {920--929} (\bibinfo {year} {2021})}\BibitemShut {NoStop}%
\bibitem [{\citenamefont {Li}\ \emph {et~al.}(2021{\natexlab{a}})\citenamefont
  {Li}, \citenamefont {Nishi}, \citenamefont {Zheng},\ and\ \citenamefont
  {Sakka}}]{D0JA00521E}%
  \BibitemOpen
  \bibfield  {author} {\bibinfo {author} {\bibfnamefont {N.}~\bibnamefont
  {Li}}, \bibinfo {author} {\bibfnamefont {N.}~\bibnamefont {Nishi}}, \bibinfo
  {author} {\bibfnamefont {R.}~\bibnamefont {Zheng}}, \ and\ \bibinfo {author}
  {\bibfnamefont {T.}~\bibnamefont {Sakka}},\ }\bibfield  {title} {\enquote
  {\bibinfo {title} {Signal enhancement in underwater long-pulse laser-induced
  breakdown spectroscopy for the analysis of bulk water},}\ }\href {\doibase
  10.1039/D0JA00521E} {\bibfield  {journal} {\bibinfo  {journal} {J. Anal. At.
  Spectrom.}\ }\textbf {\bibinfo {volume} {36}},\ \bibinfo {pages} {1170--1179}
  (\bibinfo {year} {2021}{\natexlab{a}})}\BibitemShut {NoStop}%
\bibitem [{\citenamefont {Li}\ \emph {et~al.}(2021{\natexlab{b}})\citenamefont
  {Li}, \citenamefont {Tanabe}, \citenamefont {Nishi}, \citenamefont {Zheng},\
  and\ \citenamefont {Sakka}}]{D1JA00151E}%
  \BibitemOpen
  \bibfield  {author} {\bibinfo {author} {\bibfnamefont {N.}~\bibnamefont
  {Li}}, \bibinfo {author} {\bibfnamefont {K.}~\bibnamefont {Tanabe}}, \bibinfo
  {author} {\bibfnamefont {N.}~\bibnamefont {Nishi}}, \bibinfo {author}
  {\bibfnamefont {R.}~\bibnamefont {Zheng}}, \ and\ \bibinfo {author}
  {\bibfnamefont {T.}~\bibnamefont {Sakka}},\ }\bibfield  {title} {\enquote
  {\bibinfo {title} {Simultaneous detection of a submerged cu target and bulk
  water by long-pulse laser-induced breakdown spectroscopy},}\ }\href {\doibase
  10.1039/D1JA00151E} {\bibfield  {journal} {\bibinfo  {journal} {J. Anal. At.
  Spectrom.}\ }\textbf {\bibinfo {volume} {36}},\ \bibinfo {pages} {1960--1968}
  (\bibinfo {year} {2021}{\natexlab{b}})}\BibitemShut {NoStop}%
\bibitem [{\citenamefont {Diwakar}\ \emph {et~al.}(2013)\citenamefont
  {Diwakar}, \citenamefont {Harilal}, \citenamefont {Freeman},\ and\
  \citenamefont {Hassanein}}]{DIWAKAR201365}%
  \BibitemOpen
  \bibfield  {author} {\bibinfo {author} {\bibfnamefont {P.}~\bibnamefont
  {Diwakar}}, \bibinfo {author} {\bibfnamefont {S.}~\bibnamefont {Harilal}},
  \bibinfo {author} {\bibfnamefont {J.}~\bibnamefont {Freeman}}, \ and\
  \bibinfo {author} {\bibfnamefont {A.}~\bibnamefont {Hassanein}},\ }\bibfield
  {title} {\enquote {\bibinfo {title} {Role of laser pre-pulse wavelength and
  inter-pulse delay on signal enhancement in collinear double-pulse
  laser-induced breakdown spectroscopy},}\ }\href {\doibase
  https://doi.org/10.1016/j.sab.2013.05.015} {\bibfield  {journal} {\bibinfo
  {journal} {Spectrochimica Acta Part B: Atomic Spectroscopy}\ }\textbf
  {\bibinfo {volume} {87}},\ \bibinfo {pages} {65--73} (\bibinfo {year}
  {2013})},\ \bibinfo {note} {thematic Issue: 7th International Conference on
  Laser Induced Breakdown Spectroscopy (LIBS 2012), Luxor, Egypt, 29
  September-4 October 2012}\BibitemShut {NoStop}%
\bibitem [{\citenamefont {Wang}\ \emph {et~al.}(2020)\citenamefont {Wang},
  \citenamefont {Chen}, \citenamefont {Zhang}, \citenamefont {Wang},
  \citenamefont {Li}, \citenamefont {Jiang},\ and\ \citenamefont
  {Jin}}]{Wang_POP_2020}%
  \BibitemOpen
  \bibfield  {author} {\bibinfo {author} {\bibfnamefont {Y.}~\bibnamefont
  {Wang}}, \bibinfo {author} {\bibfnamefont {A.}~\bibnamefont {Chen}}, \bibinfo
  {author} {\bibfnamefont {D.}~\bibnamefont {Zhang}}, \bibinfo {author}
  {\bibfnamefont {Q.}~\bibnamefont {Wang}}, \bibinfo {author} {\bibfnamefont
  {S.}~\bibnamefont {Li}}, \bibinfo {author} {\bibfnamefont {Y.}~\bibnamefont
  {Jiang}}, \ and\ \bibinfo {author} {\bibfnamefont {M.}~\bibnamefont {Jin}},\
  }\bibfield  {title} {\enquote {\bibinfo {title} {Enhanced optical emission in
  laser-induced breakdown spectroscopy by combining femtosecond and nanosecond
  laser pulses},}\ }\href {\doibase 10.1063/1.5131772} {\bibfield  {journal}
  {\bibinfo  {journal} {Physics of Plasmas}\ }\textbf {\bibinfo {volume}
  {27}},\ \bibinfo {pages} {023507} (\bibinfo {year} {2020})},\ \Eprint
  {http://arxiv.org/abs/https://doi.org/10.1063/1.5131772}
  {https://doi.org/10.1063/1.5131772} \BibitemShut {NoStop}%
\bibitem [{\citenamefont {Giannakaris}\ \emph {et~al.}(0)\citenamefont
  {Giannakaris}, \citenamefont {Haider}, \citenamefont {Ahamer}, \citenamefont
  {Grünberger}, \citenamefont {Trautner},\ and\ \citenamefont
  {Pedarnig}}]{Nikolaos_Applied_Spectro}%
  \BibitemOpen
  \bibfield  {author} {\bibinfo {author} {\bibfnamefont {N.}~\bibnamefont
  {Giannakaris}}, \bibinfo {author} {\bibfnamefont {A.}~\bibnamefont {Haider}},
  \bibinfo {author} {\bibfnamefont {C.~M.}\ \bibnamefont {Ahamer}}, \bibinfo
  {author} {\bibfnamefont {S.}~\bibnamefont {Grünberger}}, \bibinfo {author}
  {\bibfnamefont {S.}~\bibnamefont {Trautner}}, \ and\ \bibinfo {author}
  {\bibfnamefont {J.~D.}\ \bibnamefont {Pedarnig}},\ }\bibfield  {title}
  {\enquote {\bibinfo {title} {Femtosecond single-pulse and orthogonal
  double-pulse laser-induced breakdown spectroscopy (libs): Femtogram mass
  detection and chemical imaging with micrometer spatial resolution},}\ }\href
  {\doibase 10.1177/00037028211042398} {\bibfield  {journal} {\bibinfo
  {journal} {Applied Spectroscopy}\ }\textbf {\bibinfo {volume} {0}},\ \bibinfo
  {pages} {00037028211042398} (\bibinfo {year} {0})},\ \bibinfo {note} {pMID:
  34494912},\ \Eprint
  {http://arxiv.org/abs/https://doi.org/10.1177/00037028211042398}
  {https://doi.org/10.1177/00037028211042398} \BibitemShut {NoStop}%
\bibitem [{\citenamefont {Sivakumaran}\ \emph {et~al.}(2014)\citenamefont
  {Sivakumaran}, \citenamefont {Joshi}, \citenamefont {Singh},\ and\
  \citenamefont {Kumar}}]{Siva_POP_2014}%
  \BibitemOpen
  \bibfield  {author} {\bibinfo {author} {\bibfnamefont {V.}~\bibnamefont
  {Sivakumaran}}, \bibinfo {author} {\bibfnamefont {H.~C.}\ \bibnamefont
  {Joshi}}, \bibinfo {author} {\bibfnamefont {R.~K.}\ \bibnamefont {Singh}}, \
  and\ \bibinfo {author} {\bibfnamefont {A.}~\bibnamefont {Kumar}},\ }\bibfield
   {title} {\enquote {\bibinfo {title} {Optical time of flight studies of
  lithium plasma in double pulse laser ablation: Evidence of inverse
  bremsstrahlung absorption},}\ }\href {\doibase 10.1063/1.4885107} {\bibfield
  {journal} {\bibinfo  {journal} {Physics of Plasmas}\ }\textbf {\bibinfo
  {volume} {21}},\ \bibinfo {pages} {063110} (\bibinfo {year} {2014})},\
  \Eprint {http://arxiv.org/abs/https://doi.org/10.1063/1.4885107}
  {https://doi.org/10.1063/1.4885107} \BibitemShut {NoStop}%
\bibitem [{\citenamefont {Wang}\ \emph {et~al.}(2022)\citenamefont {Wang},
  \citenamefont {Zhao}, \citenamefont {Wang}, \citenamefont {Zhang},
  \citenamefont {Wang}, \citenamefont {Zhang}, \citenamefont {Ma},
  \citenamefont {Liu}, \citenamefont {Luo}, \citenamefont {Ma}, \citenamefont
  {Ye}, \citenamefont {Zhu}, \citenamefont {Yin},\ and\ \citenamefont
  {Jia}}]{D2JA00105E}%
  \BibitemOpen
  \bibfield  {author} {\bibinfo {author} {\bibfnamefont {J.}~\bibnamefont
  {Wang}}, \bibinfo {author} {\bibfnamefont {Y.}~\bibnamefont {Zhao}}, \bibinfo
  {author} {\bibfnamefont {G.}~\bibnamefont {Wang}}, \bibinfo {author}
  {\bibfnamefont {L.}~\bibnamefont {Zhang}}, \bibinfo {author} {\bibfnamefont
  {S.}~\bibnamefont {Wang}}, \bibinfo {author} {\bibfnamefont {W.}~\bibnamefont
  {Zhang}}, \bibinfo {author} {\bibfnamefont {X.}~\bibnamefont {Ma}}, \bibinfo
  {author} {\bibfnamefont {Z.}~\bibnamefont {Liu}}, \bibinfo {author}
  {\bibfnamefont {X.}~\bibnamefont {Luo}}, \bibinfo {author} {\bibfnamefont
  {W.}~\bibnamefont {Ma}}, \bibinfo {author} {\bibfnamefont {Z.}~\bibnamefont
  {Ye}}, \bibinfo {author} {\bibfnamefont {Z.}~\bibnamefont {Zhu}}, \bibinfo
  {author} {\bibfnamefont {W.}~\bibnamefont {Yin}}, \ and\ \bibinfo {author}
  {\bibfnamefont {S.}~\bibnamefont {Jia}},\ }\bibfield  {title} {\enquote
  {\bibinfo {title} {Theoretical study on signal enhancement of orthogonal
  double pulse induced plasma},}\ }\href {\doibase 10.1039/D2JA00105E}
  {\bibfield  {journal} {\bibinfo  {journal} {J. Anal. At. Spectrom.}\ ,\
  \bibinfo {pages} {--}} (\bibinfo {year} {2022})}\BibitemShut {NoStop}%
\bibitem [{\citenamefont {Hough}\ \emph {et~al.}(2010)\citenamefont {Hough},
  \citenamefont {McLoughlin}, \citenamefont {Harilal}, \citenamefont
  {Mosnier},\ and\ \citenamefont {Costello}}]{Hough_JAP_2010}%
  \BibitemOpen
  \bibfield  {author} {\bibinfo {author} {\bibfnamefont {P.}~\bibnamefont
  {Hough}}, \bibinfo {author} {\bibfnamefont {C.}~\bibnamefont {McLoughlin}},
  \bibinfo {author} {\bibfnamefont {S.~S.}\ \bibnamefont {Harilal}}, \bibinfo
  {author} {\bibfnamefont {J.~P.}\ \bibnamefont {Mosnier}}, \ and\ \bibinfo
  {author} {\bibfnamefont {J.~T.}\ \bibnamefont {Costello}},\ }\bibfield
  {title} {\enquote {\bibinfo {title} {Emission characteristics and dynamics of
  the stagnation layer in colliding laser produced plasmas},}\ }\href {\doibase
  10.1063/1.3282683} {\bibfield  {journal} {\bibinfo  {journal} {Journal of
  Applied Physics}\ }\textbf {\bibinfo {volume} {107}},\ \bibinfo {pages}
  {024904} (\bibinfo {year} {2010})},\ \Eprint
  {http://arxiv.org/abs/https://doi.org/10.1063/1.3282683}
  {https://doi.org/10.1063/1.3282683} \BibitemShut {NoStop}%
\bibitem [{\citenamefont {Al-Shboul}\ \emph {et~al.}(2014)\citenamefont
  {Al-Shboul}, \citenamefont {Harilal}, \citenamefont {Hassan}, \citenamefont
  {Hassanein}, \citenamefont {Costello}, \citenamefont {Yabuuchi},
  \citenamefont {Tanaka},\ and\ \citenamefont {Hirooka}}]{Shboul_POP_2014}%
  \BibitemOpen
  \bibfield  {author} {\bibinfo {author} {\bibfnamefont {K.~F.}\ \bibnamefont
  {Al-Shboul}}, \bibinfo {author} {\bibfnamefont {S.~S.}\ \bibnamefont
  {Harilal}}, \bibinfo {author} {\bibfnamefont {S.~M.}\ \bibnamefont {Hassan}},
  \bibinfo {author} {\bibfnamefont {A.}~\bibnamefont {Hassanein}}, \bibinfo
  {author} {\bibfnamefont {J.~T.}\ \bibnamefont {Costello}}, \bibinfo {author}
  {\bibfnamefont {T.}~\bibnamefont {Yabuuchi}}, \bibinfo {author}
  {\bibfnamefont {K.~A.}\ \bibnamefont {Tanaka}}, \ and\ \bibinfo {author}
  {\bibfnamefont {Y.}~\bibnamefont {Hirooka}},\ }\bibfield  {title} {\enquote
  {\bibinfo {title} {Interpenetration and stagnation in colliding laser
  plasmas},}\ }\href {\doibase 10.1063/1.4859136} {\bibfield  {journal}
  {\bibinfo  {journal} {Physics of Plasmas}\ }\textbf {\bibinfo {volume}
  {21}},\ \bibinfo {pages} {013502} (\bibinfo {year} {2014})},\ \Eprint
  {http://arxiv.org/abs/https://doi.org/10.1063/1.4859136}
  {https://doi.org/10.1063/1.4859136} \BibitemShut {NoStop}%
\bibitem [{\citenamefont {Saxena}\ \emph {et~al.}(2019)\citenamefont {Saxena},
  \citenamefont {Singh}, \citenamefont {Joshi},\ and\ \citenamefont
  {Kumar}}]{Saxena_19}%
  \BibitemOpen
  \bibfield  {author} {\bibinfo {author} {\bibfnamefont {A.~K.}\ \bibnamefont
  {Saxena}}, \bibinfo {author} {\bibfnamefont {R.~K.}\ \bibnamefont {Singh}},
  \bibinfo {author} {\bibfnamefont {H.~C.}\ \bibnamefont {Joshi}}, \ and\
  \bibinfo {author} {\bibfnamefont {A.}~\bibnamefont {Kumar}},\ }\bibfield
  {title} {\enquote {\bibinfo {title} {Spectroscopic investigation of molecular
  formation in laterally colliding laser-produced carbon plasmas},}\ }\href
  {\doibase 10.1364/AO.58.000561} {\bibfield  {journal} {\bibinfo  {journal}
  {Appl. Opt.}\ }\textbf {\bibinfo {volume} {58}},\ \bibinfo {pages} {561--570}
  (\bibinfo {year} {2019})}\BibitemShut {NoStop}%
\bibitem [{\citenamefont {Mondal}\ \emph {et~al.}(2020)\citenamefont {Mondal},
  \citenamefont {Singh}, \citenamefont {Chaudhari},\ and\ \citenamefont
  {Joshi}}]{Alam_POP_2020}%
  \BibitemOpen
  \bibfield  {author} {\bibinfo {author} {\bibfnamefont {A.}~\bibnamefont
  {Mondal}}, \bibinfo {author} {\bibfnamefont {R.~K.}\ \bibnamefont {Singh}},
  \bibinfo {author} {\bibfnamefont {V.}~\bibnamefont {Chaudhari}}, \ and\
  \bibinfo {author} {\bibfnamefont {H.~C.}\ \bibnamefont {Joshi}},\ }\bibfield
  {title} {\enquote {\bibinfo {title} {Effect of magnetic field on the lateral
  interaction of plasma plumes},}\ }\href {\doibase 10.1063/5.0006647}
  {\bibfield  {journal} {\bibinfo  {journal} {Physics of Plasmas}\ }\textbf
  {\bibinfo {volume} {27}},\ \bibinfo {pages} {093109} (\bibinfo {year}
  {2020})},\ \Eprint {http://arxiv.org/abs/https://doi.org/10.1063/5.0006647}
  {https://doi.org/10.1063/5.0006647} \BibitemShut {NoStop}%
\bibitem [{\citenamefont {Tiwari}\ \emph {et~al.}(2022)\citenamefont {Tiwari},
  \citenamefont {Behera}, \citenamefont {Singh},\ and\ \citenamefont
  {Joshi}}]{TIWARI2022106411}%
  \BibitemOpen
  \bibfield  {author} {\bibinfo {author} {\bibfnamefont {P.~K.}\ \bibnamefont
  {Tiwari}}, \bibinfo {author} {\bibfnamefont {N.}~\bibnamefont {Behera}},
  \bibinfo {author} {\bibfnamefont {R.}~\bibnamefont {Singh}}, \ and\ \bibinfo
  {author} {\bibfnamefont {H.}~\bibnamefont {Joshi}},\ }\bibfield  {title}
  {\enquote {\bibinfo {title} {Comparative study of libs signal for single and
  colliding plasma plumes in a variable magnetic field},}\ }\href {\doibase
  https://doi.org/10.1016/j.sab.2022.106411} {\bibfield  {journal} {\bibinfo
  {journal} {Spectrochimica Acta Part B: Atomic Spectroscopy}\ }\textbf
  {\bibinfo {volume} {191}},\ \bibinfo {pages} {106411} (\bibinfo {year}
  {2022})}\BibitemShut {NoStop}%
\bibitem [{\citenamefont {Delaney}\ \emph {et~al.}(2022)\citenamefont
  {Delaney}, \citenamefont {Hayden}, \citenamefont {Kelly}, \citenamefont
  {Kennedy},\ and\ \citenamefont {Costello}}]{DELANEY2022106430}%
  \BibitemOpen
  \bibfield  {author} {\bibinfo {author} {\bibfnamefont {B.}~\bibnamefont
  {Delaney}}, \bibinfo {author} {\bibfnamefont {P.}~\bibnamefont {Hayden}},
  \bibinfo {author} {\bibfnamefont {T.}~\bibnamefont {Kelly}}, \bibinfo
  {author} {\bibfnamefont {E.}~\bibnamefont {Kennedy}}, \ and\ \bibinfo
  {author} {\bibfnamefont {J.}~\bibnamefont {Costello}},\ }\bibfield  {title}
  {\enquote {\bibinfo {title} {Laser induced breakdown spectroscopy with
  annular plasmas in vacuo: Stagnation and limits of detection},}\ }\href
  {\doibase https://doi.org/10.1016/j.sab.2022.106430} {\bibfield  {journal}
  {\bibinfo  {journal} {Spectrochimica Acta Part B: Atomic Spectroscopy}\
  }\textbf {\bibinfo {volume} {193}},\ \bibinfo {pages} {106430} (\bibinfo
  {year} {2022})}\BibitemShut {NoStop}%
\bibitem [{\citenamefont {Rohwetter}\ \emph {et~al.}(2005)\citenamefont
  {Rohwetter}, \citenamefont {Stelmaszczyk}, \citenamefont {Wöste},
  \citenamefont {Ackermann}, \citenamefont {Méjean}, \citenamefont {Salmon},
  \citenamefont {Kasparian}, \citenamefont {Yu},\ and\ \citenamefont
  {Wolf}}]{ROHWETTER20051025}%
  \BibitemOpen
  \bibfield  {author} {\bibinfo {author} {\bibfnamefont {P.}~\bibnamefont
  {Rohwetter}}, \bibinfo {author} {\bibfnamefont {K.}~\bibnamefont
  {Stelmaszczyk}}, \bibinfo {author} {\bibfnamefont {L.}~\bibnamefont
  {Wöste}}, \bibinfo {author} {\bibfnamefont {R.}~\bibnamefont {Ackermann}},
  \bibinfo {author} {\bibfnamefont {G.}~\bibnamefont {Méjean}}, \bibinfo
  {author} {\bibfnamefont {E.}~\bibnamefont {Salmon}}, \bibinfo {author}
  {\bibfnamefont {J.}~\bibnamefont {Kasparian}}, \bibinfo {author}
  {\bibfnamefont {J.}~\bibnamefont {Yu}}, \ and\ \bibinfo {author}
  {\bibfnamefont {J.-P.}\ \bibnamefont {Wolf}},\ }\bibfield  {title} {\enquote
  {\bibinfo {title} {Filament-induced remote surface ablation for long range
  laser-induced breakdown spectroscopy operation},}\ }\href {\doibase
  https://doi.org/10.1016/j.sab.2005.03.017} {\bibfield  {journal} {\bibinfo
  {journal} {Spectrochimica Acta Part B: Atomic Spectroscopy}\ }\textbf
  {\bibinfo {volume} {60}},\ \bibinfo {pages} {1025--1033} (\bibinfo {year}
  {2005})},\ \bibinfo {note} {laser Induced Plasma Spectroscopy and
  Applications (LIBS 2004) Third International Conference}\BibitemShut
  {NoStop}%
\bibitem [{\citenamefont {Motto-Ros}\ \emph {et~al.}(2020)\citenamefont
  {Motto-Ros}, \citenamefont {Moncayo}, \citenamefont {Fabre},\ and\
  \citenamefont {Busser}}]{MOTTOROS2020329}%
  \BibitemOpen
  \bibfield  {author} {\bibinfo {author} {\bibfnamefont {V.}~\bibnamefont
  {Motto-Ros}}, \bibinfo {author} {\bibfnamefont {S.}~\bibnamefont {Moncayo}},
  \bibinfo {author} {\bibfnamefont {C.}~\bibnamefont {Fabre}}, \ and\ \bibinfo
  {author} {\bibfnamefont {B.}~\bibnamefont {Busser}},\ }\bibfield  {title}
  {\enquote {\bibinfo {title} {Chapter 14 - libs imaging applications},}\ }in\
  \href {\doibase https://doi.org/10.1016/B978-0-12-818829-3.00014-9} {\emph
  {\bibinfo {booktitle} {Laser-Induced Breakdown Spectroscopy (Second
  Edition)}}},\ \bibinfo {editor} {edited by\ \bibinfo {editor} {\bibfnamefont
  {J.~P.}\ \bibnamefont {Singh}}\ and\ \bibinfo {editor} {\bibfnamefont
  {S.~N.}\ \bibnamefont {Thakur}}}\ (\bibinfo  {publisher} {Elsevier},\
  \bibinfo {address} {Amsterdam},\ \bibinfo {year} {2020})\ \bibinfo {edition}
  {second edition}\ ed.,\ pp.\ \bibinfo {pages} {329--346}\BibitemShut
  {NoStop}%
\bibitem [{\citenamefont {Hu}\ \emph {et~al.}(2020)\citenamefont {Hu},
  \citenamefont {Peng}, \citenamefont {Niu},\ and\ \citenamefont
  {Zeng}}]{GIBS_2020}%
  \BibitemOpen
  \bibfield  {author} {\bibinfo {author} {\bibfnamefont {M.}~\bibnamefont
  {Hu}}, \bibinfo {author} {\bibfnamefont {J.}~\bibnamefont {Peng}}, \bibinfo
  {author} {\bibfnamefont {S.}~\bibnamefont {Niu}}, \ and\ \bibinfo {author}
  {\bibfnamefont {H.}~\bibnamefont {Zeng}},\ }\bibfield  {title} {\enquote
  {\bibinfo {title} {{Plasma-grating-induced breakdown spectroscopy}},}\ }\href
  {\doibase 10.1117/1.AP.2.6.065001} {\bibfield  {journal} {\bibinfo  {journal}
  {Advanced Photonics}\ }\textbf {\bibinfo {volume} {2}},\ \bibinfo {pages} {1
  -- 6} (\bibinfo {year} {2020})}\BibitemShut {NoStop}%
\bibitem [{\citenamefont {Huang}\ \emph {et~al.}(2020)\citenamefont {Huang},
  \citenamefont {He}, \citenamefont {Wang}, \citenamefont {Zhao},\ and\
  \citenamefont {Qiu}}]{C9JA00387H}%
  \BibitemOpen
  \bibfield  {author} {\bibinfo {author} {\bibfnamefont {W.}~\bibnamefont
  {Huang}}, \bibinfo {author} {\bibfnamefont {C.}~\bibnamefont {He}}, \bibinfo
  {author} {\bibfnamefont {Y.}~\bibnamefont {Wang}}, \bibinfo {author}
  {\bibfnamefont {W.}~\bibnamefont {Zhao}}, \ and\ \bibinfo {author}
  {\bibfnamefont {L.}~\bibnamefont {Qiu}},\ }\bibfield  {title} {\enquote
  {\bibinfo {title} {Confocal controlled libs microscopy with high spatial
  resolution and stability},}\ }\href {\doibase 10.1039/C9JA00387H} {\bibfield
  {journal} {\bibinfo  {journal} {J. Anal. At. Spectrom.}\ }\textbf {\bibinfo
  {volume} {35}},\ \bibinfo {pages} {2530--2535} (\bibinfo {year}
  {2020})}\BibitemShut {NoStop}%
\bibitem [{\citenamefont {Rezaei}\ \emph {et~al.}(2020)\citenamefont {Rezaei},
  \citenamefont {Cristoforetti}, \citenamefont {Tognoni}, \citenamefont
  {Legnaioli}, \citenamefont {Palleschi},\ and\ \citenamefont
  {Safi}}]{REZAEI2020105878}%
  \BibitemOpen
  \bibfield  {author} {\bibinfo {author} {\bibfnamefont {F.}~\bibnamefont
  {Rezaei}}, \bibinfo {author} {\bibfnamefont {G.}~\bibnamefont
  {Cristoforetti}}, \bibinfo {author} {\bibfnamefont {E.}~\bibnamefont
  {Tognoni}}, \bibinfo {author} {\bibfnamefont {S.}~\bibnamefont {Legnaioli}},
  \bibinfo {author} {\bibfnamefont {V.}~\bibnamefont {Palleschi}}, \ and\
  \bibinfo {author} {\bibfnamefont {A.}~\bibnamefont {Safi}},\ }\bibfield
  {title} {\enquote {\bibinfo {title} {A review of the current analytical
  approaches for evaluating, compensating and exploiting self-absorption in
  laser induced breakdown spectroscopy},}\ }\href {\doibase
  https://doi.org/10.1016/j.sab.2020.105878} {\bibfield  {journal} {\bibinfo
  {journal} {Spectrochimica Acta Part B: Atomic Spectroscopy}\ }\textbf
  {\bibinfo {volume} {169}},\ \bibinfo {pages} {105878} (\bibinfo {year}
  {2020})}\BibitemShut {NoStop}%
\bibitem [{\citenamefont {Yang}, \citenamefont {Hao},\ and\ \citenamefont
  {Ren}(2020)}]{YANG2020163702}%
  \BibitemOpen
  \bibfield  {author} {\bibinfo {author} {\bibfnamefont {Y.}~\bibnamefont
  {Yang}}, \bibinfo {author} {\bibfnamefont {X.}~\bibnamefont {Hao}}, \ and\
  \bibinfo {author} {\bibfnamefont {L.}~\bibnamefont {Ren}},\ }\bibfield
  {title} {\enquote {\bibinfo {title} {Correction of self-absorption effect in
  calibration-free laser-induced breakdown spectroscopy(cf-libs) by considering
  plasma temperature and electron density},}\ }\href {\doibase
  https://doi.org/10.1016/j.ijleo.2019.163702} {\bibfield  {journal} {\bibinfo
  {journal} {Optik}\ }\textbf {\bibinfo {volume} {208}},\ \bibinfo {pages}
  {163702} (\bibinfo {year} {2020})}\BibitemShut {NoStop}%
\bibitem [{\citenamefont {Touchet}\ \emph {et~al.}(2020)\citenamefont
  {Touchet}, \citenamefont {Chartier}, \citenamefont {Hermann},\ and\
  \citenamefont {Sirven}}]{TOUCHET2020105868}%
  \BibitemOpen
  \bibfield  {author} {\bibinfo {author} {\bibfnamefont {K.}~\bibnamefont
  {Touchet}}, \bibinfo {author} {\bibfnamefont {F.}~\bibnamefont {Chartier}},
  \bibinfo {author} {\bibfnamefont {J.}~\bibnamefont {Hermann}}, \ and\
  \bibinfo {author} {\bibfnamefont {J.-B.}\ \bibnamefont {Sirven}},\ }\bibfield
   {title} {\enquote {\bibinfo {title} {Laser-induced breakdown self-reversal
  isotopic spectrometry for isotopic analysis of lithium},}\ }\href {\doibase
  https://doi.org/10.1016/j.sab.2020.105868} {\bibfield  {journal} {\bibinfo
  {journal} {Spectrochimica Acta Part B: Atomic Spectroscopy}\ }\textbf
  {\bibinfo {volume} {168}},\ \bibinfo {pages} {105868} (\bibinfo {year}
  {2020})}\BibitemShut {NoStop}%
\bibitem [{\citenamefont {Kumar}, \citenamefont {Singh},\ and\ \citenamefont
  {Kumar}(2013)}]{Bhupesh_POP_2013}%
  \BibitemOpen
  \bibfield  {author} {\bibinfo {author} {\bibfnamefont {B.}~\bibnamefont
  {Kumar}}, \bibinfo {author} {\bibfnamefont {R.~K.}\ \bibnamefont {Singh}}, \
  and\ \bibinfo {author} {\bibfnamefont {A.}~\bibnamefont {Kumar}},\ }\bibfield
   {title} {\enquote {\bibinfo {title} {Dynamics of laser-blow-off induced li
  plume in confined geometry},}\ }\href {\doibase 10.1063/1.4818900} {\bibfield
   {journal} {\bibinfo  {journal} {Physics of Plasmas}\ }\textbf {\bibinfo
  {volume} {20}},\ \bibinfo {pages} {083511} (\bibinfo {year} {2013})},\
  \Eprint {http://arxiv.org/abs/https://doi.org/10.1063/1.4818900}
  {https://doi.org/10.1063/1.4818900} \BibitemShut {NoStop}%
\bibitem [{\citenamefont {Kumar}\ \emph {et~al.}(2022)\citenamefont {Kumar},
  \citenamefont {Behera}, \citenamefont {Singh},\ and\ \citenamefont
  {Joshi}}]{KUMAR2022127968}%
  \BibitemOpen
  \bibfield  {author} {\bibinfo {author} {\bibfnamefont {M.}~\bibnamefont
  {Kumar}}, \bibinfo {author} {\bibfnamefont {N.}~\bibnamefont {Behera}},
  \bibinfo {author} {\bibfnamefont {R.}~\bibnamefont {Singh}}, \ and\ \bibinfo
  {author} {\bibfnamefont {H.}~\bibnamefont {Joshi}},\ }\bibfield  {title}
  {\enquote {\bibinfo {title} {Optical time-of-flight and spectroscopic
  investigation of laser produced barium plasma in presence of magnetic field
  and ambient gas},}\ }\href {\doibase
  https://doi.org/10.1016/j.physleta.2022.127968} {\bibfield  {journal}
  {\bibinfo  {journal} {Physics Letters A}\ }\textbf {\bibinfo {volume}
  {429}},\ \bibinfo {pages} {127968} (\bibinfo {year} {2022})}\BibitemShut
  {NoStop}%
\bibitem [{\citenamefont {Urbina}\ \emph {et~al.}(2022)\citenamefont {Urbina},
  \citenamefont {Bredice}, \citenamefont {Sanchez-Aké}, \citenamefont
  {Villagrán-Muniz},\ and\ \citenamefont {Palleschi}}]{URBINA2022106489}%
  \BibitemOpen
  \bibfield  {author} {\bibinfo {author} {\bibfnamefont {I.}~\bibnamefont
  {Urbina}}, \bibinfo {author} {\bibfnamefont {F.}~\bibnamefont {Bredice}},
  \bibinfo {author} {\bibfnamefont {C.}~\bibnamefont {Sanchez-Aké}}, \bibinfo
  {author} {\bibfnamefont {M.}~\bibnamefont {Villagrán-Muniz}}, \ and\
  \bibinfo {author} {\bibfnamefont {V.}~\bibnamefont {Palleschi}},\ }\bibfield
  {title} {\enquote {\bibinfo {title} {Temporal analysis of self-reversed ag i
  resonant lines in libs experiment at different laser pulse energy and in
  different surrounding media},}\ }\href {\doibase
  https://doi.org/10.1016/j.sab.2022.106489} {\bibfield  {journal} {\bibinfo
  {journal} {Spectrochimica Acta Part B: Atomic Spectroscopy}\ }\textbf
  {\bibinfo {volume} {195}},\ \bibinfo {pages} {106489} (\bibinfo {year}
  {2022})}\BibitemShut {NoStop}%
\bibitem [{\citenamefont {Kautz}, \citenamefont {Phillips},\ and\ \citenamefont
  {Harilal}(2021)}]{Kautz_JAP_2021}%
  \BibitemOpen
  \bibfield  {author} {\bibinfo {author} {\bibfnamefont {E.~J.}\ \bibnamefont
  {Kautz}}, \bibinfo {author} {\bibfnamefont {M.~C.}\ \bibnamefont {Phillips}},
  \ and\ \bibinfo {author} {\bibfnamefont {S.~S.}\ \bibnamefont {Harilal}},\
  }\bibfield  {title} {\enquote {\bibinfo {title} {Laser-induced fluorescence
  of filament-produced plasmas},}\ }\href {\doibase 10.1063/5.0065240}
  {\bibfield  {journal} {\bibinfo  {journal} {Journal of Applied Physics}\
  }\textbf {\bibinfo {volume} {130}},\ \bibinfo {pages} {203302} (\bibinfo
  {year} {2021})},\ \Eprint
  {http://arxiv.org/abs/https://doi.org/10.1063/5.0065240}
  {https://doi.org/10.1063/5.0065240} \BibitemShut {NoStop}%
\bibitem [{\citenamefont {Mayo}, \citenamefont {Ortiz},\ and\ \citenamefont
  {Plaza}(2008)}]{Mayo_2008}%
  \BibitemOpen
  \bibfield  {author} {\bibinfo {author} {\bibfnamefont {R.}~\bibnamefont
  {Mayo}}, \bibinfo {author} {\bibfnamefont {M.}~\bibnamefont {Ortiz}}, \ and\
  \bibinfo {author} {\bibfnamefont {M.}~\bibnamefont {Plaza}},\ }\bibfield
  {title} {\enquote {\bibinfo {title} {Measured stark widths of several ni {II}
  spectral lines},}\ }\href {\doibase 10.1088/0953-4075/41/9/095702} {\bibfield
   {journal} {\bibinfo  {journal} {Journal of Physics B: Atomic, Molecular and
  Optical Physics}\ }\textbf {\bibinfo {volume} {41}},\ \bibinfo {pages}
  {095702} (\bibinfo {year} {2008})}\BibitemShut {NoStop}%
\bibitem [{\citenamefont {Aragon}, \citenamefont {Aguilera},\ and\
  \citenamefont {Manrique}(2014)}]{ARAGON201439}%
  \BibitemOpen
  \bibfield  {author} {\bibinfo {author} {\bibfnamefont {C.}~\bibnamefont
  {Aragon}}, \bibinfo {author} {\bibfnamefont {J.}~\bibnamefont {Aguilera}}, \
  and\ \bibinfo {author} {\bibfnamefont {J.}~\bibnamefont {Manrique}},\
  }\bibfield  {title} {\enquote {\bibinfo {title} {Measurement of stark
  broadening parameters of fe ii and ni ii spectral lines by laser induced
  breakdown spectroscopy using fused glass samples},}\ }\href {\doibase
  https://doi.org/10.1016/j.jqsrt.2013.10.011} {\bibfield  {journal} {\bibinfo
  {journal} {Journal of Quantitative Spectroscopy and Radiative Transfer}\
  }\textbf {\bibinfo {volume} {134}},\ \bibinfo {pages} {39--45} (\bibinfo
  {year} {2014})}\BibitemShut {NoStop}%
\bibitem [{\citenamefont {Manrique}, \citenamefont {Aguilera},\ and\
  \citenamefont {Aragon}(2016)}]{mnras_Manrique}%
  \BibitemOpen
  \bibfield  {author} {\bibinfo {author} {\bibfnamefont {J.}~\bibnamefont
  {Manrique}}, \bibinfo {author} {\bibfnamefont {J.~A.}\ \bibnamefont
  {Aguilera}}, \ and\ \bibinfo {author} {\bibfnamefont {C.}~\bibnamefont
  {Aragon}},\ }\bibfield  {title} {\enquote {\bibinfo {title} {{Experimental
  Stark widths and shifts of Ti ii spectral lines}},}\ }\href {\doibase
  10.1093/mnras/stw1641} {\bibfield  {journal} {\bibinfo  {journal} {Monthly
  Notices of the Royal Astronomical Society}\ }\textbf {\bibinfo {volume}
  {462}},\ \bibinfo {pages} {1501--1507} (\bibinfo {year} {2016})},\ \Eprint
  {http://arxiv.org/abs/https://academic.oup.com/mnras/article-pdf/462/2/1501/13773749/stw1641.pdf}
  {https://academic.oup.com/mnras/article-pdf/462/2/1501/13773749/stw1641.pdf}
  \BibitemShut {NoStop}%
\bibitem [{\citenamefont {Thomas}\ \emph {et~al.}(2018)\citenamefont {Thomas},
  \citenamefont {Joshi}, \citenamefont {Kumar},\ and\ \citenamefont
  {Philip}}]{Jinto_POP}%
  \BibitemOpen
  \bibfield  {author} {\bibinfo {author} {\bibfnamefont {J.}~\bibnamefont
  {Thomas}}, \bibinfo {author} {\bibfnamefont {H.~C.}\ \bibnamefont {Joshi}},
  \bibinfo {author} {\bibfnamefont {A.}~\bibnamefont {Kumar}}, \ and\ \bibinfo
  {author} {\bibfnamefont {R.}~\bibnamefont {Philip}},\ }\bibfield  {title}
  {\enquote {\bibinfo {title} {Effect of ambient gas pressure on nanosecond
  laser produced plasma on nickel thin film in a forward ablation geometry},}\
  }\href {\doibase 10.1063/1.5048834} {\bibfield  {journal} {\bibinfo
  {journal} {Physics of Plasmas}\ }\textbf {\bibinfo {volume} {25}},\ \bibinfo
  {pages} {103108} (\bibinfo {year} {2018})},\ \Eprint
  {http://arxiv.org/abs/https://doi.org/10.1063/1.5048834}
  {https://doi.org/10.1063/1.5048834} \BibitemShut {NoStop}%
\bibitem [{\citenamefont {{El Sherbini}}, \citenamefont {Hegazy},\ and\
  \citenamefont {{El Sherbini}}(2006)}]{ElSherbini2006}%
  \BibitemOpen
  \bibfield  {author} {\bibinfo {author} {\bibfnamefont {A.}~\bibnamefont {{El
  Sherbini}}}, \bibinfo {author} {\bibfnamefont {H.}~\bibnamefont {Hegazy}}, \
  and\ \bibinfo {author} {\bibfnamefont {T.}~\bibnamefont {{El Sherbini}}},\
  }\bibfield  {title} {\enquote {\bibinfo {title} {Measurement of electron
  density utilizing the h-alpha line from laser produced plasma in air},}\
  }\href {\doibase https://doi.org/10.1016/j.sab.2006.03.014} {\bibfield
  {journal} {\bibinfo  {journal} {Spectrochimica Acta Part B: Atomic
  Spectroscopy}\ }\textbf {\bibinfo {volume} {61}},\ \bibinfo {pages}
  {532--539} (\bibinfo {year} {2006})}\BibitemShut {NoStop}%
\bibitem [{\citenamefont {Nishijima}\ and\ \citenamefont
  {Doerner}(2015)}]{Nishijima_2015}%
  \BibitemOpen
  \bibfield  {author} {\bibinfo {author} {\bibfnamefont {D.}~\bibnamefont
  {Nishijima}}\ and\ \bibinfo {author} {\bibfnamefont {R.~P.}\ \bibnamefont
  {Doerner}},\ }\bibfield  {title} {\enquote {\bibinfo {title} {Stark width
  measurements and boltzmann plots of w i in nanosecond laser-induced
  plasmas},}\ }\href {\doibase 10.1088/0022-3727/48/32/325201} {\bibfield
  {journal} {\bibinfo  {journal} {Journal of Physics D: Applied Physics}\
  }\textbf {\bibinfo {volume} {48}},\ \bibinfo {pages} {325201} (\bibinfo
  {year} {2015})}\BibitemShut {NoStop}%
\bibitem [{\citenamefont {Liu}, \citenamefont {Truscott},\ and\ \citenamefont
  {Ashfold}(2016)}]{Liu2016}%
  \BibitemOpen
  \bibfield  {author} {\bibinfo {author} {\bibfnamefont {H.}~\bibnamefont
  {Liu}}, \bibinfo {author} {\bibfnamefont {B.~S.}\ \bibnamefont {Truscott}}, \
  and\ \bibinfo {author} {\bibfnamefont {M.~N.~R.}\ \bibnamefont {Ashfold}},\
  }\bibfield  {title} {\enquote {\bibinfo {title} {Determination of stark
  parameters by cross-calibration in a multi-element laser-induced plasma},}\
  }\href {\doibase 10.1038/srep25609} {\bibfield  {journal} {\bibinfo
  {journal} {Scientific Reports}\ }\textbf {\bibinfo {volume} {6}},\ \bibinfo
  {pages} {25609} (\bibinfo {year} {2016})}\BibitemShut {NoStop}%
\bibitem [{\citenamefont {Poggialini}\ \emph {et~al.}(2020)\citenamefont
  {Poggialini}, \citenamefont {Campanella}, \citenamefont {Jafer},
  \citenamefont {Legnaioli}, \citenamefont {Bredice}, \citenamefont {Raneri},\
  and\ \citenamefont {Palleschi}}]{POGGIALINI2020105829}%
  \BibitemOpen
  \bibfield  {author} {\bibinfo {author} {\bibfnamefont {F.}~\bibnamefont
  {Poggialini}}, \bibinfo {author} {\bibfnamefont {B.}~\bibnamefont
  {Campanella}}, \bibinfo {author} {\bibfnamefont {R.}~\bibnamefont {Jafer}},
  \bibinfo {author} {\bibfnamefont {S.}~\bibnamefont {Legnaioli}}, \bibinfo
  {author} {\bibfnamefont {F.}~\bibnamefont {Bredice}}, \bibinfo {author}
  {\bibfnamefont {S.}~\bibnamefont {Raneri}}, \ and\ \bibinfo {author}
  {\bibfnamefont {V.}~\bibnamefont {Palleschi}},\ }\bibfield  {title} {\enquote
  {\bibinfo {title} {Determination of the stark broadening coefficients of
  tantalum emission lines by time-independent extended c-sigma method},}\
  }\href {\doibase https://doi.org/10.1016/j.sab.2020.105829} {\bibfield
  {journal} {\bibinfo  {journal} {Spectrochimica Acta Part B: Atomic
  Spectroscopy}\ }\textbf {\bibinfo {volume} {167}},\ \bibinfo {pages} {105829}
  (\bibinfo {year} {2020})}\BibitemShut {NoStop}%
\bibitem [{\citenamefont {Griem}(1974)}]{Griem1974}%
  \BibitemOpen
  \bibfield  {author} {\bibinfo {author} {\bibfnamefont {H.}~\bibnamefont
  {Griem}},\ }\href@noop {} {\emph {\bibinfo {title} {Spectral Line Broadening
  by Plasmas}}}\ (\bibinfo  {publisher} {Academic Press, New York},\ \bibinfo
  {year} {1974})\BibitemShut {NoStop}%
\bibitem [{\citenamefont {Harilal}\ \emph {et~al.}(2022)\citenamefont
  {Harilal}, \citenamefont {Phillips}, \citenamefont {Froula}, \citenamefont
  {Anoop}, \citenamefont {Issac},\ and\ \citenamefont
  {Beg}}]{Hari_RevModPhys.94.035002}%
  \BibitemOpen
  \bibfield  {author} {\bibinfo {author} {\bibfnamefont {S.~S.}\ \bibnamefont
  {Harilal}}, \bibinfo {author} {\bibfnamefont {M.~C.}\ \bibnamefont
  {Phillips}}, \bibinfo {author} {\bibfnamefont {D.~H.}\ \bibnamefont
  {Froula}}, \bibinfo {author} {\bibfnamefont {K.~K.}\ \bibnamefont {Anoop}},
  \bibinfo {author} {\bibfnamefont {R.~C.}\ \bibnamefont {Issac}}, \ and\
  \bibinfo {author} {\bibfnamefont {F.~N.}\ \bibnamefont {Beg}},\ }\bibfield
  {title} {\enquote {\bibinfo {title} {Optical diagnostics of laser-produced
  plasmas},}\ }\href {\doibase 10.1103/RevModPhys.94.035002} {\bibfield
  {journal} {\bibinfo  {journal} {Rev. Mod. Phys.}\ }\textbf {\bibinfo {volume}
  {94}},\ \bibinfo {pages} {035002} (\bibinfo {year} {2022})}\BibitemShut
  {NoStop}%
\bibitem [{\citenamefont {Choudhury}\ \emph {et~al.}(2016)\citenamefont
  {Choudhury}, \citenamefont {Singh}, \citenamefont {Narayan}, \citenamefont
  {Srivastava},\ and\ \citenamefont {Kumar}}]{Choudhary_interferometry}%
  \BibitemOpen
  \bibfield  {author} {\bibinfo {author} {\bibfnamefont {K.}~\bibnamefont
  {Choudhury}}, \bibinfo {author} {\bibfnamefont {R.~K.}\ \bibnamefont
  {Singh}}, \bibinfo {author} {\bibfnamefont {S.}~\bibnamefont {Narayan}},
  \bibinfo {author} {\bibfnamefont {A.}~\bibnamefont {Srivastava}}, \ and\
  \bibinfo {author} {\bibfnamefont {A.}~\bibnamefont {Kumar}},\ }\bibfield
  {title} {\enquote {\bibinfo {title} {Time resolved interferometric study of
  the plasma plume induced shock wave in confined geometry: Two-dimensional
  mapping of the ambient and plasma density},}\ }\href {\doibase
  10.1063/1.4947032} {\bibfield  {journal} {\bibinfo  {journal} {Physics of
  Plasmas}\ }\textbf {\bibinfo {volume} {23}},\ \bibinfo {pages} {042108}
  (\bibinfo {year} {2016})},\ \Eprint
  {http://arxiv.org/abs/https://doi.org/10.1063/1.4947032}
  {https://doi.org/10.1063/1.4947032} \BibitemShut {NoStop}%
\bibitem [{\citenamefont {Muraoka}\ and\ \citenamefont
  {Kono}(2011)}]{Muraoka_2011}%
  \BibitemOpen
  \bibfield  {author} {\bibinfo {author} {\bibfnamefont {K.}~\bibnamefont
  {Muraoka}}\ and\ \bibinfo {author} {\bibfnamefont {A.}~\bibnamefont {Kono}},\
  }\bibfield  {title} {\enquote {\bibinfo {title} {Laser thomson scattering for
  low-temperature plasmas},}\ }\href {\doibase 10.1088/0022-3727/44/4/043001}
  {\bibfield  {journal} {\bibinfo  {journal} {Journal of Physics D: Applied
  Physics}\ }\textbf {\bibinfo {volume} {44}},\ \bibinfo {pages} {043001}
  (\bibinfo {year} {2011})}\BibitemShut {NoStop}%
\bibitem [{\citenamefont {Aragon}\ and\ \citenamefont
  {Aguilera}(2008)}]{ARAGON2008893}%
  \BibitemOpen
  \bibfield  {author} {\bibinfo {author} {\bibfnamefont {C.}~\bibnamefont
  {Aragon}}\ and\ \bibinfo {author} {\bibfnamefont {J.}~\bibnamefont
  {Aguilera}},\ }\bibfield  {title} {\enquote {\bibinfo {title}
  {Characterization of laser induced plasmas by optical emission spectroscopy:
  A review of experiments and methods},}\ }\href {\doibase
  https://doi.org/10.1016/j.sab.2008.05.010} {\bibfield  {journal} {\bibinfo
  {journal} {Spectrochimica Acta Part B: Atomic Spectroscopy}\ }\textbf
  {\bibinfo {volume} {63}},\ \bibinfo {pages} {893 -- 916} (\bibinfo {year}
  {2008})}\BibitemShut {NoStop}%
\bibitem [{\citenamefont {Harilal}\ \emph {et~al.}(1998)\citenamefont
  {Harilal}, \citenamefont {Bindhu}, \citenamefont {Nampoori},\ and\
  \citenamefont {Vallabhan}}]{Harilal1998}%
  \BibitemOpen
  \bibfield  {author} {\bibinfo {author} {\bibfnamefont {S.~S.}\ \bibnamefont
  {Harilal}}, \bibinfo {author} {\bibfnamefont {C.~V.}\ \bibnamefont {Bindhu}},
  \bibinfo {author} {\bibfnamefont {V.~P.~N.}\ \bibnamefont {Nampoori}}, \ and\
  \bibinfo {author} {\bibfnamefont {C.~P.~G.}\ \bibnamefont {Vallabhan}},\
  }\bibfield  {title} {\enquote {\bibinfo {title} {Time evolution of the
  electron density and temperature in laser-produced plasmas from yba2cu3o7},}\
  }\href {\doibase 10.1007/s003400050448} {\bibfield  {journal} {\bibinfo
  {journal} {Applied Physics B}\ }\textbf {\bibinfo {volume} {66}},\ \bibinfo
  {pages} {633--638} (\bibinfo {year} {1998})}\BibitemShut {NoStop}%
\bibitem [{\citenamefont {Moon}, \citenamefont {Smith},\ and\ \citenamefont
  {Omenetto}(2012)}]{MOON2012221}%
  \BibitemOpen
  \bibfield  {author} {\bibinfo {author} {\bibfnamefont {H.-Y.}\ \bibnamefont
  {Moon}}, \bibinfo {author} {\bibfnamefont {B.~W.}\ \bibnamefont {Smith}}, \
  and\ \bibinfo {author} {\bibfnamefont {N.}~\bibnamefont {Omenetto}},\
  }\bibfield  {title} {\enquote {\bibinfo {title} {Temporal behavior of
  line-to-continuum ratios and ion fractions as a means of assessing
  thermodynamic equilibrium in laser-induced breakdown spectroscopy},}\ }\href
  {\doibase https://doi.org/10.1016/j.chemphys.2011.07.002} {\bibfield
  {journal} {\bibinfo  {journal} {Chemical Physics}\ }\textbf {\bibinfo
  {volume} {398}},\ \bibinfo {pages} {221 -- 227} (\bibinfo {year} {2012})},\
  \bibinfo {note} {chemical Physics of Low-Temperature Plasmas (in honour of
  Prof Mario Capitelli)}\BibitemShut {NoStop}%
\bibitem [{\citenamefont {Cristoforetti}\ \emph {et~al.}(2010)\citenamefont
  {Cristoforetti}, \citenamefont {Giacomo}, \citenamefont {Dell'Aglio},
  \citenamefont {Legnaioli}, \citenamefont {Tognoni}, \citenamefont
  {Palleschi},\ and\ \citenamefont {Omenetto}}]{CRISTOFORETTI}%
  \BibitemOpen
  \bibfield  {author} {\bibinfo {author} {\bibfnamefont {G.}~\bibnamefont
  {Cristoforetti}}, \bibinfo {author} {\bibfnamefont {A.~D.}\ \bibnamefont
  {Giacomo}}, \bibinfo {author} {\bibfnamefont {M.}~\bibnamefont {Dell'Aglio}},
  \bibinfo {author} {\bibfnamefont {S.}~\bibnamefont {Legnaioli}}, \bibinfo
  {author} {\bibfnamefont {E.}~\bibnamefont {Tognoni}}, \bibinfo {author}
  {\bibfnamefont {V.}~\bibnamefont {Palleschi}}, \ and\ \bibinfo {author}
  {\bibfnamefont {N.}~\bibnamefont {Omenetto}},\ }\bibfield  {title} {\enquote
  {\bibinfo {title} {Local thermodynamic equilibrium in laser-induced breakdown
  spectroscopy: Beyond the mcwhirter criterion},}\ }\href {\doibase
  https://doi.org/10.1016/j.sab.2009.11.005} {\bibfield  {journal} {\bibinfo
  {journal} {Spectrochimica Acta Part B: Atomic Spectroscopy}\ }\textbf
  {\bibinfo {volume} {65}},\ \bibinfo {pages} {86 -- 95} (\bibinfo {year}
  {2010})}\BibitemShut {NoStop}%
\bibitem [{\citenamefont {Tognoni}\ \emph {et~al.}(2010)\citenamefont
  {Tognoni}, \citenamefont {Cristoforetti}, \citenamefont {Legnaioli},\ and\
  \citenamefont {Palleschi}}]{TOGNONI20101}%
  \BibitemOpen
  \bibfield  {author} {\bibinfo {author} {\bibfnamefont {E.}~\bibnamefont
  {Tognoni}}, \bibinfo {author} {\bibfnamefont {G.}~\bibnamefont
  {Cristoforetti}}, \bibinfo {author} {\bibfnamefont {S.}~\bibnamefont
  {Legnaioli}}, \ and\ \bibinfo {author} {\bibfnamefont {V.}~\bibnamefont
  {Palleschi}},\ }\bibfield  {title} {\enquote {\bibinfo {title}
  {Calibration-free laser-induced breakdown spectroscopy: State of the art},}\
  }\href {\doibase https://doi.org/10.1016/j.sab.2009.11.006} {\bibfield
  {journal} {\bibinfo  {journal} {Spectrochimica Acta Part B: Atomic
  Spectroscopy}\ }\textbf {\bibinfo {volume} {65}},\ \bibinfo {pages} {1--14}
  (\bibinfo {year} {2010})}\BibitemShut {NoStop}%
\bibitem [{\citenamefont {Aragon}\ and\ \citenamefont
  {Aguilera}(2014)}]{ARAGON201490}%
  \BibitemOpen
  \bibfield  {author} {\bibinfo {author} {\bibfnamefont {C.}~\bibnamefont
  {Aragon}}\ and\ \bibinfo {author} {\bibfnamefont {J.}~\bibnamefont
  {Aguilera}},\ }\bibfield  {title} {\enquote {\bibinfo {title} {Csigma graphs:
  A new approach for plasma characterization in laser-induced breakdown
  spectroscopy},}\ }\href {\doibase
  https://doi.org/10.1016/j.jqsrt.2014.07.026} {\bibfield  {journal} {\bibinfo
  {journal} {Journal of Quantitative Spectroscopy and Radiative Transfer}\
  }\textbf {\bibinfo {volume} {149}},\ \bibinfo {pages} {90--102} (\bibinfo
  {year} {2014})}\BibitemShut {NoStop}%
\bibitem [{\citenamefont {Sun}\ and\ \citenamefont
  {Yu}(2009)}]{Sun2009CorrectionOS}%
  \BibitemOpen
  \bibfield  {author} {\bibinfo {author} {\bibfnamefont {L.}~\bibnamefont
  {Sun}}\ and\ \bibinfo {author} {\bibfnamefont {H.}~\bibnamefont {Yu}},\
  }\bibfield  {title} {\enquote {\bibinfo {title} {Correction of
  self-absorption effect in calibration-free laser-induced breakdown
  spectroscopy by an internal reference method.}}\ }\href@noop {} {\bibfield
  {journal} {\bibinfo  {journal} {Talanta}\ }\textbf {\bibinfo {volume} {79
  2}},\ \bibinfo {pages} {388--95} (\bibinfo {year} {2009})}\BibitemShut
  {NoStop}%
\bibitem [{\citenamefont {Hu}\ \emph {et~al.}(2021)\citenamefont {Hu},
  \citenamefont {Chen}, \citenamefont {Zhang}, \citenamefont {Chu},
  \citenamefont {Wang}, \citenamefont {Tang},\ and\ \citenamefont
  {Guo}}]{HU2021339008}%
  \BibitemOpen
  \bibfield  {author} {\bibinfo {author} {\bibfnamefont {Z.}~\bibnamefont
  {Hu}}, \bibinfo {author} {\bibfnamefont {F.}~\bibnamefont {Chen}}, \bibinfo
  {author} {\bibfnamefont {D.}~\bibnamefont {Zhang}}, \bibinfo {author}
  {\bibfnamefont {Y.}~\bibnamefont {Chu}}, \bibinfo {author} {\bibfnamefont
  {W.}~\bibnamefont {Wang}}, \bibinfo {author} {\bibfnamefont {Y.}~\bibnamefont
  {Tang}}, \ and\ \bibinfo {author} {\bibfnamefont {L.}~\bibnamefont {Guo}},\
  }\bibfield  {title} {\enquote {\bibinfo {title} {A method for improving the
  accuracy of calibration-free laser-induced breakdown spectroscopy by
  exploiting self-absorption},}\ }\href {\doibase
  https://doi.org/10.1016/j.aca.2021.339008} {\bibfield  {journal} {\bibinfo
  {journal} {Analytica Chimica Acta}\ }\textbf {\bibinfo {volume} {1183}},\
  \bibinfo {pages} {339008} (\bibinfo {year} {2021})}\BibitemShut {NoStop}%
\bibitem [{\citenamefont {Hai}\ \emph {et~al.}(2019)\citenamefont {Hai},
  \citenamefont {He}, \citenamefont {Wu}, \citenamefont {Tong}, \citenamefont
  {Sattar}, \citenamefont {Imran},\ and\ \citenamefont {Ding}}]{C9JA00261H}%
  \BibitemOpen
  \bibfield  {author} {\bibinfo {author} {\bibfnamefont {R.}~\bibnamefont
  {Hai}}, \bibinfo {author} {\bibfnamefont {Z.}~\bibnamefont {He}}, \bibinfo
  {author} {\bibfnamefont {D.}~\bibnamefont {Wu}}, \bibinfo {author}
  {\bibfnamefont {W.}~\bibnamefont {Tong}}, \bibinfo {author} {\bibfnamefont
  {H.}~\bibnamefont {Sattar}}, \bibinfo {author} {\bibfnamefont
  {M.}~\bibnamefont {Imran}}, \ and\ \bibinfo {author} {\bibfnamefont
  {H.}~\bibnamefont {Ding}},\ }\bibfield  {title} {\enquote {\bibinfo {title}
  {Influence of sample temperature on the laser-induced breakdown spectroscopy
  of a molybdenum–tungsten alloy},}\ }\href {\doibase 10.1039/C9JA00261H}
  {\bibfield  {journal} {\bibinfo  {journal} {J. Anal. At. Spectrom.}\ }\textbf
  {\bibinfo {volume} {34}},\ \bibinfo {pages} {2378--2384} (\bibinfo {year}
  {2019})}\BibitemShut {NoStop}%
\bibitem [{\citenamefont {Lednev}\ \emph {et~al.}(2019)\citenamefont {Lednev},
  \citenamefont {Grishin}, \citenamefont {Sdvizhenskii}, \citenamefont
  {Asyutin}, \citenamefont {Tretyakov}, \citenamefont {Stavertiy},\ and\
  \citenamefont {Pershin}}]{C8JA00348C}%
  \BibitemOpen
  \bibfield  {author} {\bibinfo {author} {\bibfnamefont {V.~N.}\ \bibnamefont
  {Lednev}}, \bibinfo {author} {\bibfnamefont {M.~Y.}\ \bibnamefont {Grishin}},
  \bibinfo {author} {\bibfnamefont {P.~A.}\ \bibnamefont {Sdvizhenskii}},
  \bibinfo {author} {\bibfnamefont {R.~D.}\ \bibnamefont {Asyutin}}, \bibinfo
  {author} {\bibfnamefont {R.~S.}\ \bibnamefont {Tretyakov}}, \bibinfo {author}
  {\bibfnamefont {A.~Y.}\ \bibnamefont {Stavertiy}}, \ and\ \bibinfo {author}
  {\bibfnamefont {S.~M.}\ \bibnamefont {Pershin}},\ }\bibfield  {title}
  {\enquote {\bibinfo {title} {Sample temperature effect on laser ablation and
  analytical capabilities of laser induced breakdown spectroscopy},}\ }\href
  {\doibase 10.1039/C8JA00348C} {\bibfield  {journal} {\bibinfo  {journal} {J.
  Anal. At. Spectrom.}\ }\textbf {\bibinfo {volume} {34}},\ \bibinfo {pages}
  {607--615} (\bibinfo {year} {2019})}\BibitemShut {NoStop}%
\bibitem [{\citenamefont {Guo}\ \emph {et~al.}(2019)\citenamefont {Guo},
  \citenamefont {Chen}, \citenamefont {Xu}, \citenamefont {Zhang},\ and\
  \citenamefont {Jin}}]{Guo_AIP_Adva_2019}%
  \BibitemOpen
  \bibfield  {author} {\bibinfo {author} {\bibfnamefont {K.}~\bibnamefont
  {Guo}}, \bibinfo {author} {\bibfnamefont {A.}~\bibnamefont {Chen}}, \bibinfo
  {author} {\bibfnamefont {W.}~\bibnamefont {Xu}}, \bibinfo {author}
  {\bibfnamefont {D.}~\bibnamefont {Zhang}}, \ and\ \bibinfo {author}
  {\bibfnamefont {M.}~\bibnamefont {Jin}},\ }\bibfield  {title} {\enquote
  {\bibinfo {title} {Effect of sample temperature on time-resolved
  laser-induced breakdown spectroscopy},}\ }\href {\doibase 10.1063/1.5097301}
  {\bibfield  {journal} {\bibinfo  {journal} {AIP Advances}\ }\textbf {\bibinfo
  {volume} {9}},\ \bibinfo {pages} {065214} (\bibinfo {year} {2019})},\ \Eprint
  {http://arxiv.org/abs/https://doi.org/10.1063/1.5097301}
  {https://doi.org/10.1063/1.5097301} \BibitemShut {NoStop}%
\bibitem [{\citenamefont {Tavassoli}\ and\ \citenamefont
  {Gragossian}(2009)}]{TAVASSOLI2009481}%
  \BibitemOpen
  \bibfield  {author} {\bibinfo {author} {\bibfnamefont {S.}~\bibnamefont
  {Tavassoli}}\ and\ \bibinfo {author} {\bibfnamefont {A.}~\bibnamefont
  {Gragossian}},\ }\bibfield  {title} {\enquote {\bibinfo {title} {Effect of
  sample temperature on laser-induced breakdown spectroscopy},}\ }\href
  {\doibase https://doi.org/10.1016/j.optlastec.2008.07.010} {\bibfield
  {journal} {\bibinfo  {journal} {Optics \& Laser Technology}\ }\textbf
  {\bibinfo {volume} {41}},\ \bibinfo {pages} {481--485} (\bibinfo {year}
  {2009})}\BibitemShut {NoStop}%
\bibitem [{\citenamefont {Rai}\ \emph {et~al.}(2003)\citenamefont {Rai},
  \citenamefont {Zhang}, \citenamefont {Yueh}, \citenamefont {Singh},\ and\
  \citenamefont {Kumar}}]{Rai_2003}%
  \BibitemOpen
  \bibfield  {author} {\bibinfo {author} {\bibfnamefont {V.~N.}\ \bibnamefont
  {Rai}}, \bibinfo {author} {\bibfnamefont {H.}~\bibnamefont {Zhang}}, \bibinfo
  {author} {\bibfnamefont {F.~Y.}\ \bibnamefont {Yueh}}, \bibinfo {author}
  {\bibfnamefont {J.~P.}\ \bibnamefont {Singh}}, \ and\ \bibinfo {author}
  {\bibfnamefont {A.}~\bibnamefont {Kumar}},\ }\bibfield  {title} {\enquote
  {\bibinfo {title} {Effect of steady magnetic field on laser-induced breakdown
  spectroscopy},}\ }\href {\doibase 10.1364/AO.42.003662} {\bibfield  {journal}
  {\bibinfo  {journal} {Appl. Opt.}\ }\textbf {\bibinfo {volume} {42}},\
  \bibinfo {pages} {3662--3669} (\bibinfo {year} {2003})}\BibitemShut {NoStop}%
\bibitem [{\citenamefont {Joshi}\ \emph {et~al.}(2010)\citenamefont {Joshi},
  \citenamefont {Kumar}, \citenamefont {Singh},\ and\ \citenamefont
  {Prahlad}}]{JOSHI2010415}%
  \BibitemOpen
  \bibfield  {author} {\bibinfo {author} {\bibfnamefont {H.}~\bibnamefont
  {Joshi}}, \bibinfo {author} {\bibfnamefont {A.}~\bibnamefont {Kumar}},
  \bibinfo {author} {\bibfnamefont {R.}~\bibnamefont {Singh}}, \ and\ \bibinfo
  {author} {\bibfnamefont {V.}~\bibnamefont {Prahlad}},\ }\bibfield  {title}
  {\enquote {\bibinfo {title} {Effect of a transverse magnetic field on the
  plume emission in laser-produced plasma: An atomic analysis},}\ }\href
  {\doibase https://doi.org/10.1016/j.sab.2010.04.018} {\bibfield  {journal}
  {\bibinfo  {journal} {Spectrochimica Acta Part B: Atomic Spectroscopy}\
  }\textbf {\bibinfo {volume} {65}},\ \bibinfo {pages} {415--419} (\bibinfo
  {year} {2010})}\BibitemShut {NoStop}%
\bibitem [{\citenamefont {Ahmed}\ \emph {et~al.}(2020)\citenamefont {Ahmed},
  \citenamefont {Jabbar}, \citenamefont {Akhtar}, \citenamefont {Umar},\ and\
  \citenamefont {Baig}}]{Ahmed2020}%
  \BibitemOpen
  \bibfield  {author} {\bibinfo {author} {\bibfnamefont {R.}~\bibnamefont
  {Ahmed}}, \bibinfo {author} {\bibfnamefont {A.}~\bibnamefont {Jabbar}},
  \bibinfo {author} {\bibfnamefont {M.}~\bibnamefont {Akhtar}}, \bibinfo
  {author} {\bibfnamefont {Z.~A.}\ \bibnamefont {Umar}}, \ and\ \bibinfo
  {author} {\bibfnamefont {M.~A.}\ \bibnamefont {Baig}},\ }\bibfield  {title}
  {\enquote {\bibinfo {title} {Amelioration in the detection of chlorine using
  electric field assisted libs},}\ }\href {\doibase 10.1007/s11090-020-10072-4}
  {\bibfield  {journal} {\bibinfo  {journal} {Plasma Chemistry and Plasma
  Processing}\ }\textbf {\bibinfo {volume} {40}},\ \bibinfo {pages} {809--818}
  (\bibinfo {year} {2020})}\BibitemShut {NoStop}%
\bibitem [{\citenamefont {AHMED}\ \emph {et~al.}(2021)\citenamefont {AHMED},
  \citenamefont {JABBAR}, \citenamefont {UMAR},\ and\ \citenamefont
  {BAIG}}]{AHMED_2021}%
  \BibitemOpen
  \bibfield  {author} {\bibinfo {author} {\bibfnamefont {R.}~\bibnamefont
  {AHMED}}, \bibinfo {author} {\bibfnamefont {A.}~\bibnamefont {JABBAR}},
  \bibinfo {author} {\bibfnamefont {Z.~A.}\ \bibnamefont {UMAR}}, \ and\
  \bibinfo {author} {\bibfnamefont {M.~A.}\ \bibnamefont {BAIG}},\ }\bibfield
  {title} {\enquote {\bibinfo {title} {Electric-field induced fluctuations in
  laser generated plasma plume},}\ }\href {\doibase 10.1088/2058-6272/abea70}
  {\bibfield  {journal} {\bibinfo  {journal} {Plasma Science and Technology}\
  }\textbf {\bibinfo {volume} {23}},\ \bibinfo {pages} {045505} (\bibinfo
  {year} {2021})}\BibitemShut {NoStop}%
\bibitem [{\citenamefont {Asamoah}\ \emph {et~al.}(2021)\citenamefont
  {Asamoah}, \citenamefont {Jiawei}, \citenamefont {Ayepah}, \citenamefont
  {Wilson}, \citenamefont {Hongbing}, \citenamefont {Weihua}, \citenamefont
  {Lin}, \citenamefont {Pengyu},\ and\ \citenamefont {Asamoah}}]{Asamoah2021}%
  \BibitemOpen
  \bibfield  {author} {\bibinfo {author} {\bibfnamefont {E.}~\bibnamefont
  {Asamoah}}, \bibinfo {author} {\bibfnamefont {C.}~\bibnamefont {Jiawei}},
  \bibinfo {author} {\bibfnamefont {K.}~\bibnamefont {Ayepah}}, \bibinfo
  {author} {\bibfnamefont {S.}~\bibnamefont {Wilson}}, \bibinfo {author}
  {\bibfnamefont {Y.}~\bibnamefont {Hongbing}}, \bibinfo {author}
  {\bibfnamefont {Z.}~\bibnamefont {Weihua}}, \bibinfo {author} {\bibfnamefont
  {Z.}~\bibnamefont {Lin}}, \bibinfo {author} {\bibfnamefont {W.}~\bibnamefont
  {Pengyu}}, \ and\ \bibinfo {author} {\bibfnamefont {A.}~\bibnamefont
  {Asamoah}},\ }\bibfield  {title} {\enquote {\bibinfo {title} {Investigating a
  laser-induced titanium plasma under an applied static electric field},}\
  }\href
  {https://www.spectroscopyonline.com/view/investigating-a-laser-induced-titanium-plasma-under-an-applied-static-electric-field}
  {\bibfield  {journal} {\bibinfo  {journal} {Spectroscopy (Santa Monica)}\
  }\textbf {\bibinfo {volume} {36}} (\bibinfo {year} {2021})},\ \bibinfo {note}
  {cited By 1}\BibitemShut {NoStop}%
\bibitem [{\citenamefont {Tereszchuk}, \citenamefont {Vadillo},\ and\
  \citenamefont {Laserna}(2008)}]{Tereszchuk_2008}%
  \BibitemOpen
  \bibfield  {author} {\bibinfo {author} {\bibfnamefont {K.~A.}\ \bibnamefont
  {Tereszchuk}}, \bibinfo {author} {\bibfnamefont {J.~M.}\ \bibnamefont
  {Vadillo}}, \ and\ \bibinfo {author} {\bibfnamefont {J.~J.}\ \bibnamefont
  {Laserna}},\ }\bibfield  {title} {\enquote {\bibinfo {title}
  {Glow-discharge-assisted laser-induced breakdown spectroscopy: Increased
  sensitivity in solid analysis},}\ }\href
  {http://www.osapublishing.org/as/abstract.cfm?URI=as-62-11-1262} {\bibfield
  {journal} {\bibinfo  {journal} {Appl. Spectrosc.}\ }\textbf {\bibinfo
  {volume} {62}},\ \bibinfo {pages} {1262--1267} (\bibinfo {year}
  {2008})}\BibitemShut {NoStop}%
\bibitem [{\citenamefont {Tereszchuk}, \citenamefont {Vadillo},\ and\
  \citenamefont {Laserna}(2009)}]{TERESZCHUK2009378}%
  \BibitemOpen
  \bibfield  {author} {\bibinfo {author} {\bibfnamefont {K.}~\bibnamefont
  {Tereszchuk}}, \bibinfo {author} {\bibfnamefont {J.}~\bibnamefont {Vadillo}},
  \ and\ \bibinfo {author} {\bibfnamefont {J.}~\bibnamefont {Laserna}},\
  }\bibfield  {title} {\enquote {\bibinfo {title} {Depth profile analysis of
  layered samples using glow discharge assisted laser-induced breakdown
  spectrometry (gd-libs)},}\ }\href {\doibase
  https://doi.org/10.1016/j.sab.2009.04.005} {\bibfield  {journal} {\bibinfo
  {journal} {Spectrochimica Acta Part B: Atomic Spectroscopy}\ }\textbf
  {\bibinfo {volume} {64}},\ \bibinfo {pages} {378--383} (\bibinfo {year}
  {2009})}\BibitemShut {NoStop}%
\bibitem [{\citenamefont {Liu}\ \emph {et~al.}(2014)\citenamefont {Liu},
  \citenamefont {Li}, \citenamefont {He}, \citenamefont {Huang}, \citenamefont
  {Zhang}, \citenamefont {Fan}, \citenamefont {Wang}, \citenamefont {Zhou},
  \citenamefont {Chen}, \citenamefont {Jiang}, \citenamefont {Silvain},\ and\
  \citenamefont {Lu}}]{Liu_2014}%
  \BibitemOpen
  \bibfield  {author} {\bibinfo {author} {\bibfnamefont {L.}~\bibnamefont
  {Liu}}, \bibinfo {author} {\bibfnamefont {S.}~\bibnamefont {Li}}, \bibinfo
  {author} {\bibfnamefont {X.~N.}\ \bibnamefont {He}}, \bibinfo {author}
  {\bibfnamefont {X.}~\bibnamefont {Huang}}, \bibinfo {author} {\bibfnamefont
  {C.~F.}\ \bibnamefont {Zhang}}, \bibinfo {author} {\bibfnamefont {L.~S.}\
  \bibnamefont {Fan}}, \bibinfo {author} {\bibfnamefont {M.~X.}\ \bibnamefont
  {Wang}}, \bibinfo {author} {\bibfnamefont {Y.~S.}\ \bibnamefont {Zhou}},
  \bibinfo {author} {\bibfnamefont {K.}~\bibnamefont {Chen}}, \bibinfo {author}
  {\bibfnamefont {L.}~\bibnamefont {Jiang}}, \bibinfo {author} {\bibfnamefont
  {J.~F.}\ \bibnamefont {Silvain}}, \ and\ \bibinfo {author} {\bibfnamefont
  {Y.~F.}\ \bibnamefont {Lu}},\ }\bibfield  {title} {\enquote {\bibinfo {title}
  {Flame-enhanced laser-induced breakdown spectroscopy},}\ }\href {\doibase
  10.1364/OE.22.007686} {\bibfield  {journal} {\bibinfo  {journal} {Opt.
  Express}\ }\textbf {\bibinfo {volume} {22}},\ \bibinfo {pages} {7686--7693}
  (\bibinfo {year} {2014})}\BibitemShut {NoStop}%
\bibitem [{\citenamefont {Liu}, \citenamefont {Baudelet},\ and\ \citenamefont
  {Richardson}(2010)}]{Liu_2010}%
  \BibitemOpen
  \bibfield  {author} {\bibinfo {author} {\bibfnamefont {Y.}~\bibnamefont
  {Liu}}, \bibinfo {author} {\bibfnamefont {M.}~\bibnamefont {Baudelet}}, \
  and\ \bibinfo {author} {\bibfnamefont {M.}~\bibnamefont {Richardson}},\
  }\bibfield  {title} {\enquote {\bibinfo {title} {Elemental analysis by
  microwave-assisted laser-induced breakdown spectroscopy: Evaluation on
  ceramics},}\ }\href {\doibase 10.1039/C003304A} {\bibfield  {journal}
  {\bibinfo  {journal} {J. Anal. At. Spectrom.}\ }\textbf {\bibinfo {volume}
  {25}},\ \bibinfo {pages} {1316--1323} (\bibinfo {year} {2010})}\BibitemShut
  {NoStop}%
\bibitem [{\citenamefont {Oba}\ \emph {et~al.}(2020)\citenamefont {Oba},
  \citenamefont {Miyabe}, \citenamefont {Akaoka},\ and\ \citenamefont
  {Wakaida}}]{Oba_2020}%
  \BibitemOpen
  \bibfield  {author} {\bibinfo {author} {\bibfnamefont {M.}~\bibnamefont
  {Oba}}, \bibinfo {author} {\bibfnamefont {M.}~\bibnamefont {Miyabe}},
  \bibinfo {author} {\bibfnamefont {K.}~\bibnamefont {Akaoka}}, \ and\ \bibinfo
  {author} {\bibfnamefont {I.}~\bibnamefont {Wakaida}},\ }\bibfield  {title}
  {\enquote {\bibinfo {title} {Development of microwave-assisted, laser-induced
  breakdown spectroscopy without a microwave cavity or waveguide},}\ }\href
  {\doibase 10.35848/1347-4065/ab8b3f} {\bibfield  {journal} {\bibinfo
  {journal} {Japanese Journal of Applied Physics}\ }\textbf {\bibinfo {volume}
  {59}},\ \bibinfo {pages} {062001} (\bibinfo {year} {2020})}\BibitemShut
  {NoStop}%
\bibitem [{\citenamefont {Ikeda}\ and\ \citenamefont
  {Soriano}(2022)}]{Ikeda_22}%
  \BibitemOpen
  \bibfield  {author} {\bibinfo {author} {\bibfnamefont {Y.}~\bibnamefont
  {Ikeda}}\ and\ \bibinfo {author} {\bibfnamefont {J.~K.}\ \bibnamefont
  {Soriano}},\ }\bibfield  {title} {\enquote {\bibinfo {title}
  {Microwave-enhanced laser-induced air plasma at atmospheric pressure},}\
  }\href {\doibase 10.1364/OE.470072} {\bibfield  {journal} {\bibinfo
  {journal} {Opt. Express}\ }\textbf {\bibinfo {volume} {30}},\ \bibinfo
  {pages} {33756--33766} (\bibinfo {year} {2022})}\BibitemShut {NoStop}%
\bibitem [{\citenamefont {Yu}\ \emph {et~al.}(2019)\citenamefont {Yu},
  \citenamefont {Yao}, \citenamefont {Zhang}, \citenamefont {Lu}, \citenamefont
  {Lie},\ and\ \citenamefont {Lu}}]{C8JA00347E}%
  \BibitemOpen
  \bibfield  {author} {\bibinfo {author} {\bibfnamefont {Z.}~\bibnamefont
  {Yu}}, \bibinfo {author} {\bibfnamefont {S.}~\bibnamefont {Yao}}, \bibinfo
  {author} {\bibfnamefont {L.}~\bibnamefont {Zhang}}, \bibinfo {author}
  {\bibfnamefont {Z.}~\bibnamefont {Lu}}, \bibinfo {author} {\bibfnamefont
  {Z.~S.}\ \bibnamefont {Lie}}, \ and\ \bibinfo {author} {\bibfnamefont
  {J.}~\bibnamefont {Lu}},\ }\bibfield  {title} {\enquote {\bibinfo {title}
  {Surface-enhanced laser-induced breakdown spectroscopy utilizing metallic
  target for direct analysis of particle flow},}\ }\href {\doibase
  10.1039/C8JA00347E} {\bibfield  {journal} {\bibinfo  {journal} {J. Anal. At.
  Spectrom.}\ }\textbf {\bibinfo {volume} {34}},\ \bibinfo {pages} {172--179}
  (\bibinfo {year} {2019})}\BibitemShut {NoStop}%
\bibitem [{\citenamefont {Yang}\ \emph {et~al.}(2020)\citenamefont {Yang},
  \citenamefont {Li}, \citenamefont {Cui}, \citenamefont {Yao}, \citenamefont
  {Zhou},\ and\ \citenamefont {Li}}]{Yang2020}%
  \BibitemOpen
  \bibfield  {author} {\bibinfo {author} {\bibfnamefont {X.}~\bibnamefont
  {Yang}}, \bibinfo {author} {\bibfnamefont {X.}~\bibnamefont {Li}}, \bibinfo
  {author} {\bibfnamefont {Z.}~\bibnamefont {Cui}}, \bibinfo {author}
  {\bibfnamefont {G.}~\bibnamefont {Yao}}, \bibinfo {author} {\bibfnamefont
  {Z.}~\bibnamefont {Zhou}}, \ and\ \bibinfo {author} {\bibfnamefont
  {K.}~\bibnamefont {Li}},\ }\bibfield  {title} {\enquote {\bibinfo {title}
  {Improving the sensitivity of surface-enhanced laser-induced breakdown
  spectroscopy by repeating sample preparation},}\ }\href {\doibase
  10.3389/fphy.2020.00194} {\bibfield  {journal} {\bibinfo  {journal}
  {Frontiers in Physics}\ }\textbf {\bibinfo {volume} {8}} (\bibinfo {year}
  {2020}),\ 10.3389/fphy.2020.00194},\ \bibinfo {note} {cited By 5}\BibitemShut
  {NoStop}%
\bibitem [{\citenamefont {Elhamdaoui}\ \emph {et~al.}(2022)\citenamefont
  {Elhamdaoui}, \citenamefont {Mohamed}, \citenamefont {Selmani}, \citenamefont
  {Bouchard}, \citenamefont {Sabsabi}, \citenamefont {Constantin},\ and\
  \citenamefont {Vidal}}]{D2JA00120A}%
  \BibitemOpen
  \bibfield  {author} {\bibinfo {author} {\bibfnamefont {I.}~\bibnamefont
  {Elhamdaoui}}, \bibinfo {author} {\bibfnamefont {N.}~\bibnamefont {Mohamed}},
  \bibinfo {author} {\bibfnamefont {S.}~\bibnamefont {Selmani}}, \bibinfo
  {author} {\bibfnamefont {P.}~\bibnamefont {Bouchard}}, \bibinfo {author}
  {\bibfnamefont {M.}~\bibnamefont {Sabsabi}}, \bibinfo {author} {\bibfnamefont
  {M.}~\bibnamefont {Constantin}}, \ and\ \bibinfo {author} {\bibfnamefont
  {F.}~\bibnamefont {Vidal}},\ }\bibfield  {title} {\enquote {\bibinfo {title}
  {Rapid quantitative analysis of palladium in ores using laser-induced
  breakdown spectroscopy assisted with laser-induced fluorescence
  (libs-lif)},}\ }\href {\doibase 10.1039/D2JA00120A} {\bibfield  {journal}
  {\bibinfo  {journal} {J. Anal. At. Spectrom.}\ ,\ \bibinfo {pages} {--}}
  (\bibinfo {year} {2022})}\BibitemShut {NoStop}%
\bibitem [{\citenamefont {Zhou}\ \emph {et~al.}(2021)\citenamefont {Zhou},
  \citenamefont {Liu}, \citenamefont {Tang}, \citenamefont {Gao}, \citenamefont
  {Yan},\ and\ \citenamefont {Li}}]{Zhou21_LIBS_LIF}%
  \BibitemOpen
  \bibfield  {author} {\bibinfo {author} {\bibfnamefont {R.}~\bibnamefont
  {Zhou}}, \bibinfo {author} {\bibfnamefont {K.}~\bibnamefont {Liu}}, \bibinfo
  {author} {\bibfnamefont {Z.}~\bibnamefont {Tang}}, \bibinfo {author}
  {\bibfnamefont {P.}~\bibnamefont {Gao}}, \bibinfo {author} {\bibfnamefont
  {J.}~\bibnamefont {Yan}}, \ and\ \bibinfo {author} {\bibfnamefont
  {X.}~\bibnamefont {Li}},\ }\bibfield  {title} {\enquote {\bibinfo {title}
  {High-sensitivity determination of available cobalt in soil using
  laser-induced breakdown spectroscopy assisted with laser-induced
  fluorescence},}\ }\href {\doibase 10.1364/AO.433538} {\bibfield  {journal}
  {\bibinfo  {journal} {Appl. Opt.}\ }\textbf {\bibinfo {volume} {60}},\
  \bibinfo {pages} {9062--9066} (\bibinfo {year} {2021})}\BibitemShut {NoStop}%
\bibitem [{\citenamefont {{De Giacomo}}\ \emph {et~al.}(2014)\citenamefont {{De
  Giacomo}}, \citenamefont {Gaudiuso}, \citenamefont {Koral}, \citenamefont
  {Dell'Aglio},\ and\ \citenamefont {{De Pascale}}}]{DEGIACOMO201419}%
  \BibitemOpen
  \bibfield  {author} {\bibinfo {author} {\bibfnamefont {A.}~\bibnamefont {{De
  Giacomo}}}, \bibinfo {author} {\bibfnamefont {R.}~\bibnamefont {Gaudiuso}},
  \bibinfo {author} {\bibfnamefont {C.}~\bibnamefont {Koral}}, \bibinfo
  {author} {\bibfnamefont {M.}~\bibnamefont {Dell'Aglio}}, \ and\ \bibinfo
  {author} {\bibfnamefont {O.}~\bibnamefont {{De Pascale}}},\ }\bibfield
  {title} {\enquote {\bibinfo {title} {Nanoparticle enhanced laser induced
  breakdown spectroscopy: Effect of nanoparticles deposited on sample surface
  on laser ablation and plasma emission},}\ }\href {\doibase
  https://doi.org/10.1016/j.sab.2014.05.010} {\bibfield  {journal} {\bibinfo
  {journal} {Spectrochimica Acta Part B: Atomic Spectroscopy}\ }\textbf
  {\bibinfo {volume} {98}},\ \bibinfo {pages} {19--27} (\bibinfo {year}
  {2014})}\BibitemShut {NoStop}%
\bibitem [{\citenamefont {Tang}\ \emph {et~al.}(2021)\citenamefont {Tang},
  \citenamefont {Liu}, \citenamefont {Hao}, \citenamefont {Liu}, \citenamefont
  {Zhang}, \citenamefont {Li}, \citenamefont {Zhu}, \citenamefont {Chen},\ and\
  \citenamefont {Li}}]{D0JA00528B}%
  \BibitemOpen
  \bibfield  {author} {\bibinfo {author} {\bibfnamefont {Z.}~\bibnamefont
  {Tang}}, \bibinfo {author} {\bibfnamefont {K.}~\bibnamefont {Liu}}, \bibinfo
  {author} {\bibfnamefont {Z.}~\bibnamefont {Hao}}, \bibinfo {author}
  {\bibfnamefont {K.}~\bibnamefont {Liu}}, \bibinfo {author} {\bibfnamefont
  {W.}~\bibnamefont {Zhang}}, \bibinfo {author} {\bibfnamefont
  {Q.}~\bibnamefont {Li}}, \bibinfo {author} {\bibfnamefont {C.}~\bibnamefont
  {Zhu}}, \bibinfo {author} {\bibfnamefont {J.}~\bibnamefont {Chen}}, \ and\
  \bibinfo {author} {\bibfnamefont {X.}~\bibnamefont {Li}},\ }\bibfield
  {title} {\enquote {\bibinfo {title} {The validity of nanoparticle enhanced
  molecular laser-induced breakdown spectroscopy},}\ }\href {\doibase
  10.1039/D0JA00528B} {\bibfield  {journal} {\bibinfo  {journal} {J. Anal. At.
  Spectrom.}\ }\textbf {\bibinfo {volume} {36}},\ \bibinfo {pages} {1034--1040}
  (\bibinfo {year} {2021})}\BibitemShut {NoStop}%
\bibitem [{\citenamefont {Fortes}, \citenamefont {Fernández-Bravo},\ and\
  \citenamefont {{Javier Laserna}}(2014)}]{FORTES201478}%
  \BibitemOpen
  \bibfield  {author} {\bibinfo {author} {\bibfnamefont {F.~J.}\ \bibnamefont
  {Fortes}}, \bibinfo {author} {\bibfnamefont {A.}~\bibnamefont
  {Fernández-Bravo}}, \ and\ \bibinfo {author} {\bibfnamefont
  {J.}~\bibnamefont {{Javier Laserna}}},\ }\bibfield  {title} {\enquote
  {\bibinfo {title} {Chemical characterization of single micro- and
  nano-particles by optical catapulting–optical trapping–laser-induced
  breakdown spectroscopy},}\ }\href {\doibase
  https://doi.org/10.1016/j.sab.2014.08.023} {\bibfield  {journal} {\bibinfo
  {journal} {Spectrochimica Acta Part B: Atomic Spectroscopy}\ }\textbf
  {\bibinfo {volume} {100}},\ \bibinfo {pages} {78--85} (\bibinfo {year}
  {2014})},\ \bibinfo {note} {dedicated To Nicolo Omenetto On the Occasion of
  his 75th Birthday}\BibitemShut {NoStop}%
\bibitem [{\citenamefont {Purohit}, \citenamefont {Fortes},\ and\ \citenamefont
  {Laserna}(2021)}]{acs.analchem.0c04827}%
  \BibitemOpen
  \bibfield  {author} {\bibinfo {author} {\bibfnamefont {P.}~\bibnamefont
  {Purohit}}, \bibinfo {author} {\bibfnamefont {F.~J.}\ \bibnamefont {Fortes}},
  \ and\ \bibinfo {author} {\bibfnamefont {J.~J.}\ \bibnamefont {Laserna}},\
  }\bibfield  {title} {\enquote {\bibinfo {title} {Optical trapping as a
  morphologically selective tool for in situ libs elemental characterization of
  single nanoparticles generated by laser ablation of bulk targets in air},}\
  }\href {\doibase 10.1021/acs.analchem.0c04827} {\bibfield  {journal}
  {\bibinfo  {journal} {Analytical Chemistry}\ }\textbf {\bibinfo {volume}
  {93}},\ \bibinfo {pages} {2635--2643} (\bibinfo {year} {2021})},\ \bibinfo
  {note} {pMID: 33400487},\ \Eprint
  {http://arxiv.org/abs/https://doi.org/10.1021/acs.analchem.0c04827}
  {https://doi.org/10.1021/acs.analchem.0c04827} \BibitemShut {NoStop}%
\bibitem [{\citenamefont {Purohit}, \citenamefont {Fortes},\ and\ \citenamefont
  {Laserna}(2017)}]{PUROHIT201775}%
  \BibitemOpen
  \bibfield  {author} {\bibinfo {author} {\bibfnamefont {P.}~\bibnamefont
  {Purohit}}, \bibinfo {author} {\bibfnamefont {F.~J.}\ \bibnamefont {Fortes}},
  \ and\ \bibinfo {author} {\bibfnamefont {J.~J.}\ \bibnamefont {Laserna}},\
  }\bibfield  {title} {\enquote {\bibinfo {title} {Atomization efficiency and
  photon yield in laser-induced breakdown spectroscopy analysis of single
  nanoparticles in an optical trap},}\ }\href {\doibase
  https://doi.org/10.1016/j.sab.2017.02.009} {\bibfield  {journal} {\bibinfo
  {journal} {Spectrochimica Acta Part B: Atomic Spectroscopy}\ }\textbf
  {\bibinfo {volume} {130}},\ \bibinfo {pages} {75--81} (\bibinfo {year}
  {2017})}\BibitemShut {NoStop}%
\bibitem [{\citenamefont {Wubetu}\ \emph
  {et~al.}(2017{\natexlab{a}})\citenamefont {Wubetu}, \citenamefont
  {Fiedorowicz}, \citenamefont {Costello},\ and\ \citenamefont
  {Kelly}}]{POP_anisotropy}%
  \BibitemOpen
  \bibfield  {author} {\bibinfo {author} {\bibfnamefont {G.~A.}\ \bibnamefont
  {Wubetu}}, \bibinfo {author} {\bibfnamefont {H.}~\bibnamefont {Fiedorowicz}},
  \bibinfo {author} {\bibfnamefont {J.~T.}\ \bibnamefont {Costello}}, \ and\
  \bibinfo {author} {\bibfnamefont {T.~J.}\ \bibnamefont {Kelly}},\ }\bibfield
  {title} {\enquote {\bibinfo {title} {Time resolved anisotropic emission from
  an aluminium laser produced plasma},}\ }\href {\doibase 10.1063/1.4973444}
  {\bibfield  {journal} {\bibinfo  {journal} {Physics of Plasmas}\ }\textbf
  {\bibinfo {volume} {24}},\ \bibinfo {pages} {013105} (\bibinfo {year}
  {2017}{\natexlab{a}})},\ \Eprint
  {http://arxiv.org/abs/https://doi.org/10.1063/1.4973444}
  {https://doi.org/10.1063/1.4973444} \BibitemShut {NoStop}%
\bibitem [{\citenamefont {Aghababaei~Nejad}, \citenamefont {Soltanolkotabi},\
  and\ \citenamefont {Eslami~Majd}(2018)}]{JLA_Aghaba}%
  \BibitemOpen
  \bibfield  {author} {\bibinfo {author} {\bibfnamefont {M.}~\bibnamefont
  {Aghababaei~Nejad}}, \bibinfo {author} {\bibfnamefont {M.}~\bibnamefont
  {Soltanolkotabi}}, \ and\ \bibinfo {author} {\bibfnamefont {A.}~\bibnamefont
  {Eslami~Majd}},\ }\bibfield  {title} {\enquote {\bibinfo {title}
  {Polarization investigation of laser-induced breakdown plasma emission from
  al, cu, mo, w, and pb elements using nongated detector},}\ }\href {\doibase
  10.2351/1.5012507} {\bibfield  {journal} {\bibinfo  {journal} {Journal of
  Laser Applications}\ }\textbf {\bibinfo {volume} {30}},\ \bibinfo {pages}
  {022005} (\bibinfo {year} {2018})},\ \Eprint
  {http://arxiv.org/abs/https://doi.org/10.2351/1.5012507}
  {https://doi.org/10.2351/1.5012507} \BibitemShut {NoStop}%
\bibitem [{\citenamefont {Wubetu}\ \emph {et~al.}(2020)\citenamefont {Wubetu},
  \citenamefont {Kelly}, \citenamefont {Hayden}, \citenamefont {Fiedorowicz},
  \citenamefont {Skrzeczanowski},\ and\ \citenamefont
  {Costello}}]{Wubetu_2020}%
  \BibitemOpen
  \bibfield  {author} {\bibinfo {author} {\bibfnamefont {G.~A.}\ \bibnamefont
  {Wubetu}}, \bibinfo {author} {\bibfnamefont {T.~J.}\ \bibnamefont {Kelly}},
  \bibinfo {author} {\bibfnamefont {P.}~\bibnamefont {Hayden}}, \bibinfo
  {author} {\bibfnamefont {H.}~\bibnamefont {Fiedorowicz}}, \bibinfo {author}
  {\bibfnamefont {W.}~\bibnamefont {Skrzeczanowski}}, \ and\ \bibinfo {author}
  {\bibfnamefont {J.~T.}\ \bibnamefont {Costello}},\ }\bibfield  {title}
  {\enquote {\bibinfo {title} {Recombination contributions to the anisotropic
  emission from a laser produced copper plasma},}\ }\href {\doibase
  10.1088/1361-6455/ab66d3} {\bibfield  {journal} {\bibinfo  {journal} {Journal
  of Physics B: Atomic, Molecular and Optical Physics}\ }\textbf {\bibinfo
  {volume} {53}},\ \bibinfo {pages} {065701} (\bibinfo {year}
  {2020})}\BibitemShut {NoStop}%
\bibitem [{\citenamefont {Sharma}\ and\ \citenamefont
  {Thareja}(2007)}]{SHARMA20073113}%
  \BibitemOpen
  \bibfield  {author} {\bibinfo {author} {\bibfnamefont {A.}~\bibnamefont
  {Sharma}}\ and\ \bibinfo {author} {\bibfnamefont {R.}~\bibnamefont
  {Thareja}},\ }\bibfield  {title} {\enquote {\bibinfo {title} {Anisotropic
  emission in laser-produced aluminum plasma in ambient nitrogen},}\ }\href
  {\doibase https://doi.org/10.1016/j.apsusc.2006.07.014} {\bibfield  {journal}
  {\bibinfo  {journal} {Applied Surface Science}\ }\textbf {\bibinfo {volume}
  {253}},\ \bibinfo {pages} {3113--3121} (\bibinfo {year} {2007})}\BibitemShut
  {NoStop}%
\bibitem [{\citenamefont {Zhao}\ \emph {et~al.}(2014)\citenamefont {Zhao},
  \citenamefont {Farid}, \citenamefont {Hai}, \citenamefont {Wu},\ and\
  \citenamefont {Ding}}]{Zhao_2014}%
  \BibitemOpen
  \bibfield  {author} {\bibinfo {author} {\bibfnamefont {D.}~\bibnamefont
  {Zhao}}, \bibinfo {author} {\bibfnamefont {N.}~\bibnamefont {Farid}},
  \bibinfo {author} {\bibfnamefont {R.}~\bibnamefont {Hai}}, \bibinfo {author}
  {\bibfnamefont {D.}~\bibnamefont {Wu}}, \ and\ \bibinfo {author}
  {\bibfnamefont {H.}~\bibnamefont {Ding}},\ }\bibfield  {title} {\enquote
  {\bibinfo {title} {Diagnostics of first wall materials in a magnetically
  confined fusion device by polarization-resolved laser-induced breakdown
  spectroscopy},}\ }\href {\doibase 10.1088/1009-0630/16/2/11} {\bibfield
  {journal} {\bibinfo  {journal} {Plasma Science and Technology}\ }\textbf
  {\bibinfo {volume} {16}},\ \bibinfo {pages} {149--154} (\bibinfo {year}
  {2014})}\BibitemShut {NoStop}%
\bibitem [{\citenamefont {Wubetu}\ \emph
  {et~al.}(2017{\natexlab{b}})\citenamefont {Wubetu}, \citenamefont
  {Fiedorowicz}, \citenamefont {Costello},\ and\ \citenamefont
  {Kelly}}]{Anisotropy_2017}%
  \BibitemOpen
  \bibfield  {author} {\bibinfo {author} {\bibfnamefont {G.~A.}\ \bibnamefont
  {Wubetu}}, \bibinfo {author} {\bibfnamefont {H.}~\bibnamefont {Fiedorowicz}},
  \bibinfo {author} {\bibfnamefont {J.~T.}\ \bibnamefont {Costello}}, \ and\
  \bibinfo {author} {\bibfnamefont {T.~J.}\ \bibnamefont {Kelly}},\ }\bibfield
  {title} {\enquote {\bibinfo {title} {Time resolved anisotropic emission from
  an aluminium laser produced plasma},}\ }\href {\doibase 10.1063/1.4973444}
  {\bibfield  {journal} {\bibinfo  {journal} {Physics of Plasmas}\ }\textbf
  {\bibinfo {volume} {24}},\ \bibinfo {pages} {013105} (\bibinfo {year}
  {2017}{\natexlab{b}})},\ \Eprint
  {http://arxiv.org/abs/https://doi.org/10.1063/1.4973444}
  {https://doi.org/10.1063/1.4973444} \BibitemShut {NoStop}%
\bibitem [{\citenamefont {Rifai}\ \emph {et~al.}(2013)\citenamefont {Rifai},
  \citenamefont {Vidal}, \citenamefont {Chaker},\ and\ \citenamefont
  {Sabsabi}}]{C3JA30308J}%
  \BibitemOpen
  \bibfield  {author} {\bibinfo {author} {\bibfnamefont {K.}~\bibnamefont
  {Rifai}}, \bibinfo {author} {\bibfnamefont {F.}~\bibnamefont {Vidal}},
  \bibinfo {author} {\bibfnamefont {M.}~\bibnamefont {Chaker}}, \ and\ \bibinfo
  {author} {\bibfnamefont {M.}~\bibnamefont {Sabsabi}},\ }\bibfield  {title}
  {\enquote {\bibinfo {title} {Resonant laser-induced breakdown spectroscopy
  (rlibs) analysis of traces through selective excitation of aluminum in
  aluminum alloys},}\ }\href {\doibase 10.1039/C3JA30308J} {\bibfield
  {journal} {\bibinfo  {journal} {J. Anal. At. Spectrom.}\ }\textbf {\bibinfo
  {volume} {28}},\ \bibinfo {pages} {388--395} (\bibinfo {year}
  {2013})}\BibitemShut {NoStop}%
\bibitem [{\citenamefont {Liu}\ \emph {et~al.}(2021)\citenamefont {Liu},
  \citenamefont {Tang}, \citenamefont {Zhou}, \citenamefont {Zhang},
  \citenamefont {Li}, \citenamefont {Zhu}, \citenamefont {He}, \citenamefont
  {Liu},\ and\ \citenamefont {Li}}]{D1JA00250C}%
  \BibitemOpen
  \bibfield  {author} {\bibinfo {author} {\bibfnamefont {K.}~\bibnamefont
  {Liu}}, \bibinfo {author} {\bibfnamefont {Z.}~\bibnamefont {Tang}}, \bibinfo
  {author} {\bibfnamefont {R.}~\bibnamefont {Zhou}}, \bibinfo {author}
  {\bibfnamefont {W.}~\bibnamefont {Zhang}}, \bibinfo {author} {\bibfnamefont
  {Q.}~\bibnamefont {Li}}, \bibinfo {author} {\bibfnamefont {C.}~\bibnamefont
  {Zhu}}, \bibinfo {author} {\bibfnamefont {C.}~\bibnamefont {He}}, \bibinfo
  {author} {\bibfnamefont {K.}~\bibnamefont {Liu}}, \ and\ \bibinfo {author}
  {\bibfnamefont {X.}~\bibnamefont {Li}},\ }\bibfield  {title} {\enquote
  {\bibinfo {title} {Determination of lead in aqueous solutions using resonant
  surface-enhanced libs},}\ }\href {\doibase 10.1039/D1JA00250C} {\bibfield
  {journal} {\bibinfo  {journal} {J. Anal. At. Spectrom.}\ }\textbf {\bibinfo
  {volume} {36}},\ \bibinfo {pages} {2480--2484} (\bibinfo {year}
  {2021})}\BibitemShut {NoStop}%
\bibitem [{\citenamefont {Abdel-Harith}\ \emph {et~al.}(2021)\citenamefont
  {Abdel-Harith}, \citenamefont {Elhassan}, \citenamefont {Abdel-Salam},\ and\
  \citenamefont {Ali}}]{ABDELHARITH2021339024}%
  \BibitemOpen
  \bibfield  {author} {\bibinfo {author} {\bibfnamefont {M.}~\bibnamefont
  {Abdel-Harith}}, \bibinfo {author} {\bibfnamefont {A.}~\bibnamefont
  {Elhassan}}, \bibinfo {author} {\bibfnamefont {Z.}~\bibnamefont
  {Abdel-Salam}}, \ and\ \bibinfo {author} {\bibfnamefont {M.~F.}\ \bibnamefont
  {Ali}},\ }\bibfield  {title} {\enquote {\bibinfo {title}
  {Back-reflection-enhanced laser-induced breakdown spectroscopy (brelibs) on
  transparent materials: Application on archaeological glass},}\ }\href
  {\doibase https://doi.org/10.1016/j.aca.2021.339024} {\bibfield  {journal}
  {\bibinfo  {journal} {Analytica Chimica Acta}\ }\textbf {\bibinfo {volume}
  {1184}},\ \bibinfo {pages} {339024} (\bibinfo {year} {2021})}\BibitemShut
  {NoStop}%
\bibitem [{\citenamefont {Sharma}\ \emph {et~al.}(2007)\citenamefont {Sharma},
  \citenamefont {Misra}, \citenamefont {Lucey}, \citenamefont {Wiens},\ and\
  \citenamefont {Clegg}}]{SHARMA20071036}%
  \BibitemOpen
  \bibfield  {author} {\bibinfo {author} {\bibfnamefont {S.}~\bibnamefont
  {Sharma}}, \bibinfo {author} {\bibfnamefont {A.}~\bibnamefont {Misra}},
  \bibinfo {author} {\bibfnamefont {P.}~\bibnamefont {Lucey}}, \bibinfo
  {author} {\bibfnamefont {R.}~\bibnamefont {Wiens}}, \ and\ \bibinfo {author}
  {\bibfnamefont {S.}~\bibnamefont {Clegg}},\ }\bibfield  {title} {\enquote
  {\bibinfo {title} {Combined remote libs and raman spectroscopy at 8.6m of
  sulfur-containing minerals, and minerals coated with hematite or covered with
  basaltic dust},}\ }\href {\doibase https://doi.org/10.1016/j.saa.2007.06.046}
  {\bibfield  {journal} {\bibinfo  {journal} {Spectrochimica Acta Part A:
  Molecular and Biomolecular Spectroscopy}\ }\textbf {\bibinfo {volume} {68}},\
  \bibinfo {pages} {1036--1045} (\bibinfo {year} {2007})},\ \bibinfo {note}
  {seventh International Conference on Raman Spectroscopy Applied to the Earth
  and Planetary Sciences}\BibitemShut {NoStop}%
\bibitem [{\citenamefont {Clegg}\ \emph {et~al.}(2014)\citenamefont {Clegg},
  \citenamefont {Wiens}, \citenamefont {Misra}, \citenamefont {Sharma},
  \citenamefont {Lambert}, \citenamefont {Bender}, \citenamefont {Newell},
  \citenamefont {Nowak-Lovato}, \citenamefont {Smrekar}, \citenamefont {Dyar},\
  and\ \citenamefont {Maurice}}]{Samuel_appl_spec}%
  \BibitemOpen
  \bibfield  {author} {\bibinfo {author} {\bibfnamefont {S.~M.}\ \bibnamefont
  {Clegg}}, \bibinfo {author} {\bibfnamefont {R.}~\bibnamefont {Wiens}},
  \bibinfo {author} {\bibfnamefont {A.~K.}\ \bibnamefont {Misra}}, \bibinfo
  {author} {\bibfnamefont {S.~K.}\ \bibnamefont {Sharma}}, \bibinfo {author}
  {\bibfnamefont {J.}~\bibnamefont {Lambert}}, \bibinfo {author} {\bibfnamefont
  {S.}~\bibnamefont {Bender}}, \bibinfo {author} {\bibfnamefont
  {R.}~\bibnamefont {Newell}}, \bibinfo {author} {\bibfnamefont
  {K.}~\bibnamefont {Nowak-Lovato}}, \bibinfo {author} {\bibfnamefont
  {S.}~\bibnamefont {Smrekar}}, \bibinfo {author} {\bibfnamefont {M.~D.}\
  \bibnamefont {Dyar}}, \ and\ \bibinfo {author} {\bibfnamefont
  {S.}~\bibnamefont {Maurice}},\ }\bibfield  {title} {\enquote {\bibinfo
  {title} {Planetary geochemical investigations using raman and laser-induced
  breakdown spectroscopy},}\ }\href {\doibase 10.1366/13-07386} {\bibfield
  {journal} {\bibinfo  {journal} {Applied Spectroscopy}\ }\textbf {\bibinfo
  {volume} {68}},\ \bibinfo {pages} {925--936} (\bibinfo {year} {2014})},\
  \bibinfo {note} {pMID: 25226246},\ \Eprint
  {http://arxiv.org/abs/https://doi.org/10.1366/13-07386}
  {https://doi.org/10.1366/13-07386} \BibitemShut {NoStop}%
\bibitem [{\citenamefont {Lin}\ \emph {et~al.}(2013)\citenamefont {Lin},
  \citenamefont {Niu}, \citenamefont {Wang}, \citenamefont {Yu},\ and\
  \citenamefont {Duan}}]{Qingyu_ASR}%
  \BibitemOpen
  \bibfield  {author} {\bibinfo {author} {\bibfnamefont {Q.}~\bibnamefont
  {Lin}}, \bibinfo {author} {\bibfnamefont {G.}~\bibnamefont {Niu}}, \bibinfo
  {author} {\bibfnamefont {Q.}~\bibnamefont {Wang}}, \bibinfo {author}
  {\bibfnamefont {Q.}~\bibnamefont {Yu}}, \ and\ \bibinfo {author}
  {\bibfnamefont {Y.}~\bibnamefont {Duan}},\ }\bibfield  {title} {\enquote
  {\bibinfo {title} {Combined laser-induced breakdown with raman spectroscopy:
  Historical technology development and recent applications},}\ }\href
  {\doibase 10.1080/05704928.2012.751028} {\bibfield  {journal} {\bibinfo
  {journal} {Applied Spectroscopy Reviews}\ }\textbf {\bibinfo {volume} {48}},\
  \bibinfo {pages} {487--508} (\bibinfo {year} {2013})},\ \Eprint
  {http://arxiv.org/abs/https://doi.org/10.1080/05704928.2012.751028}
  {https://doi.org/10.1080/05704928.2012.751028} \BibitemShut {NoStop}%
\bibitem [{\citenamefont {S}\ \emph {et~al.}(2021{\natexlab{b}})\citenamefont
  {S}, \citenamefont {George}, \citenamefont {Kartha}, \citenamefont
  {Chidangil},\ and\ \citenamefont {K}}]{ASR_hybrid_LIBS_RAMAN}%
  \BibitemOpen
  \bibfield  {author} {\bibinfo {author} {\bibfnamefont {D.~V.}\ \bibnamefont
  {S}}, \bibinfo {author} {\bibfnamefont {S.~D.}\ \bibnamefont {George}},
  \bibinfo {author} {\bibfnamefont {V.~B.}\ \bibnamefont {Kartha}}, \bibinfo
  {author} {\bibfnamefont {S.}~\bibnamefont {Chidangil}}, \ and\ \bibinfo
  {author} {\bibfnamefont {U.~V.}\ \bibnamefont {K}},\ }\bibfield  {title}
  {\enquote {\bibinfo {title} {Hybrid libs-raman-lif systems for multi-modal
  spectroscopic applications: a topical review},}\ }\href {\doibase
  10.1080/05704928.2020.1800486} {\bibfield  {journal} {\bibinfo  {journal}
  {Applied Spectroscopy Reviews}\ }\textbf {\bibinfo {volume} {56}},\ \bibinfo
  {pages} {463--491} (\bibinfo {year} {2021}{\natexlab{b}})},\ \Eprint
  {http://arxiv.org/abs/https://doi.org/10.1080/05704928.2020.1800486}
  {https://doi.org/10.1080/05704928.2020.1800486} \BibitemShut {NoStop}%
\bibitem [{\citenamefont {Lednev}\ \emph {et~al.}(2018)\citenamefont {Lednev},
  \citenamefont {Pershin}, \citenamefont {Sdvizhenskii}, \citenamefont
  {Grishin}, \citenamefont {Fedorov}, \citenamefont {Bukin}, \citenamefont
  {Oshurko},\ and\ \citenamefont {Shchegolikhin}}]{Lednev2018}%
  \BibitemOpen
  \bibfield  {author} {\bibinfo {author} {\bibfnamefont {V.~N.}\ \bibnamefont
  {Lednev}}, \bibinfo {author} {\bibfnamefont {S.~M.}\ \bibnamefont {Pershin}},
  \bibinfo {author} {\bibfnamefont {P.~A.}\ \bibnamefont {Sdvizhenskii}},
  \bibinfo {author} {\bibfnamefont {M.~Y.}\ \bibnamefont {Grishin}}, \bibinfo
  {author} {\bibfnamefont {A.~N.}\ \bibnamefont {Fedorov}}, \bibinfo {author}
  {\bibfnamefont {V.~V.}\ \bibnamefont {Bukin}}, \bibinfo {author}
  {\bibfnamefont {V.~B.}\ \bibnamefont {Oshurko}}, \ and\ \bibinfo {author}
  {\bibfnamefont {A.~N.}\ \bibnamefont {Shchegolikhin}},\ }\bibfield  {title}
  {\enquote {\bibinfo {title} {Combining raman and laser induced breakdown
  spectroscopy by double pulse lasing},}\ }\href {\doibase
  10.1007/s00216-017-0719-6} {\bibfield  {journal} {\bibinfo  {journal}
  {Analytical and Bioanalytical Chemistry}\ }\textbf {\bibinfo {volume}
  {410}},\ \bibinfo {pages} {277--286} (\bibinfo {year} {2018})}\BibitemShut
  {NoStop}%
\bibitem [{\citenamefont {{Muhammed Shameem}}\ \emph
  {et~al.}(2022)\citenamefont {{Muhammed Shameem}}, \citenamefont {Dhanada},
  \citenamefont {George}, \citenamefont {Kartha}, \citenamefont {Santhosh},\
  and\ \citenamefont {Unnikrishnan}}]{MUHAMMEDSHAMEEM2022108264}%
  \BibitemOpen
  \bibfield  {author} {\bibinfo {author} {\bibfnamefont {K.}~\bibnamefont
  {{Muhammed Shameem}}}, \bibinfo {author} {\bibfnamefont {V.}~\bibnamefont
  {Dhanada}}, \bibinfo {author} {\bibfnamefont {S.~D.}\ \bibnamefont {George}},
  \bibinfo {author} {\bibfnamefont {V.}~\bibnamefont {Kartha}}, \bibinfo
  {author} {\bibfnamefont {C.}~\bibnamefont {Santhosh}}, \ and\ \bibinfo
  {author} {\bibfnamefont {V.}~\bibnamefont {Unnikrishnan}},\ }\bibfield
  {title} {\enquote {\bibinfo {title} {Assessing the feasibility of a
  low-throughput gated echelle spectrograph for laser-induced breakdown
  spectroscopy (libs)-raman measurements at standoff distances},}\ }\href
  {\doibase https://doi.org/10.1016/j.optlastec.2022.108264} {\bibfield
  {journal} {\bibinfo  {journal} {Optics \& Laser Technology}\ }\textbf
  {\bibinfo {volume} {153}},\ \bibinfo {pages} {108264} (\bibinfo {year}
  {2022})}\BibitemShut {NoStop}%
\bibitem [{\citenamefont {Meissner}\ \emph {et~al.}(2004)\citenamefont
  {Meissner}, \citenamefont {Lippert}, \citenamefont {Wokaun},\ and\
  \citenamefont {Guenther}}]{MEISSNER2004316}%
  \BibitemOpen
  \bibfield  {author} {\bibinfo {author} {\bibfnamefont {K.}~\bibnamefont
  {Meissner}}, \bibinfo {author} {\bibfnamefont {T.}~\bibnamefont {Lippert}},
  \bibinfo {author} {\bibfnamefont {A.}~\bibnamefont {Wokaun}}, \ and\ \bibinfo
  {author} {\bibfnamefont {D.}~\bibnamefont {Guenther}},\ }\bibfield  {title}
  {\enquote {\bibinfo {title} {Analysis of trace metals in comparison of
  laser-induced breakdown spectroscopy with la-icp-ms},}\ }\href {\doibase
  https://doi.org/10.1016/j.tsf.2003.11.174} {\bibfield  {journal} {\bibinfo
  {journal} {Thin Solid Films}\ }\textbf {\bibinfo {volume} {453-454}},\
  \bibinfo {pages} {316--322} (\bibinfo {year} {2004})},\ \bibinfo {note}
  {proceedings of Symposium H on Photonic Processing of Surfaces, Thin Films
  and Devices, of the E-MRS 2003 Spring Conference}\BibitemShut {NoStop}%
\bibitem [{\citenamefont {Oropeza}\ \emph {et~al.}(2019)\citenamefont
  {Oropeza}, \citenamefont {Gonz\'{a}lez}, \citenamefont {Chirinos},
  \citenamefont {Zorba}, \citenamefont {Rogel}, \citenamefont {Ovalles},\ and\
  \citenamefont {L\'{o}pez-Linares}}]{Asphaltenes_LIBS}%
  \BibitemOpen
  \bibfield  {author} {\bibinfo {author} {\bibfnamefont {D.}~\bibnamefont
  {Oropeza}}, \bibinfo {author} {\bibfnamefont {J.}~\bibnamefont
  {Gonz\'{a}lez}}, \bibinfo {author} {\bibfnamefont {J.}~\bibnamefont
  {Chirinos}}, \bibinfo {author} {\bibfnamefont {V.}~\bibnamefont {Zorba}},
  \bibinfo {author} {\bibfnamefont {E.}~\bibnamefont {Rogel}}, \bibinfo
  {author} {\bibfnamefont {C.}~\bibnamefont {Ovalles}}, \ and\ \bibinfo
  {author} {\bibfnamefont {F.}~\bibnamefont {L\'{o}pez-Linares}},\ }\bibfield
  {title} {\enquote {\bibinfo {title} {Elemental analysis of asphaltenes using
  simultaneous laser-induced breakdown spectroscopy (libs)\&\#x2013;laser
  ablation inductively coupled plasma optical emission spectrometry
  (la-icp-oes)},}\ }\href {\doibase 10.1364/AS.73.000540} {\bibfield  {journal}
  {\bibinfo  {journal} {Appl. Spectrosc.}\ }\textbf {\bibinfo {volume} {73}},\
  \bibinfo {pages} {540--549} (\bibinfo {year} {2019})}\BibitemShut {NoStop}%
\bibitem [{\citenamefont {Chirinos}\ \emph {et~al.}(2014)\citenamefont
  {Chirinos}, \citenamefont {Oropeza}, \citenamefont {Gonzalez}, \citenamefont
  {Hou}, \citenamefont {Morey}, \citenamefont {Zorba},\ and\ \citenamefont
  {Russo}}]{LA-ICP-MS_JAAS}%
  \BibitemOpen
  \bibfield  {author} {\bibinfo {author} {\bibfnamefont {J.~R.}\ \bibnamefont
  {Chirinos}}, \bibinfo {author} {\bibfnamefont {D.~D.}\ \bibnamefont
  {Oropeza}}, \bibinfo {author} {\bibfnamefont {J.~J.}\ \bibnamefont
  {Gonzalez}}, \bibinfo {author} {\bibfnamefont {H.}~\bibnamefont {Hou}},
  \bibinfo {author} {\bibfnamefont {M.}~\bibnamefont {Morey}}, \bibinfo
  {author} {\bibfnamefont {V.}~\bibnamefont {Zorba}}, \ and\ \bibinfo {author}
  {\bibfnamefont {R.~E.}\ \bibnamefont {Russo}},\ }\bibfield  {title} {\enquote
  {\bibinfo {title} {Simultaneous 3-dimensional elemental imaging with libs and
  la-icp-ms},}\ }\href {\doibase 10.1039/C4JA00066H} {\bibfield  {journal}
  {\bibinfo  {journal} {J. Anal. At. Spectrom.}\ }\textbf {\bibinfo {volume}
  {29}},\ \bibinfo {pages} {1292--1298} (\bibinfo {year} {2014})}\BibitemShut
  {NoStop}%
\bibitem [{\citenamefont {Brunnbauer}\ \emph {et~al.}(2020)\citenamefont
  {Brunnbauer}, \citenamefont {Mayr}, \citenamefont {Larisegger}, \citenamefont
  {Nelhiebel}, \citenamefont {Pagnin}, \citenamefont {Wiesinger}, \citenamefont
  {Schreiner},\ and\ \citenamefont {Limbeck}}]{LA_ICP_MS_LIBS_SR}%
  \BibitemOpen
  \bibfield  {author} {\bibinfo {author} {\bibfnamefont {L.}~\bibnamefont
  {Brunnbauer}}, \bibinfo {author} {\bibfnamefont {M.}~\bibnamefont {Mayr}},
  \bibinfo {author} {\bibfnamefont {S.}~\bibnamefont {Larisegger}}, \bibinfo
  {author} {\bibfnamefont {M.}~\bibnamefont {Nelhiebel}}, \bibinfo {author}
  {\bibfnamefont {L.}~\bibnamefont {Pagnin}}, \bibinfo {author} {\bibfnamefont
  {R.}~\bibnamefont {Wiesinger}}, \bibinfo {author} {\bibfnamefont
  {M.}~\bibnamefont {Schreiner}}, \ and\ \bibinfo {author} {\bibfnamefont
  {A.}~\bibnamefont {Limbeck}},\ }\bibfield  {title} {\enquote {\bibinfo
  {title} {Combined la-icp-ms/libs: powerful analytical tools for the
  investigation of polymer alteration after treatment under corrosive
  conditions},}\ }\href {\doibase 10.1038/s41598-020-69210-9} {\bibfield
  {journal} {\bibinfo  {journal} {Scientific Reports}\ }\textbf {\bibinfo
  {volume} {10}},\ \bibinfo {pages} {12513} (\bibinfo {year}
  {2020})}\BibitemShut {NoStop}%
\bibitem [{\citenamefont {Alberghina}\ \emph {et~al.}(2009)\citenamefont
  {Alberghina}, \citenamefont {Barraco}, \citenamefont {Brai}, \citenamefont
  {Schillaci},\ and\ \citenamefont {Tranchina}}]{Spie_LIBS_XRF}%
  \BibitemOpen
  \bibfield  {author} {\bibinfo {author} {\bibfnamefont {M.~F.}\ \bibnamefont
  {Alberghina}}, \bibinfo {author} {\bibfnamefont {R.}~\bibnamefont {Barraco}},
  \bibinfo {author} {\bibfnamefont {M.}~\bibnamefont {Brai}}, \bibinfo {author}
  {\bibfnamefont {T.}~\bibnamefont {Schillaci}}, \ and\ \bibinfo {author}
  {\bibfnamefont {L.}~\bibnamefont {Tranchina}},\ }\bibfield  {title} {\enquote
  {\bibinfo {title} {{Double laser LIBS and micro-XRF spectroscopy applied to
  characterize materials coming from the Greek-Roman theater of Taormina}},}\
  }in\ \href {\doibase 10.1117/12.827777} {\emph {\bibinfo {booktitle} {O3A:
  Optics for Arts, Architecture, and Archaeology II}}},\ Vol.\ \bibinfo
  {volume} {7391},\ \bibinfo {editor} {edited by\ \bibinfo {editor}
  {\bibfnamefont {L.}~\bibnamefont {Pezzati}}\ and\ \bibinfo {editor}
  {\bibfnamefont {R.}~\bibnamefont {Salimbeni}}},\ \bibinfo {organization}
  {International Society for Optics and Photonics}\ (\bibinfo  {publisher}
  {SPIE},\ \bibinfo {year} {2009})\ p.\ \bibinfo {pages} {739107}\BibitemShut
  {NoStop}%
\bibitem [{\citenamefont {Rao}\ \emph {et~al.}(2022)\citenamefont {Rao},
  \citenamefont {Jenkins}, \citenamefont {Auxier}, \citenamefont {Shattan},\
  and\ \citenamefont {Patnaik}}]{D1JA00404B}%
  \BibitemOpen
  \bibfield  {author} {\bibinfo {author} {\bibfnamefont {A.~P.}\ \bibnamefont
  {Rao}}, \bibinfo {author} {\bibfnamefont {P.~R.}\ \bibnamefont {Jenkins}},
  \bibinfo {author} {\bibfnamefont {J.~D.}\ \bibnamefont {Auxier}}, \bibinfo
  {author} {\bibfnamefont {M.~B.}\ \bibnamefont {Shattan}}, \ and\ \bibinfo
  {author} {\bibfnamefont {A.~K.}\ \bibnamefont {Patnaik}},\ }\bibfield
  {title} {\enquote {\bibinfo {title} {Analytical comparisons of handheld libs
  and xrf devices for rapid quantification of gallium in a plutonium surrogate
  matrix},}\ }\href {\doibase 10.1039/D1JA00404B} {\bibfield  {journal}
  {\bibinfo  {journal} {J. Anal. At. Spectrom.}\ }\textbf {\bibinfo {volume}
  {37}},\ \bibinfo {pages} {1090--1098} (\bibinfo {year} {2022})}\BibitemShut
  {NoStop}%
\end{thebibliography}
 
%

\end{document}